\documentclass{article}

\usepackage{arxiv}

\usepackage[utf8]{inputenc} 
\usepackage[T1]{fontenc}    
\usepackage{hyperref}       
\usepackage{url}            
\usepackage{booktabs}       
\usepackage{amsfonts}       
\usepackage{nicefrac}       
\usepackage{microtype}      
\usepackage{lipsum}		
\usepackage{graphicx}
\usepackage{natbib}
\usepackage{doi}

\usepackage{appendix}
\usepackage{amsmath}
\usepackage{lineno}
\usepackage{array}
\usepackage{longtable}
\usepackage{physics}
\usepackage{multirow}
\setcitestyle{aysep={}}
\nolinenumbers
\usepackage{xcolor}
\definecolor{amber}{rgb}{1.0, 0.75, 0.0}
\definecolor{brickred}{rgb}{0.7960, 0.2550, 0.3290}
\definecolor{steelblue}{rgb}{0.27, 0.51, 0.71}
\definecolor{mitgrey}{rgb}{0.539, 0.543, 0.546}
\definecolor{glaucous}{rgb}{0.38, 0.51, 0.71}
\definecolor{mediumtealblue}{rgb}{0.0, 0.33, 0.71}
\definecolor{eg_green}{rgb}{0.5703125, 0.8125, 0.3125}
\definecolor{mit_red}{rgb}{0.45703125, 0.0, 0.078125}
\definecolor{royalfuchsia}{rgb}{0.79, 0.17, 0.57}
\definecolor{burntsienna}{rgb}{0.91, 0.45, 0.32}

\newcommand*\patchAmsMathEnvironmentForLineno[1]{%
\expandafter\let\csname old#1\expandafter\endcsname\csname #1\endcsname
\expandafter\let\csname oldend#1\expandafter\endcsname\csname end#1\endcsname
\renewenvironment{#1}%
{\linenomath\csname old#1\endcsname}%
{\csname oldend#1\endcsname\endlinenomath}}%
\newcommand*\patchBothAmsMathEnvironmentsForLineno[1]{%
\patchAmsMathEnvironmentForLineno{#1}%
\patchAmsMathEnvironmentForLineno{#1*}}%
\AtBeginDocument{%
\patchBothAmsMathEnvironmentsForLineno{equation}%
\patchBothAmsMathEnvironmentsForLineno{align}%
\patchBothAmsMathEnvironmentsForLineno{flalign}%
\patchBothAmsMathEnvironmentsForLineno{alignat}%
\patchBothAmsMathEnvironmentsForLineno{gather}%
\patchBothAmsMathEnvironmentsForLineno{multline}%
}

\newcommand{\subfigimgthree}[3][,]{%
  \setbox1=\hbox{\includegraphics[#1]{#3}}
  \leavevmode\rlap{\usebox1}
  \rlap{\hspace*{-10pt}\raisebox{\dimexpr\ht1-1\baselineskip}{#2}}
  \phantom{\usebox1}
}

\usepackage{tikz}

\usepackage{subcaption}
\usepackage{mathtools}
\usepackage{bm}
\usepackage[utf8]{inputenc}
\usepackage{placeins}
\usepackage[export]{adjustbox}

\title{Accelerated Bayesian calibration and uncertainty quantification of RANS turbulence model parameters for stratified atmospheric boundary layer flows}

\author{ Ethan YoungIn Shin \\
	Department of Civil and Environmental Engineering\\
	Massachusetts Institute of Technology\\
	Cambridge, MA 02139 \\
	\texttt{youngin@mit.edu} \\
	\And
	Michael F. Howland\thanks{Corresponding author}\\
	Department of Civil and Environmental Engineering\\
	Massachusetts Institute of Technology\\
	Cambridge, MA 02139 \\
	\texttt{mhowland@mit.edu} \\
}

\date{}

\hypersetup{
pdftitle={Accelerated Bayesian calibration and uncertainty quantification of RANS turbulence model parameters for stratified atmospheric boundary layer flows},
pdfsubject={Boundary layer meteorology; RANS turbulence modeling; Bayesian calibration; uncertainty quantification},
pdfauthor={Ethan Y.~Shin, Michael F.~Howland},
pdfkeywords={Bayesian inversion, Uncertainty quantification, Turbulence parameterizations, Stratified atmospheric flows},
}

\begin{document}
\maketitle

\begin{abstract}
In operational weather models, the effects of turbulence in the atmospheric boundary layer (ABL) on the resolved flow are modeled using turbulence parameterizations.
These parameterizations typically use a predetermined set of model parameters that are tuned to limited data from canonical flows.
Using these fixed parameters results in deterministic predictions that neglect uncertainty in the unresolved turbulence processes.
In this study, we perform a machine learning-accelerated Bayesian inversion of a single-column model of the ABL.
This approach is used to calibrate and quantify uncertainty in model parameters of Reynolds-averaged Navier-Stokes turbulence models.
To verify the data-driven uncertainty quantification methodology, we test in an idealized setup in which a prescribed but unobserved set of parameters is learned from noisy approximations of the model output.
Following this verification, we learn the parameters and their uncertainties in two different turbulence models conditioned on scale-resolving large-eddy simulation data over a range of ABL stabilities.
We show how Bayesian inversion of a numerical model improves flow predictions by investigating the underlying mean momentum budgets.
Further, we show that uncertainty quantification based on neutral ABL surface layer data recovers the relationships between parameters that have been predicted using theoretical modeling, but that learning the parameters based on stable ABL data or data from outside the surface layer can lead to different parameter relationships than neutral surface layer theory.
Efforts to systematically reduce parameter uncertainty reveal that (1) sampling wind speed up to the ABL height can reduce uncertainty in key model parameters by up to 84\%, and (2) assimilating fluid flow quantities beyond first-order moment statistics can further reduce uncertainty in ways that baseline wind speed assimilation alone cannot achieve.
The parameters learned using Bayesian uncertainty quantification generally yield lower error than standard deterministic parameters in out-of-sample tests and also provide uncertainty intervals on predictions.
\end{abstract}


\section{Introduction}
\label{intro}

The atmospheric boundary layer (ABL) is the lowest portion of the troposphere, typically extending from hundreds to a few thousand meters above the surface, that is directly influenced by interactions with the Earth’s surface \citep{stull2012introduction}. 
Models that predict flow in this region of the atmosphere are used to support everyday decision making in various applications, including numerical weather prediction \citep{shin2011intercomparison, cohen2015review}, wind engineering \citep{blocken2007cfd, parente2011comprehensive}, and wind energy systems \citep{calaf2010large, howland2020influence}.
In this study, we refer to these atmospheric models that are sufficiently fast-running for everyday decision making as operational models.
To accurately predict flow quantities, these models must take into account the effects of turbulence, one of the primary transport processes in the ABL.

Because turbulence is an inherently chaotic phenomenon that spans multiple spatial and temporal scales, resolving all the scales at very high Reynolds numbers is computationally prohibitive \citep{wyngaard2010turbulence}.
Although the growth in computational power over the years has enabled direct numerical simulations (DNS) to resolve down to the smallest scales of flows \citep{flores2011analysis} and large-eddy simulations (LES) to resolve most scales, the scale separation that exists between the largest and smallest eddies in high-Reynolds number ABL flows renders predictions to remain dependent on lower-fidelity approaches that represent the effects of turbulence through parameterizations of turbulence \citep{parente2011comprehensive, nakanishi2009development}. 
Historically, atmospheric models, both mesoscale and microscale, have primarily relied on solving the unsteady Reynolds-averaged Navier-Stokes (RANS) equations to capture the evolution of mean flow statistics \citep{mellor1982development, hong2006new}. 
These models provide first-order moment statistics that convey the quantities of interest relevant to operational decision-making in many geophysical applications.

In the RANS equations, the effects of turbulence to the mean flow are parameterized.
However, these parameterizations of turbulence are a key source of error and epistemic, or reducible, uncertainty in predictions \citep{xiao2019quantification}.
Hereafter, the terms ``turbulence model" and ``turbulence parameterization" will be used interchangeably.
Traditionally, the parameters in these turbulence models have been deterministically calibrated using limited data from wind tunnel experiments, field campaigns, or simulations, many from canonical engineering flows and some from ABL flows.
This approach results in deterministic parameters that correspondingly result in deterministic predictions of mean statistics (with sufficient averaging in the chaotic system); these predictions neglect the uncertainties that arise from the unresolved turbulence processes \citep{launder1983numerical, freedman2002transport}.
The fixed parameters, often applied out-of-the-box in most cases, lead to varying levels of generalizability \citep{lopez2010optimization, zhao2022generalizability} and introduce unquantified uncertainties \citep{edeling2014bayesian}.
Quantifying uncertainties in these parameters is a critical first step towards identifying areas for model improvement and systematically exploring methods to reduce model uncertainties \citep{duraisamy2019turbulence}.

Within the broader field of computational fluid dynamics applied to boundary layer meteorology, advances in modern computing have significantly increased the availability of scale-resolving simulations for ABL flows \citep{kumar2006large, stoll2020large}.
LES resolves most of the turbulence (large-scale eddies), leaving only the effect of small-scale eddies to be modeled using a subgrid-scale turbulence model \citep{lilly1967representation, deardorff1970numerical, nicoud1999subgrid}.
LES's accurate representation of increasingly complex ABL physics has made it a widely used tool for modeling ABL dynamics \citep{beare2006intercomparison, stoll2020large, liu2021geostrophic}.
\cite{stoll2020large} provides a comprehensive review of the history and progress for LES applications.
However, the high computational cost of LES remains a significant barrier to applying the method to large-scale regions (on the order of tens to hundreds of kilometers) in which RANS-based operational models operate.
Instead, LES are typically performed at targeted sites with boundary conditions provided by mesoscale RANS simulations or reanalysis data \citep{schalkwijk2015year, wehrle2024introducing}.
Yet, it is this availability of scale-resolving simulation data that has opened exciting pathways for using data-driven approaches to better understand and address uncertainties in the representation of turbulence \citep{singh2016using, cotteleer2024flow}.

Bayesian inference provides a statistically rigorous framework for integrating models with data while explicitly representing prior knowledge and data uncertainties as probability distributions \citep{bolstad2016introduction}.
The process of inferring model parameters from indirect observations of quantities of interest constitutes an inverse problem \citep{evans2002inverse, biegler2011large}.
Observations of the ABL flow are noisy both because they are finite time averages and because they could also be affected by sensor noise. 
Finite time averaging is particularly pertinent to the ABL where small-scale turbulence appears as noise in observations, but longer time-averaging to reduce the influence of turbulence on the measurement are affected by averaging over transient mesoscale and diurnal variations.
In the Bayesian approach to the inverse problem, these model parameters are considered random variables that are inputs to a simulation of a physical system, commonly referred to as a \textit{forward model}, which generates the model predictions \citep{kennedy2001bayesian}.
This probabilistic inference of model parameters in the forward model is also known as Bayesian inversion, Bayesian calibration \citep{kennedy2001bayesian, trucano2006calibration} and inverse uncertainty quantification (UQ).
The solution to the Bayesian inverse problem is a posterior probability density function (PDF) of the model parameters conditioned on the observed data, from which statistical measures may be computed to provide details on the parameter uncertainties.

However, a computational challenge lies in characterizing the posterior distribution, for which simple analytical solutions exist only for a very limited set of specific distributions \citep{biegler2011large}.
In many cases, traditional sampling methods such as Markov chain Monte Carlo (MCMC) are used in which repeated evaluations of the forward model are required to capture the posterior distribution \citep{tierney1994markov}.
Accurate sampling of the posterior in a multidimensional parameter space requires a large number of model evaluations, often on the order of $\mathcal{O}(10^5)$ runs \citep{geyer2011introduction}.
This presents a clear, practical challenge in most computational fluid dynamics applications.
Therefore, previous studies have focused on accelerating the Bayesian inversion process in the directions of more efficient sampling methods \citep{andrieu2010particle, cui2016dimension} and surrogate modeling \citep{marzouk2009stochastic, cleary2021calibrate}.

The development of efficient UQ methods has in turn increased interest in applying Bayesian UQ in various fields, including climate modeling \citep{dunbar2021calibration,howland2022parameter}, wind engineering \citep{lamberti2019uncertainty,giacomini2024quantification}, and turbulence modeling \citep{oliver2011bayesian, guillas2014bayesian}.
In \cite{edeling2014bayesian} and \cite{singh2016using}, the Bayesian approach has been adopted to quantify uncertainties in both the turbulence model parameters and the model form error for engineering flows including channel flows and flows with curvature and separation.
In the analysis of wind flow over urban or plant canopies, \cite{lamberti2019uncertainty} and \cite{giacomini2024quantification} applied the Bayesian framework to quantifying uncertainties in inflow conditions, such as wind speed and direction.
However, uncertainties in turbulence parameterizations for stratified ABL flows remain relatively underexplored.
Nevertheless, a long trail of literature suggest these turbulence parameterizations are important in predicting numerous quantities of interest in the ABL and still have significant potential for improvement \citep{lupkes1996modelling, machado2015effect, doubrawa2020simulating}.

Motivated by the aforementioned advances and limitations, in this study, we develop a computationally-efficient Bayesian framework for quantifying uncertainty in RANS turbulence model parameters based on scale-resolving data.
Given the extensive validation conducted for LES in ABL applications \citep{kumar_largeeddy_2006, beare2006intercomparison, stoll2020large} and the limited availability of field observations, LES is regarded as the ground truth in this study and is hereafter interchangeably referred to as `scale-resolving data' to emphasize the model fidelity difference from the learning RANS model or `truth data' in the context of the inverse problem.
However, the methods developed in this study are general and naturally suited for UQ based on DNS, experimental, or field data.
We quantify uncertainty in a RANS-based single-column model (SCM), often used in meteorological models with idealized forcings to better understand and parameterize the ABL \citep{holtslag2013stable, christensen2018forcing}.
To address the computational bottleneck in the Bayesian inverse problem, we use a machine learning-accelerated UQ methodology that combines stochastic optimization and surrogate modeling with sampling \citep{cleary2021calibrate, oliver2024calibrateemulatesample}.
In doing so, the framework significantly reduces the computational cost of the sampling process compared to traditional requirements of approximately $\mathcal{O}(10^5)$ evaluations.
This speed-up of Bayesian inversion using only $\mathcal{O}(10^2)$ evaluations enables the inverse UQ of RANS-based ABL models in this study.
Our work is directed at two main objectives: (1) facilitating a computationally efficient probabilistic approach to RANS ABL predictions based on observationally quantified uncertainties in turbulence parameterizations and (2) exploring opportunities to systematically reduce the epistemic uncertainty in these parameterizations, such as including a targeted selection of the fluid flow quantities that are used for assimilation and identifying promising sensing locations that maximally reduce uncertainty.

The remainder of this paper is structured as follows.
Section~\ref{sec:methods} introduces the formulation of the inverse problem, outlining its main components and application-specific details on the use of the Bayesian framework in the learning process.
Section~\ref{sec:results_verification} demonstrates the verification of the UQ methodology used, and Sec.~\ref{sec:results_uncertaintyquantification} presents the results of the parameter uncertainty inferred from scale-resolving simulation data.
Section~\ref{sec:results_uncertaintyreduction} presents the results of uncertainty reduction leveraging the Bayesian inversion framework.
Key conclusions from our findings are drawn and summarized in Sec.~\ref{sec:conclusions}.

\section{Methods}
\label{sec:methods}

This section provides the technical details of each component in the framework to quantify the uncertainty of the turbulence model parameters in modeling ABL flows.
It provides a description of the forward model, a SCM in which the turbulence models are implemented, followed by a description of the LES methodology used to generate truth data for the inverse problem.
Next, it formulates the inverse problem that allows inference of model parameters $\boldsymbol{\theta}$ in the turbulence models from the inherently noisy truth data $\boldsymbol{y}$.
This is followed by a description of the UQ methodology that facilitates accelerated, approximate Bayesian learning in the inverse problem framework.
Lastly, the choices of prior distributions and observed statistics are described.

\subsection{Forward model}
\label{sec:forwardmodel}

We use a single-column model (SCM) with implementations of RANS turbulence models as the forward model in the UQ framework.
Assuming a horizontally homogeneous, dry, high Reynolds number atmosphere in which the radiative flux divergence is neglected, the Reynolds-averaged governing equations for the ABL are

\begin{equation}
    \frac{\partial \overline{u}}{\partial t} = f(\bar{v} - v_G) - \frac{\partial \overline{u'w'}}{\partial z}
    \label{eq:xmomentum}
\end{equation}
\begin{equation}
    \frac{\partial \overline{v}}{\partial t} = f(u_G - \bar{u}) - \frac{\partial \overline{v'w'}}{\partial z}
    \label{eq:ymomentum}
\end{equation}
\begin{equation}
    \frac{\partial \overline{\theta}}{\partial t} = - \frac{\partial \overline{w'\theta'}}{\partial z},
\end{equation}
where the overbar denotes ensemble averaging, $\overline{u}$ and $\overline{v}$ are the velocity components, $u_G$ and $v_G$ are the geostrophic wind components, $\overline{\theta}$ is the potential temperature, $f$ is the Coriolis parameter, and $\overline{u'w'}$, $\overline{v'w'}$, and $\overline{w'\theta'}$ are the turbulent momentum and heat fluxes.
The turbulent fluxes are parameterized with the eddy viscosity $\nu_t$ using K-theory \citep{stull2012introduction}:
\begin{equation}
    \overline{u'w'}=-\nu_t \frac{\partial \overline{u}}{\partial z}
\end{equation}
\begin{equation}
    \overline{v'w'}=-\nu_t \frac{\partial \overline{v}}{\partial z}
\end{equation}
\begin{equation}
    \overline{w'\theta'}=-\frac{\nu_t}{\mathrm{Pr_t}} \frac{\partial \overline{\theta}}{\partial z},
\end{equation}
where $\mathrm{Pr_t}$ is the turbulent Prandtl number. 

The wall treatment, formulated on the basis of Monin-Obhukov similarity theory (MOST), provides a shear-stress surface boundary condition \citep{basu2008inconvenient}.
A fourth-order Runge-Kutta method is used for time integration, while a second-order accurate central finite difference scheme is used for spatial derivatives.

\begin{table}
\caption{Summary of Selected Turbulence Models and Parameters}
\label{tab:turbulencemodels}     
\centering
\begin{tabular}{lllp{4cm}}
\hline\noalign{\smallskip}
Model & Formulation & Parameters & Reference \\
\noalign{\smallskip}\hline\noalign{\smallskip}
STD \(k - \epsilon\) & \multirow{2}{*}{\( \nu_t = C_\mu \frac{k^2}{\epsilon} \)} & \( C_\mu, C_1, C_2, C_3, \sigma_k, \sigma_\epsilon \) & \cite{LAUNDER1974269} \\
MOST \(k - \epsilon\) & & \( C_\mu, C_1, C_2, \sigma_k, \sigma_\epsilon \) & \cite{van2017new} \\
\noalign{\smallskip}\hline
\end{tabular}
\end{table}

Table~\ref{tab:turbulencemodels} shows the two different eddy viscosity-based RANS turbulence models tested in the study: the standard $k$-$\epsilon$ model \citep{LAUNDER1974269}, hereafter referred to as the STD $k$-$\epsilon$ model, and the modified $k$-$\epsilon$ model \citep{van2017new}, hereafter referred to as the MOST $k$-$\epsilon$ model.
The details and implementation verification of the two turbulence models are provided in \ref{sec:appendix1}.
The primary difference between the two models lies in their intended applications.
The STD $k$-$\epsilon$ model was developed and calibrated for a broad range of engineering flows, excluding effects of stratification or rotation.
In contrast, the MOST $k$-$\epsilon$ model is specifically adapted for the atmospheric surface layer, for a broad range of stability conditions.
The selection of turbulence models is guided by (1) the widespread use of $k$-$\epsilon$ models in industry and academia and (2) the consequent availability of verification data.
Because variants of the two-equation $k$-$\epsilon$ model have also been integrated into operational weather forecasting models as planetary boundary layer schemes, we expect that our insights from microscale ABL simulations are extensible to larger-scale atmospheric models \citep{zhang2020evaluation, zonato2022new}, but this should be explicitly investigated in future work.
Typical wall boundary conditions of $k$ and $\epsilon$ used for the neutral surface layer are used for both models, as commonly recommended in previous studies \citep{detering1985application, richards1993appropriate}.

The target model parameters that are learned in the Bayesian inversion are listed in the third column of Table~\ref{tab:turbulencemodels}.
Both turbulence models use the same parameters with the exception of $C_3$, which is explicitly modeled by the MOST $k$-$\epsilon$ model.
$C_\mu$ is the coefficient of turbulent eddy viscosity, $C_1, C_2, C_3$ are coefficients in the turbulence kinetic energy (TKE) dissipation rate equation associated with the shear production, dissipation and buoyancy production/destruction terms, respectively, and $\sigma_k$ and $\sigma_\epsilon$ are the coefficients associated with turbulent transport of TKE and TKE dissipation rate, respectively.
We adopt the following default parameter sets: for the STD $k$–$\epsilon$ model \citep{LAUNDER1974269},
\[
C_\mu = 0.09,\quad 
C_1 = 1.44,\quad 
C_2 = 1.92,\quad 
C_3 = 1.0,\quad 
\sigma_k = 1.0,\quad 
\sigma_\epsilon = 1.3
\]
 and for the MOST $k$–$\epsilon$ model \citep{van2017new},
 \[
C_\mu = 0.03,\quad 
C_1 = 1.21,\quad 
C_2 = 1.92,\quad
\sigma_k = 1.0,\quad 
\sigma_\epsilon = 1.3.
\]

\subsection{Scale-resolving data from large-eddy simulations}
\label{sec:turbulencecapturingdata}

In this study, an open-source pseudospectral flow solver Pad\text{\'{e}}Ops \citep{ghate2017subfilter, howland2020influence, heck2025coriolis} is used to generate the LES data, which serves as the ground truth for learning parameters in RANS turbulence models.
It solves the filtered, incompressible Navier-Stokes equations with the Boussinesq approximation for buoyancy, using Fourier collocation in horizontal directions and a sixth-order staggered compact finite-difference scheme in the vertical direction. 
The subgrid scale stresses are modeled using the sigma model \citep{nicoud2011using}. 
The equations are marched forward using a fourth-order strong stability-preserving Runge-Kutta method. 
The readers are directed to \cite{heck2025coriolis} for further details on the LES methodology and the validation.

\subsection{Atmospheric boundary layer case details}
\label{sec:ablcasedetails}

\begin{table}
\caption{Summary of Computational Setup for the Stability Cases}
\label{tab:stabilitycases}
\centering
\begin{tabular}{lccc}
\hline\noalign{\smallskip}
Stability case & TNBL & SBL(W) & SBL(M) \\
\noalign{\smallskip}\hline\noalign{\smallskip}
Domain height $z_{top}$ (m) & 5000 & 400 & 400 \\
Grid points $N_z$ (-)  & 360 & 64 & 64 \\
Grid size $\Delta z$ (m) & 13.89 & 6.25 & 6.25 \\
Geostrophic wind speed $u_G$ (m s$^{-1}$) & 12.0 & 8.0 & 8.0 \\
Initial potential temperature profile $\theta(z)$ (K) & $\theta_0$ & (Eq.~\ref{eq:stablepotentialtemperature}) & (Eq.~\ref{eq:stablepotentialtemperature}) \\
Free-atmosphere lapse rate $\Gamma$ (K m$^{-1}$) & 0 & 0.01 & 0.01 \\
Latitude $\phi$ (°) & 70 & 73 & 73 \\
Roughness length $z_0$ (m) & 0.1 & 0.1 & 0.1 \\
Surface cooling rate (K h$^{-1}$) & 0 & 0.05 & 0.25 \\
Spin-up time (h) & 25 & 7 & 7 \\
\noalign{\smallskip}\hline
\end{tabular}
\end{table}

This section details the ABL cases simulated by both LES and the SCM to facilitate the Bayesian learning process.
The truly neutral boundary layer and stable boundary layer with two different cooling rates are simulated for a total of three ABL cases.
The unstable boundary layer is excluded due to its dominant nonlocal mixing, which is not well represented by a local turbulence closure \citep{holtslag1993local}, but can be investigated using the same framework in future work with a non-local turbulence model.
The computational setups of the different cases are summarized in Table~\ref{tab:stabilitycases}.
10-minute averaging of the data is used, as commonly observed in meteorological measurements.
This choice should average over turbulent fluctuations \citep{trevino2000averaging}, but avoid averaging submesoscale motions, which, in principle, should be treated separately from turbulent contributions to the flow \citep{mahrt2009characteristics, mahrt2020non}.

\subsubsection{Truly neutral boundary layer initialization}
The truly neutral boundary layer (TNBL) is a canonical ABL flow with statically neutral stratification for which numerous parameterizations have traditionally been calibrated \citep{detering1985application, weng2003modelling}.
We use a computational setup derived from the conventionally neutral case in \cite{liu2021geostrophic}.
The geostrophic wind speed is $u_G=12~\mathrm{m~s^{-1}}$, the Coriolis parameter is $f=2\omega_c \sin\phi=1.37\times10^{-4}$, which corresponds to latitude $\phi=70^\circ$, and the aerodynamic roughness length is $z_0=0.1~\mathrm{m}$.
A constant potential temperature profile of $\theta_0=300~\mathrm{K}$ and zero surface heat flux are used.
The domain height $z_{top}$ is modified according to a constraint of $z_{top} > 2.5 h_E$ where $h_E$ is the Rossby-Montgomery equilibrium height \citep{zilitinkevich2007further}.
10-minute time-averaged vertical profiles of flow statistics are sampled between the time interval of $\Delta t=2\pi/f  \approx 13~\mathrm{hr}$ after an initial spin-up period of $t=12/f \approx 25~\mathrm{hr}$ to reduce the effect of inertial oscillations.

\subsubsection{Stable boundary layer initialization}
The stable boundary layer (SBL), characterized by smaller eddies, intermittent turbulence, and often modeled using a prescribed surface cooling rate, poses significant challenges to models due to its strong sensitivity to turbulence parameterizations \citep{van2003intermittent, cuxart2006single}.
We explore two different stability strengths, based on the well-documented setup of the moderately stable ABL in the GABLS intercomparison \citep{beare2006intercomparison}.
An initial profile for the potential temperature is prescribed as
\begin{equation}
  \theta(z) =
  \begin{cases}
    \theta_0 + 0.01\,(z - 100), & z \ge 100,\\
    265\ \mathrm{K},            & z < 100,
  \end{cases}
  \quad
  \theta_0 = 265\ \mathrm{K}
  \label{eq:stablepotentialtemperature}
\end{equation}
where $\theta_0$ is the reference potential temperature.
With all other configurations fixed, we vary the prescribed surface cooling rate, [$0.05, 0.25]~\mathrm{K}~\mathrm{h}^{-1}$, which represent the weakly stable (SBL(W)) and moderately stable (SBL(M)) boundary layers, respectively.
This design choice for the scale-resolving data and the SCM facilitates a data-driven investigation into the adequacy of the constant turbulence model parameters across different stability strengths.
In a similar manner to the collection of TNBL samples, 10-minute time-averaged vertical profiles of flow statistics are sampled between the time interval of $\Delta t \approx [7,9]~\mathrm{h}$ after an initial spin-up period of $t=7~\mathrm{h}$ to achieve statistical quasi-stationarity.

\subsection{Inverse problem}
\label{sec:inverseproblem}

\begin{figure}
    \centering
        \includegraphics[width=0.9\textwidth]{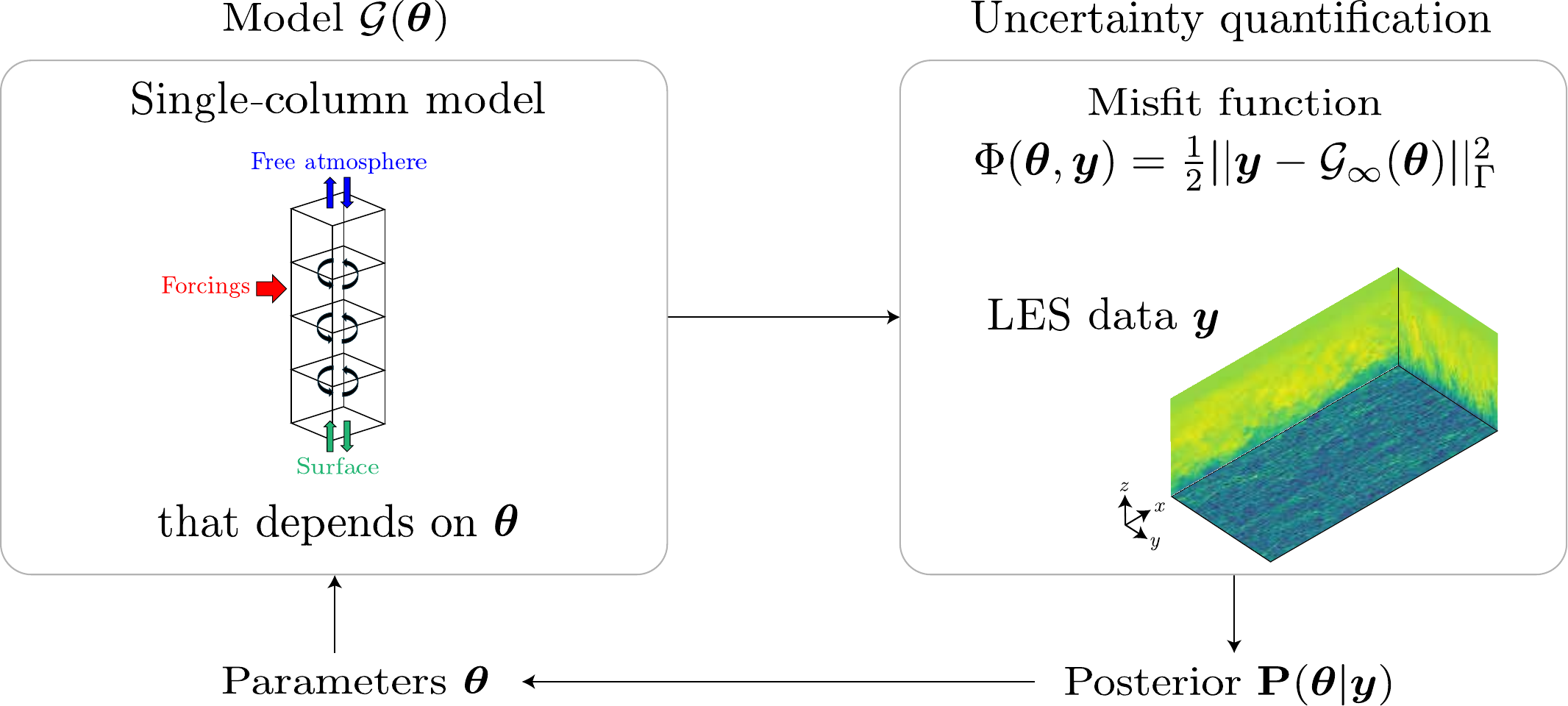}
    \caption{Bayesian inverse problem setup for learning RANS turbulence model parameters $\boldsymbol{\theta}$ and their uncertainties in the SCM from LES data $\boldsymbol{y}$. $\Phi(\boldsymbol{\theta}, \boldsymbol{y})$ is the misfit function, $\mathcal{G}_\infty(\boldsymbol{\theta})$ is the infinite mean of the forward model, and $\Gamma$ is the observational noise covariance matrix.}
    \label{fig:inverseproblem}
\end{figure}

The general formulation of the inverse problem poses the learning of parameter uncertainty from scale-resolving data
\begin{equation}
    \boldsymbol{y}=\mathcal{G}(\boldsymbol{\theta}, \xi)+\eta
\end{equation}
where $\boldsymbol{y}$ is the data, $\boldsymbol{\theta}$ are the parameters to be estimated, $\xi$ are the initial conditions, $\mathcal{G}(\boldsymbol{\theta}, \xi)$ is the forward model, and $\eta$ is the measurement error in $\boldsymbol{y}$.
In this study, the SCM RANS model is the forward model and data $\boldsymbol{y}$ are generated using LES.
However, this formulation of the inverse problem is inherently difficult in the need to jointly estimate the state (e.g., initial conditions $\xi$) and parameters $\boldsymbol{\theta}$ in the forward model. 
To avoid this challenge, an infinite average of the forward model $\mathcal{G}_\infty (\boldsymbol{\theta})$ is posed such that the joint estimation may be avoided \citep{dunbar2021calibration}:
\begin{equation}
    \mathcal{G}_\infty (\boldsymbol{\theta})=\lim_{M\rightarrow\infty} \frac{1}{M} \sum_{i=1}^{M} \mathcal{G}(\boldsymbol{\theta}, \xi_i).
    \label{eq:inverse}
\end{equation}
The updated inverse problem formulation based on the infinite average is
\begin{equation}
    \boldsymbol{y}=\mathcal{G}_\infty(\boldsymbol{\theta})+\gamma
    \label{eq:inverse}
\end{equation}
where the observational noise $\gamma \sim \mathcal{N}(0,\Gamma)$ is a realization of the internal variability in the chaotic, dynamical ABL system.
While an infinite time average is available as a steady-state RANS solution, it is computationally infeasible to compute for statistics of LES.
The key advantage of the proposed UQ approach (Section~\ref{sec:uqmethodology}) is that the emulator approximates the infinite time-averaged statistics of $\mathcal{G}_\infty (\boldsymbol{\theta})$.
The observational noise covariance matrix $\Gamma$ is assumed to be Gaussian based on the central limit theorem with finite-time averaged data $\boldsymbol{y}$ \citep{cleary2021calibrate}.
Note that this does not assume that microscale turbulence is Gaussian but is instead a description of the finite time averaged data.
Statistical measures shown in \ref{sec:appendix2} indicate that this approximation is justified for 10-minute averaged ABL quantities of interest.

Figure~\ref{fig:inverseproblem} schematically illustrates the workflow of the inverse problem. The model predictions $\mathcal{G}(\boldsymbol{\theta})$ from the SCM, which depend on turbulence model parameters $\boldsymbol{\theta}$, are combined with reference data $\boldsymbol{y}$ within a probabilistic framework.
The solution to this inverse problem minimizes the least-squares objective:
\begin{equation}
\Phi( \boldsymbol{\theta}, \boldsymbol{y}) = \frac{1}{2}\|\boldsymbol{y} - \mathcal{G}_{\infty}(\boldsymbol{\theta})\|_{\Gamma}^2,
\end{equation}
where $\Phi( \boldsymbol{\theta}, \boldsymbol{y})$ is interpreted to be the negative log density in the case of a Gaussian likelihood
\[
\begin{aligned}
P(\boldsymbol{\theta} \mid \boldsymbol{y}) &\propto P(\boldsymbol{y} \mid \boldsymbol{\theta})\,P(\boldsymbol{\theta}) \\
&\propto e^{-\frac{1}{2}\|\boldsymbol{y} - \mathcal{G}_{\infty}(\boldsymbol{\theta})\|_{\Gamma}^2}\,P(\boldsymbol{\theta}) \\
&\propto e^{-\Phi(\boldsymbol{\theta},\boldsymbol{y})}\,P(\boldsymbol{\theta}),
\end{aligned}
\]
and drives the solution to the \textit{maximum a posteriori} (MAP) estimate.

An assumption in this formulation is that the error and uncertainty in the model are attributed to the model parameters $\boldsymbol{\theta}$.
We evaluate the role of model-form error through an $\textit{a posteriori}$ analysis for structural discrepancies between SCM predictions and LES data after uncertainty quantification in Sec.~\ref{sec:results_uncertaintyquantification}.

\subsection{Uncertainty quantification methodology}
\label{sec:uqmethodology}

Sampling methods like MCMC used to estimate the posterior often require a prohibitively large number of likelihood evaluations for computational fluids applications \citep{brooks2011handbook}.
In this study, the UQ process is significantly accelerated through the use of the open-source calibrate-emulate-sample (CES) approach, a methodology that combines data assimilation and machine learning-based surrogate modeling to efficiently enable the sampling process \citep{cleary2021calibrate, oliver2024calibrateemulatesample}.

For a detailed explanation on the CES methodology and its applications, readers are referred to \cite{cleary2021calibrate}, \cite{dunbar2021calibration}, and \cite{howland2022parameter}.
Here, CES is summarized in its three sequential steps:

\begin{enumerate}
    \item Calibrate: Calibration of parameters is performed using the gradient-free ensemble Kalman inversion (EKI), a data assimilation-based optimization method that drives an ensemble of parameters to the solution of the inverse problem \citep{iglesias2013ensemble}.
    This serves as an efficient method of generating parameter–model output pairs that can be used as training data for supervised learning.
    \item Emulate: A Gaussian process emulator is built using the training data to approximate the parameter-to-data map.
    The accuracy of the emulator in the parameter region of interest is made possible by the high concentration of training data around the solution of the inverse problem.
    \item Sample: MCMC sampling of the computationally cheap emulator instead of the true forward model is performed to approximate the posterior distribution of model parameters. 
\end{enumerate}

The CES methodology offers several benefits for tackling inverse problems with the complexity typically encountered in chaotic atmospheric flows.
First, the model calibration process is gradient-free and non-intrusive, allowing ease in application to different numerical solvers of atmospheric dynamics.
Referring back to Eq.~\ref{eq:inverse}, the CES method enables the training of the emulator on parameter-model output pairs subject to independent realizations of noise in the chaotic system, thereby generating a surrogate model of the infinite time averaged model statistics $\mathcal{G}_\infty (\boldsymbol{\theta})$. 
And most importantly, sampling this emulator instead of the true forward model significantly reduces the computational cost, leading to approximately a thousand-fold speedup in the sampling process.
This is valuable because even the relatively lower-fidelity RANS models used in modeling the ABL require computational effort.
This observationally quantified parameters posterior may then be forward-propagated into the model to generate model predictions with credible intervals.

In this study, exogenous parameters and settings required by the methodology are determined based on previous studies \citep{cleary2021calibrate, howland2022parameter, shin2024multi}.
In the calibration phase, the EKI is run using a 100-member ensemble for 5 iterations.
The squared exponential kernel is used for the Gaussian process emulation, and the random-walk Metropolis-Hastings algorithm is used in the MCMC sampling process to take 200000 samples after 2000 burn-in samples.
Each of the three steps in the CES methodology is modular and can be substituted with methods that fit the target application.

\subsection{Choice of prior distributions}
\label{sec:priordistributions}

\begin{table}
\centering
\caption{Prior Distributions of Parameters for the STD $k$-$\epsilon$ Model}
\label{tab:priors_STD}
\begin{tabular}{ccc}
\hline\noalign{\smallskip}
Parameter & Default value & Prior distribution \\
\noalign{\smallskip}\hline\noalign{\smallskip}
\( C_\mu \) & 0.09 & \( \mathcal{U}(0.045, 0.135) \) \\
\( C_1 \) & 1.44 & \( \mathcal{U}(0.72, 2.16) \)    \\
\( C_2 \) & 1.92 & \( \mathcal{U}(0.96, 2.88) \)  \\
\( C_3 \) & 1.0 & \( \mathcal{U}(0.5, 1.5) \)    \\
\( \sigma_k \) & 1.0 & \( \mathcal{U}(0.5, 1.5) \)    \\
\( \sigma_\epsilon \) & 1.3 & \( \mathcal{U}(0.65, 1.95) \) \\
\noalign{\smallskip}\hline
\end{tabular}
\end{table}

\begin{table}
\centering
\caption{Prior Distributions of Parameters for the MOST $k$-$\epsilon$ Model}
\label{tab:priors_MOST}
\begin{tabular}{ccc}
\hline\noalign{\smallskip}
Parameter & Default value & Prior distribution \\
\noalign{\smallskip}\hline\noalign{\smallskip}
\( C_\mu \) & 0.03 & \( \mathcal{U}(0.015, 0.045) \) \\
\( C_1 \) & 1.21 & \( \mathcal{U}(0.605, 1.815) \)    \\
\( C_2 \) & 1.92 & \( \mathcal{U}(0.96, 2.88) \)  \\
\( \sigma_k \) & 1.0 & \( \mathcal{U}(0.5, 1.5) \)    \\
\( \sigma_\epsilon \) & 1.3 & \( \mathcal{U}(0.65, 1.95) \) \\
\noalign{\smallskip}\hline
\end{tabular}
\end{table}

Prior distributions $P(\boldsymbol{\theta})$ play a pivotal role in Bayesian inference.
They reflect the knowledge of the target parameters prior to evaluating the forward model \citep{van2021bayesian}.
The prior distributions of the target model parameters for the STD $k$-$\epsilon$ model (MOST $k$-$\epsilon$ model) are shown in Table~\ref{tab:priors_STD} (Table~\ref{tab:priors_MOST}).
Informed by a previous study \citep{edeling2014bayesian}, we use independent uniform distributions with $\pm 50\%$ bounds from the default values for each model.
However, unlike many previous studies, we do not enforce a constraint between $C_1$ and $C_2$. 
Although this flexibility can yield more frequent non-physical parameter combinations and model failures, these issues are mitigated by the robustness of the ensemble-based calibration in the CES methodology.
This unconstrained exploration enables a data-driven analysis of the $C_1$-$C_2$ relationship (Sec.~\ref{sec:uq_datainformedlearning}).

As the EKI update equation serves as the \textit{maximum a posteriori} (MAP) estimator for the Bayesian inverse problem with Gaussian priors, normal distributions are used for the EKI update in the physically unconstrained space, but then are transformed to the uniform distributions to accommodate any necessary parameter constraints in the physically constrained space for the forward model.
For example, all parameters in Table~\ref{tab:priors_MOST} and Table~\ref{tab:priors_STD} must be positive and within certain ranges to preserve a physically meaningful balance of turbulent contributions to the mean flow.
This is made possible through a sampling method called the probability integral transform \citep{fisher1970statistical}.
To transform an unconstrained parameter $\vartheta$ of a normal distribution to a physically constrained parameter $\theta$ of uniform distribution with specified bounds, we apply the cumulative distribution function (CDF) of the normal distribution. 
The CDF of $\vartheta$ is given by:
\[
F(\vartheta) = \frac{1}{2} \left[ 1 + \mathrm{erf}\left( \frac{\vartheta - \mu}{\sigma \sqrt{2}} \right) \right],
\]
where $\mathrm{erf}(\cdot)$ is the error function, $\mu$ is the mean, and $\sigma$ is the standard deviation.

Using the transformation
\[
\theta = F(\vartheta),
\]
the physical parameter $\theta$ follows a uniform distribution, $\theta \sim \mathcal{U}(0, 1)$, which can be further manipulated to have specified upper and lower bounds.

\subsection{Choice of observed statistics}
\label{sec:observedstatistics}

Here, we present the types of statistics that serve as observed data $\boldsymbol{y}$ in the Bayesian inversion and describe our selection criteria.
Traditionally, turbulence model parameters are deterministically calibrated using mean-flow statistics and analytical assumptions.
For example, the estimation of $C_\mu$ in $k$-$\epsilon$ models typically depends on surface-layer ratios of TKE to friction velocity ($k/u_*^2$), derived under the assumption of surface TKE equilibrium \citep{hagen1981simulation, temel2017two}.
For ABL flows, statistics from the surface layer are predominantly used, as much of the model development is initiated from flows for which similarity theories exist.
In a more general definition of fitting, parameters are chosen to show good alignment in model predictions with profiles of wind speed \citep{yang2009new}, velocity components \citep{detering1985application, xu1997turbulence}, eddy diffusivity \citep{detering1985application, troen1986simple}, TKE \citep{xu1997turbulence, yang2009new, parente2011comprehensive}, turbulent momentum flux profiles \citep{menter1994two} and surface shear stresses \citep{hagen1981simulation}.
The use of direct numerical simulation data \citep{kim1987turbulence} and experimental campaigns \citep{lettau1962theoretical} in calibration indicates the meticulous attention that has been paid to the selection of data used in fitting the parameters.
Such manual selection inherently limits the flow physics to which the parameters are tuned, and despite extensive historical calibration, it remains unclear which statistics optimally inform turbulence parameter learning.
As a baseline, we assimilate vertical profiles of 10-minute averaged wind speed within the ABL, addressing whether robust parameter inference is feasible from noisy, first-order moment statistics alone.

The degree of information in the observed data can improve the accuracy of parameter estimation.
This is often examined within the framework of information theory and in applications of optimal experimental design \citep{huan2013simulation}.
Here, we utilize the ensemble Kalman calibration in CES to systematically evaluate how assimilated statistics reduce parameter uncertainty \citep{cleary2021calibrate}.
In Sec.~\ref{sec:height_experiment}, we investigate how targeted vertical measurement locations influence parameter uncertainty.
This allows us to ask whether the acquisition of measurements at higher elevations provides additional information gain in calibrating turbulence parameterizations.
In Sec.~\ref{sec:informative_statistics}, we seek to identify the flow quantities that maximally reduce parameter uncertainty.
We hypothesize that assimilating flow quantities directly associated with modeling the turbulent fluxes will be most effective in reducing parameter uncertainty. 
In focusing on two turbulence parameterizations that use a modeled turbulent kinetic energy (TKE) $k$ and a modeled TKE dissipation rate $\epsilon$, these two quantities are expected to be most informative.
Based on their availability in field measurement campaigns, wind direction and TKE are incorporated to quantify their incremental contribution to information gain.

\section{Results: Verification of UQ methodology for RANS turbulence models}
\label{sec:results_verification}

To verify the utility of the CES method to learn parameters in the RANS turbulece model, we leverage a perfect-model test where synthetic wind speed data are generated using the SCM with prescribed Gaussian parameter distributions, ensuring the absence of both aleatoric and model-form uncertainties.
These prescribed model parameters remain unobserved throughout the CES process.
Therefore, their recovery provides a basis for assessing the effectiveness of the CES methodology.

\begin{figure}
    \centering
        \includegraphics[width=0.9\textwidth]{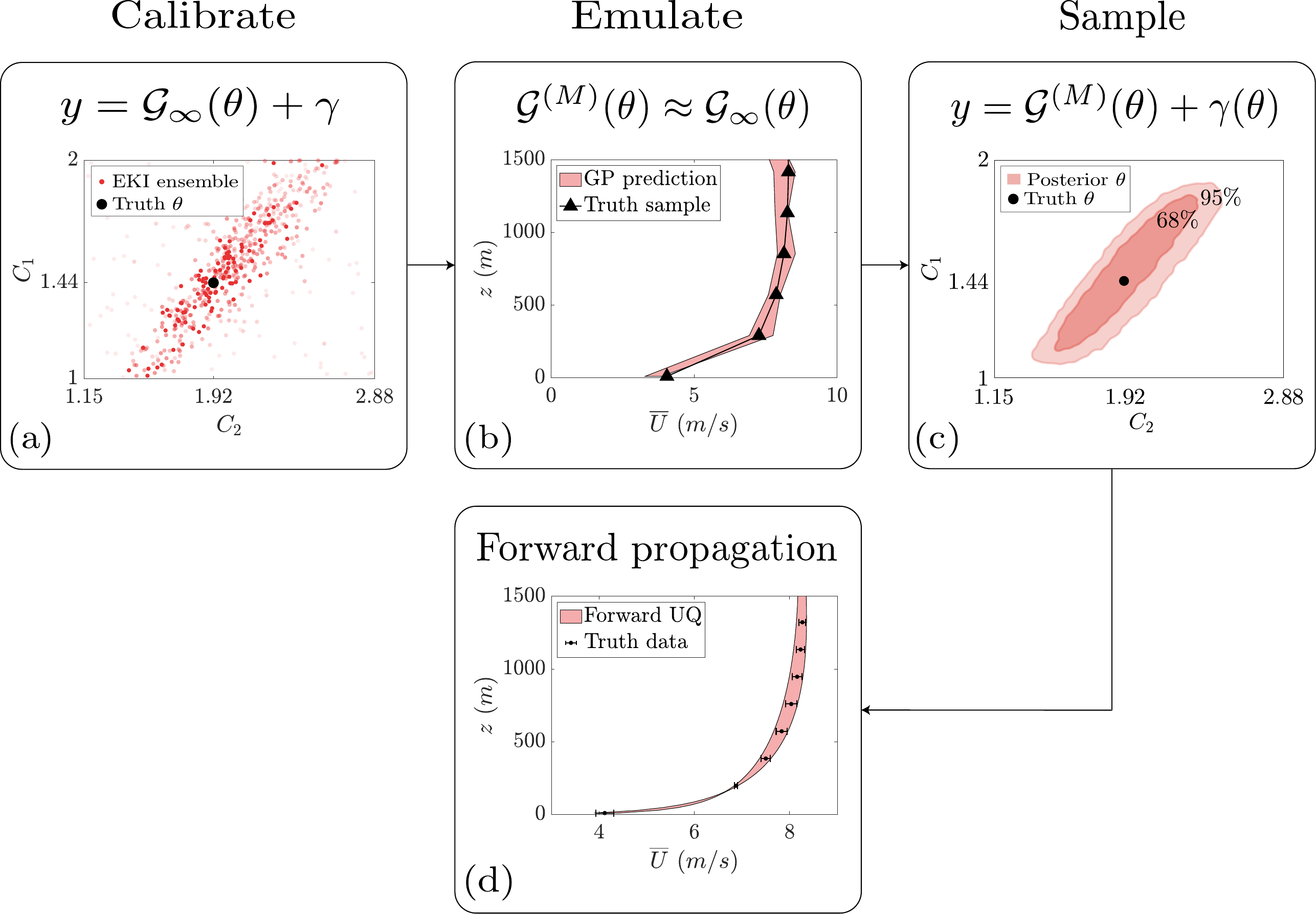}
    \caption{Verification results in the perfect-model test using the STD $k$-$\epsilon$ model for (a) Calibrate using the Ensemble Kalman Inversion (EKI), (b) Emulate using Gaussian Process (GP) Regression, (c) Sample using the Metropolis-Hastings algorithm, with (d) Forward propagation of random samples from the posterior}
    \label{fig:ces_pipeline}
\end{figure}

Figure~\ref{fig:ces_pipeline} presents the results in the perfect-model test at each stage within the CES pipeline using the STD $k$-$\epsilon$ model.
Analysis focuses on \( C_1 \) and \( C_2 \), the two dominant parameters that influence the TKE balance.
In the prognostic transport equation for the TKE dissipation rate $\epsilon$, parameter $C_1$ is related to the shear production term, while the parameter $C_2$ corresponds to the destruction by dissipation term.
The calibration results, shown in Fig.~\ref{fig:ces_pipeline}a, illustrate the parameter space concentrated near the data-model output misfit. 
The parameter ensemble concentrated around the prescribed true parameters, $(C_1,C_2)=(1.44,1.92)$ shown as a black circle, is mapped to model predictions in the calibration process, resulting in labeled pairs for supervised learning.
Figure~\ref{fig:ces_pipeline}b illustrates the training of the Gaussian Process emulator conditioned on synthetic Gaussian noise in the perfect-model test.
Figure~\ref{fig:ces_pipeline}c shows the approximate Bayesian learning performed through MCMC sampling from the GP emulator, instead of the more expensive forward model, to estimate the parameter posterior.
The 68\% and 95\% highest posterior density regions (HPDR) are shown for the $C_1$-$C_2$ joint posterior distribution in the dark red and light red shadings, respectively.
These represent the smallest credible regions that contain the specified percentages of the posterior mass.
In relating the statistical measure to a physical interpretation of the uncertainty, the term ``area" is hereafter used interchangeably with the HPDR.
The prescribed truth parameters are captured in the 68\% area of the posterior, verifying the accuracy of the CES methodology.
Furthermore, the positive correlation between $C_1$ and $C_2$ despite the lack of an enforced parameter constraint provides physical insight into the balance between the mechanisms of production and destruction in the TKE dissipation rate given the observed flow.
The three-step pipeline illustrated in Fig.~\ref{fig:ces_pipeline}a-c is the full inverse UQ process.

In addition, Fig.~\ref{fig:ces_pipeline}d shows the forward propagation of a 100-member ensemble of random samples from the posterior, which enables the generation of probabilistic predictions.
These observationally quantified credible intervals of uncertainty in model predictions are made possible through the inverse UQ of model parameters.
Because of the absence of model-form error in the perfect-model test, the forward propagation of the learned parameter uncertainty accurately recovers the noise in the truth data, verifying the accuracy of the posterior estimated in the CES methodology.

\section{Results: Parameter uncertainty quantification}
\label{sec:results_uncertaintyquantification}

\begin{figure}
    \centering
        \includegraphics[width=0.9\textwidth]{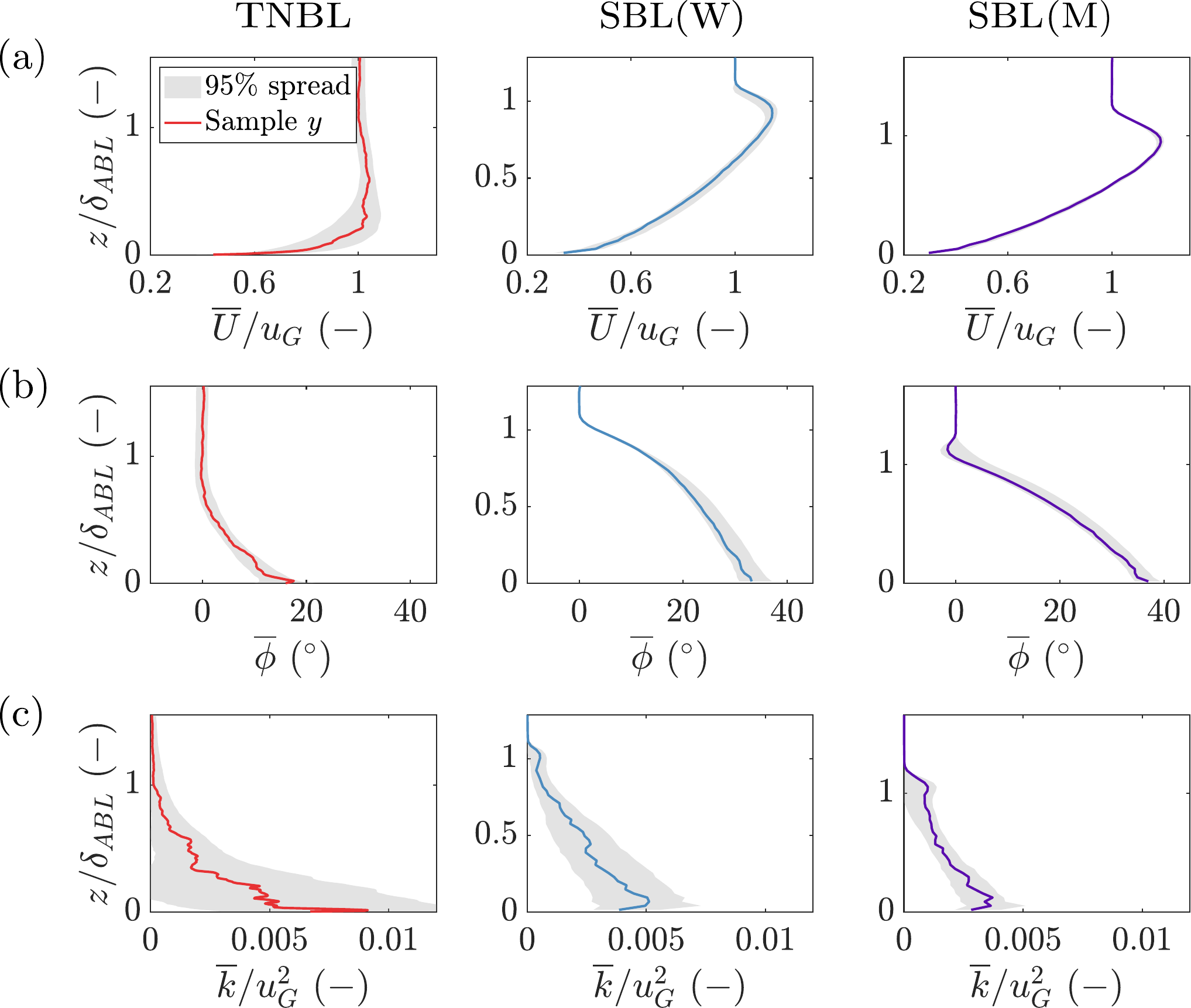}
    \caption{LES data 95\% spread with a 10-minute averaged sample profile for (a) non-dimensional wind speed $\overline{U}/u_G$, (b) wind direction $\overline{\phi}$, and (c) non-dimensional TKE $\overline{k}/u^2_G$, for the three stability regimes. Variables: $z$ is height, $\delta_{ABL}$ is ABL height, $\overline{U}$ is time-averaged wind speed, and $u_G$ is geostrophic wind speed.}
    \label{fig:les_data}
\end{figure}

Following the verification of the CES UQ methodology using the perfect model test, we transition to the inference of RANS turbulence model parameters conditioned on scale-resolving data.
Figure~\ref{fig:les_data} shows the LES data with its 95\% spread and a 10-minute averaged sample profile for (a) wind speed $\overline{U}$, (b) wind direction $\overline{\phi}$, and (c) TKE $\overline{k}$, for the three stability regimes.
The wind speed and TKE profiles are non-dimensionalized using the respective geostrophic wind speed $u_G$ while the vertical axis is non-dimensionalized using the ABL height $\delta_{ABL}$ of the respective stability regime.
The ABL height is estimated by identifying the height $\delta_{0.05}$ where the vertical momentum flux reaches 5\% of its surface value, followed by a linear extrapolation to $\delta_{ABL} = \delta_{0.05}/0.95$ \citep{kosovic2000large}.

\subsection{Learning from LES data of single stability regimes}
\label{sec:uq_single}

\begin{figure}
    \centering
        \includegraphics[width=0.9\textwidth]{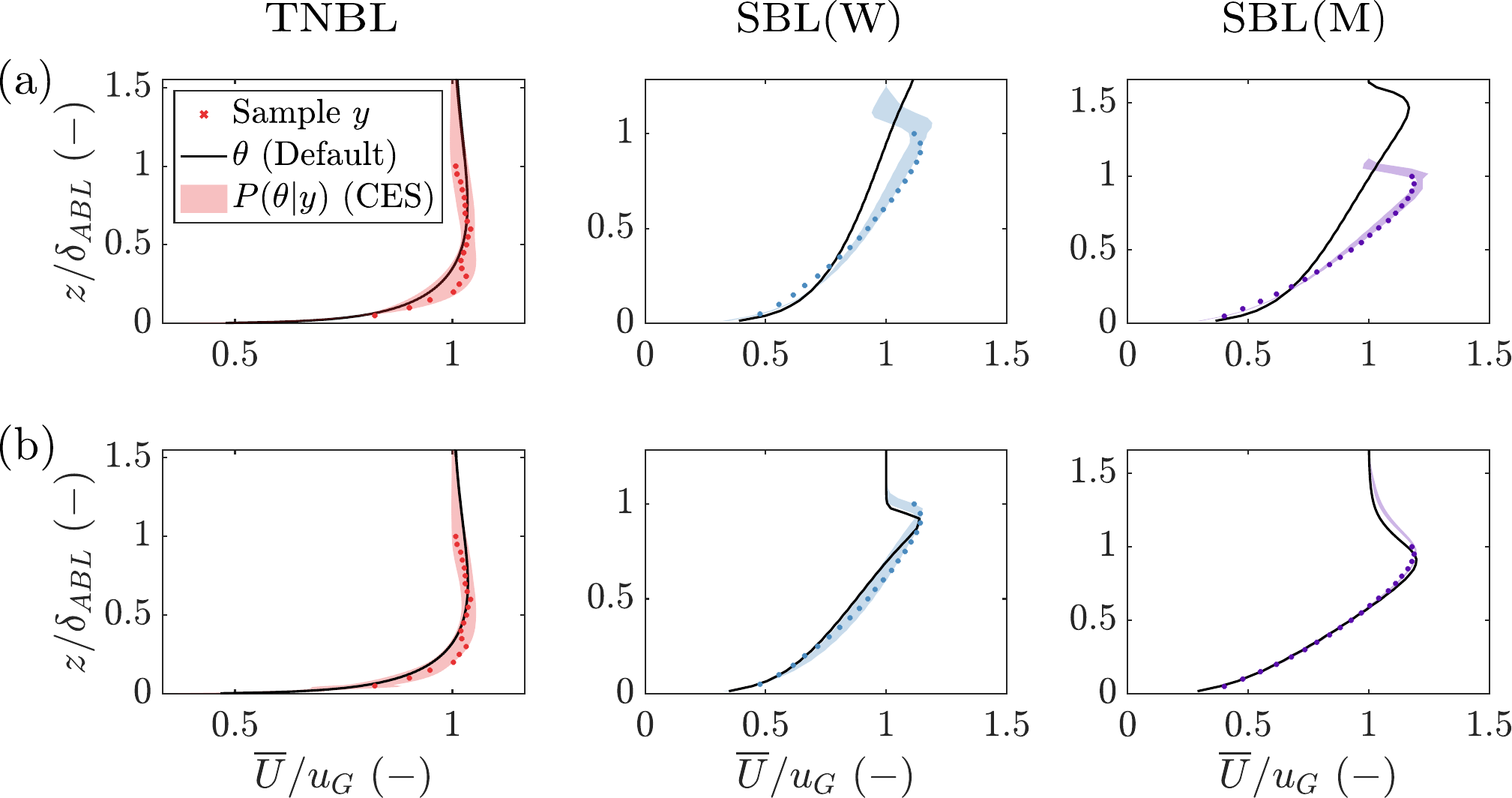}
    \caption{Comparison of wind speed predictions for each calibration ABL case, including the observed sample $\boldsymbol{y}$, the default-parameter SCM prediction (black line), and the forward UQ prediction ($95\%$ spread shading) for (a) STD $k-\epsilon$ model and (b) MOST $k-\epsilon$ model. Variables: $z$ is height, $\delta_{ABL}$ is ABL height, $\overline{U}$ is time-averaged wind speed, and $u_G$ is geostrophic wind speed.}
    \label{fig:credibleintervals}
\end{figure}

First, the turbulence model parameters are inferred from wind speed data of the three stability regimes independently.
This inverse problem setup provides useful information on the parameter uncertainty based on data of a specific stability condition, allowing the identification of the parameter dependence on stability.

Instead of visualizing each of the parameter posterior distributions, we focus on the credible intervals on predictions.
The credible bounds are each generated from a 100-member ensemble of model predictions by propagating forward random samples from the respective parameter posterior into the forward model.
The ensemble size is empirically determined as a balance between a sufficiently large ensemble size for a representative distribution of quantities of interest (QOIs) and computational efficiency.
Future work can use efficient forward UQ methods \citep{sullivan2015introduction, garcia2018uncertainty, lamberti2018uncertainty}.
The ensemble prediction resulting from the forward uncertainty propagation process is hereafter referred to as a forward UQ prediction.

The 95\% credible intervals (shaded regions) for wind speed predictions are shown for the STD $k$-$\epsilon$ model (Fig.~\ref{fig:credibleintervals}a) and the MOST $k$-$\epsilon$ model (Fig.~\ref{fig:credibleintervals}a), in comparison to the respective default-parameter predictions (black line) and the observed sample (crosses).
The effectiveness of the UQ approach is demonstrated by the observed sample $\boldsymbol{y}$ falling within the 95\% credible intervals across all six cases.
The default-parameter predictions from the two turbulence models clearly illustrate the stability-dependent behavior captured by the MOST $k$-$\epsilon$ model and not by the STD $k$-$\epsilon$ model. 
In the case of the MOST $k$-$\epsilon$ model, a close alignment between the default-parameter predictions and the observed sample $\boldsymbol{y}$ demonstrates that the MOST $k$-$\epsilon$ model performs with a significantly lower error prior to calibration in all three stability cases.
This is attributed to its stability-dependent formulation of $C_3(\zeta)$ (Eq.~\ref{eq:C3}), which accounts for buoyancy effects in stable stratification.
We note that even in these cases where reductions in mean error for a given turbulence model and stability regime may be relatively small, Bayesian UQ provides the benefit of the observationally quantified uncertainty bounds of these parameters, which can then be used in decision-making under uncertainty.
In contrast, forward UQ predictions by the STD $k$-$\epsilon$ model for the two stable cases (Fig.~\ref{fig:credibleintervals}a, middle and right) display improvements, notably a one-third reduction in the ABL height for the SBL(M) case in calibration to the LES sample.
The STD $k$-$\epsilon$ model's overprediction of the ABL height in stable stratification is associated with deeper turbulent mixing, which has been reported in previous studies \citep{weng2003modelling}.
The primary reason for this phenomenon is that the model parameters are calibrated to engineering flows, which typically do not carry effects of stratification.
Bayesian model calibration identifies model discrepancies and, in this case, corrects for stable stratification by adjusting the model parameters.

\begin{figure}
    \centering
    \begin{tabular}
    {@{}p{0.4\linewidth}@{\quad}p{0.4\linewidth}}
        \subfigimgthree[width=\linewidth,valign=t]
        {(a)}
        {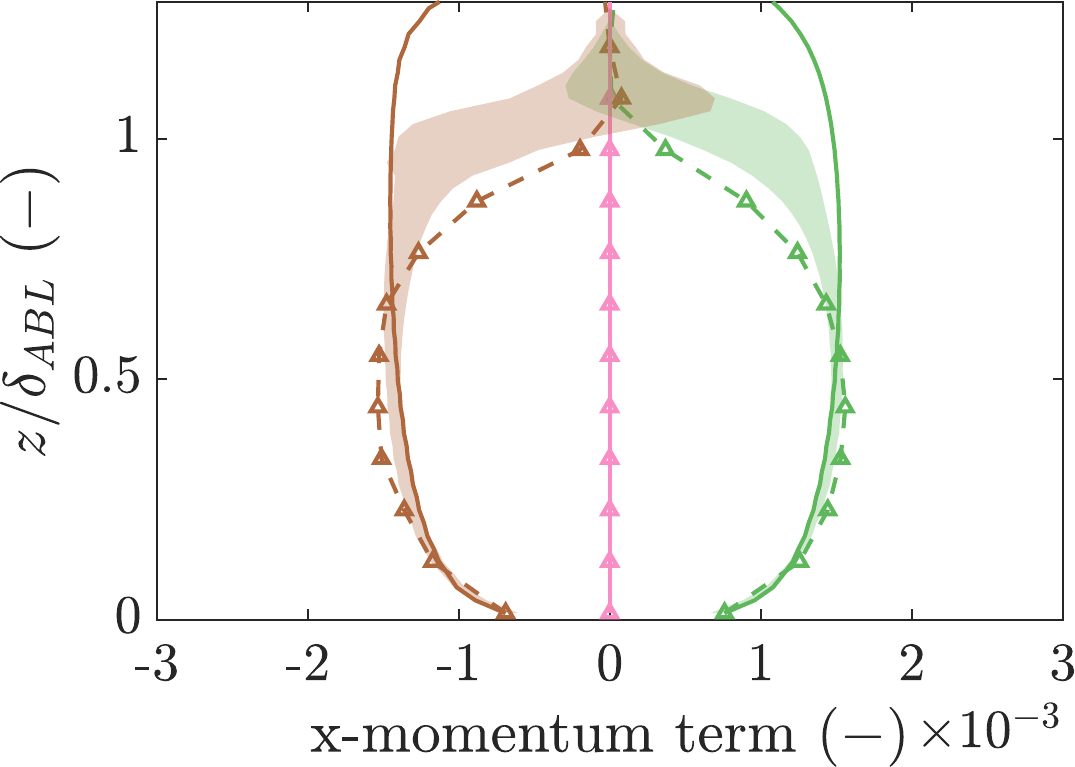} &
        \subfigimgthree[width=\linewidth,valign=t]
        {(b)}
        {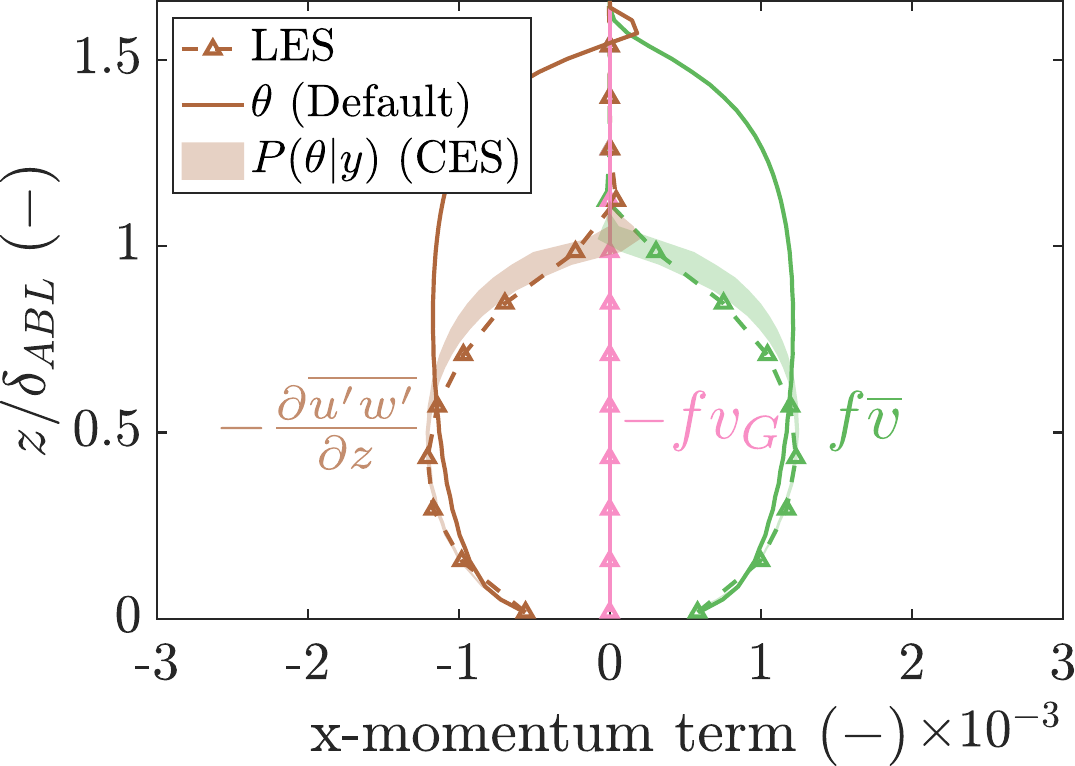} 
    \end{tabular}
    \caption{Non-dimensional streamwise momentum terms for (a) SBL(W) case and (b) SBL(M) case: LES data (dashed line with triangle marker) vs. default-parameter STD $k$-$\epsilon$ prediction (solid line) vs. forward UQ STD $k$-$\epsilon$ prediction (95\% spread shading).}
    \label{fig:xmomentumbudgets}
\end{figure}

\begin{figure}
    \centering
    \begin{tabular}
    {@{}p{0.4\linewidth}@{\quad}p{0.4\linewidth}}
        \subfigimgthree[width=\linewidth,valign=t]
        {(a)}
        {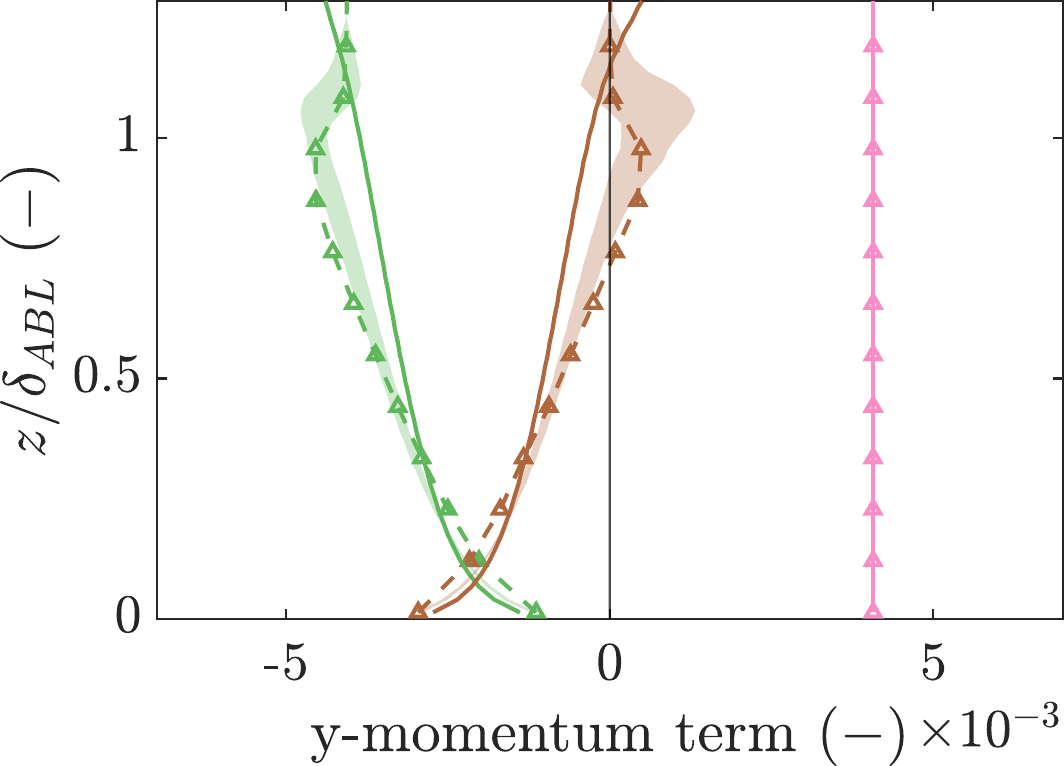} &
        \subfigimgthree[width=\linewidth,valign=t]
        {(b)}
        {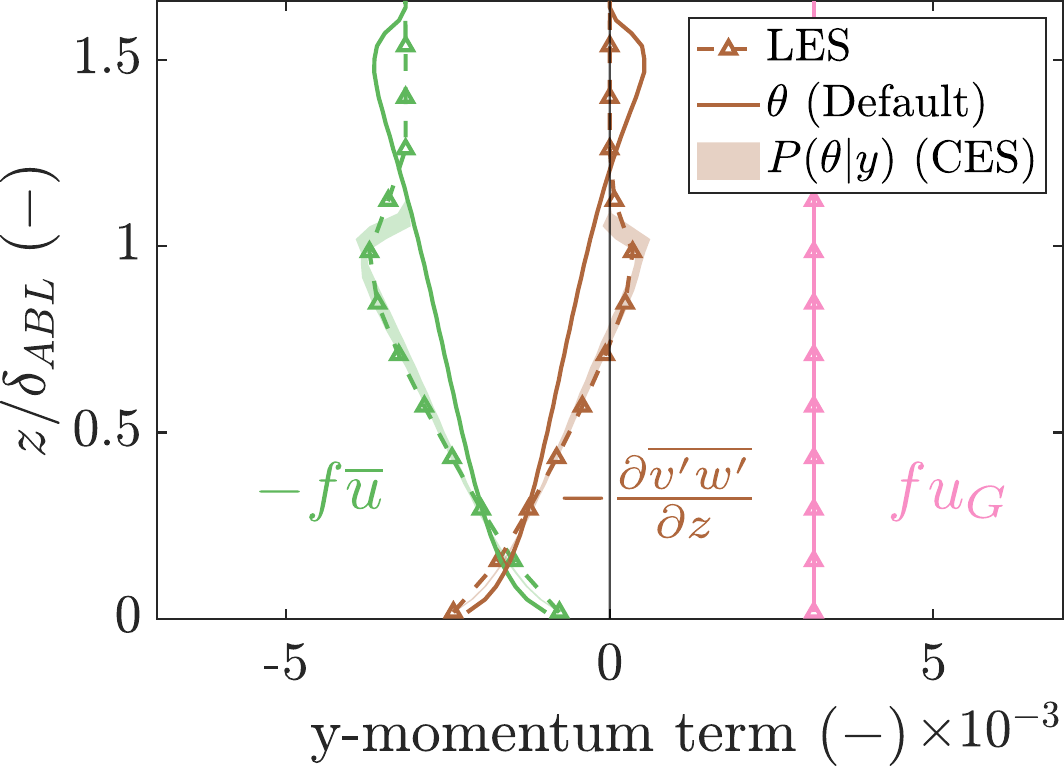} 
    \end{tabular}
    \caption{Non-dimensional cross-stream momentum terms for (a) SBL(W) case and (b) SBL(M) case: LES data (dashed line with triangle marker) vs. default-parameter STD $k$-$\epsilon$ prediction (solid line) vs. forward UQ STD $k$-$\epsilon$ prediction (95\% spread shading).}
    \label{fig:ymomentumbudgets}
\end{figure}

The improved wind speed predictions in stable stratification motivate an analysis of the mechanisms of error reduction in the STD $k$-$\epsilon$ model.
We compare budgets of the streamwise and cross-stream momentum equations (Eqs.~\ref{eq:xmomentum}, \ref{eq:ymomentum}).
Limited to this analysis, 1-hour and horizontally averaged mean statistics of LES data are used for comparison with SCM model predictions because the convergence of statistics for the divergence of the Reynolds stresses typically requires averaging periods much longer than 10 minutes.
For both of the stable regime cases, we show a comparison of streamwise (Fig.~\ref{fig:xmomentumbudgets}) and cross-stream (Fig.~\ref{fig:ymomentumbudgets}) momentum budgets for LES, the default-parameter SCM predictions and forward UQ predictions using the STD $k$-$\epsilon$ model.
All terms are non-dimensionalized by $u_G^2/\delta_{ABL}$.
In both momentum balances, we observe that the STD $k$-$\epsilon$ predictions show an improvement in the balance between the Coriolis term (green) and the divergence of the Reynold stress (brown) for the forward UQ prediction (95\% spread shading) with respect to the corresponding LES terms (dashed line with triangle marker).
The ABL height, which is overpredicted by approximately 50\% by default parameters in the SBL(M) case, aligns accurately with the LES data at $z/\delta_{ABL}=1$ after Bayesian calibration.
This connection between improved wind speed predictions and corrections to terms in the mean momentum balance demonstrates that (1) adjusting physically meaningful model parameters (linked to terms in the TKE dissipation rate equation) can sufficiently improve the accuracy of general-purpose engineering turbulence models for ABL predictions, and (2) improvements in mean momentum balance forcings suggest the possibility of physics-based generalization using Bayesian model calibration.

\subsection{Data-informed learning of parameter relationships}
\label{sec:uq_datainformedlearning}

In contrast to purely data-driven approaches, the CES methodology offers a framework that leverages data to drive learning within physics-based numerical models.
Similarly, recent work haved used scale-resolving simulation data in the investigation of higher-fidelity physics to drive the development of reduced order modeling for turbulent flows \citep{xiao2019reduced, schmelzer2020discovery}.
This prospect is first tested using the CES methodology to determine whether existing analytical parameter relationships in models can be recovered from data.
This framework can evaluate the validity of theoretical modeling assumptions.
Parameter uncertainty quantification enables a quantitative assessment of the validity of parameter relationships in their comparison to statistically significant posterior regions.

An analytically derived relationship between four parameters, $C_1$, $C_2$, $\sigma_\epsilon$ and $C_\mu$ used in $k$-$\epsilon$ models is derived for a steady, horizontally homogeneous neutrally stratified atmospheric surface layer \citep{richards1993appropriate, temel2017two}.
The full derivation is beyond the scope of this work, and readers are directed to \cite{temel2017two} for further details.
Here, we summarize the components necessary to understand the assumptions built into the analytically derived parameter relationship.

The transport equation for the TKE dissipation rate $\epsilon$ is simplified using assumptions of steady state, horizontal homogeneity, and neutral stratification:
\begin{equation}
    0 = C_{1} \frac{\epsilon}{k} \nu_t \left(\frac{\partial \overline{U}}{\partial z} \right)^2 + \frac{\partial}{\partial z} \left( \frac{\nu_t}{\sigma_\epsilon} \frac{\partial \epsilon}{\partial z} \right) - C_{2} \frac{\epsilon^2}{k}
    \label{eq:tkedissipationequation}
\end{equation}

Within the neutral surface layer, the constant-flux assumption leads to the simplification of the eddy viscosity $\nu_t$ and TKE dissipation rate $\epsilon$ to functions of friction velocity $u_*$ and distance $z$ from the surface \citep{tennekes1973similarity}:
\begin{equation}
    \nu_t = u_* \kappa z
    \label{eq:nu_t}
\end{equation}
\begin{equation}
    \epsilon = \frac{u_*^3} {\kappa z}.
    \label{eq:epsilon}
\end{equation}

Then, a local equilibrium of shear production and dissipation of TKE in neutral stratification is assumed:
\begin{equation}
    \nu_t \left(\frac{\partial \overline{U}}{\partial z} \right)^2=\epsilon = \frac{u_*^3} {\kappa z},
    \label{eq:localequilibirum}
\end{equation}
Eq.~\ref{eq:nu_t}, \ref{eq:epsilon}, and \ref{eq:localequilibirum} may be inserted into Eq.~\ref{eq:tkedissipationequation} to derive a relationship between four turbulence model parameters:
\begin{equation}
    C_2 = C_1 + \frac{\kappa^2}{\sigma_{\epsilon} \sqrt{C_\mu}}.
    \label{eq:C1C2}
\end{equation}

Previous work that have performed UQ on $k$-$\epsilon$ models have either (1) used a parameter constraint that fixes the relationship between $C_1$ and $C_2$ \citep{platteeuw2008uncertainty, edeling2014bayesian} or (2) used a subset of the model parameters \citep{turutoglu2023improvement}, which has left the validity of the four-parameter relationship beyond the surface layer and neutral stratification unexplored.
Other studies indicate this parameter relationship does not hold above the surface layer or under stable stratification and investigate methods of varying either $C_1$ or $C_2$ over height or stability to get better model predictions \citep{detering1985application, freedman2003modification}.
This motivates our data-informed approach, using the CES methodology, to verify regions in which this analytically derived parameter relationship remains valid and to investigate how it breaks down beyond these regions.

\begin{figure}
    \centering
    \begin{tabular}
    {@{}p{0.45\linewidth}@{\quad}p{0.45\linewidth}}
        \subfigimgthree[width=\linewidth,valign=t]
        {(a)}
        {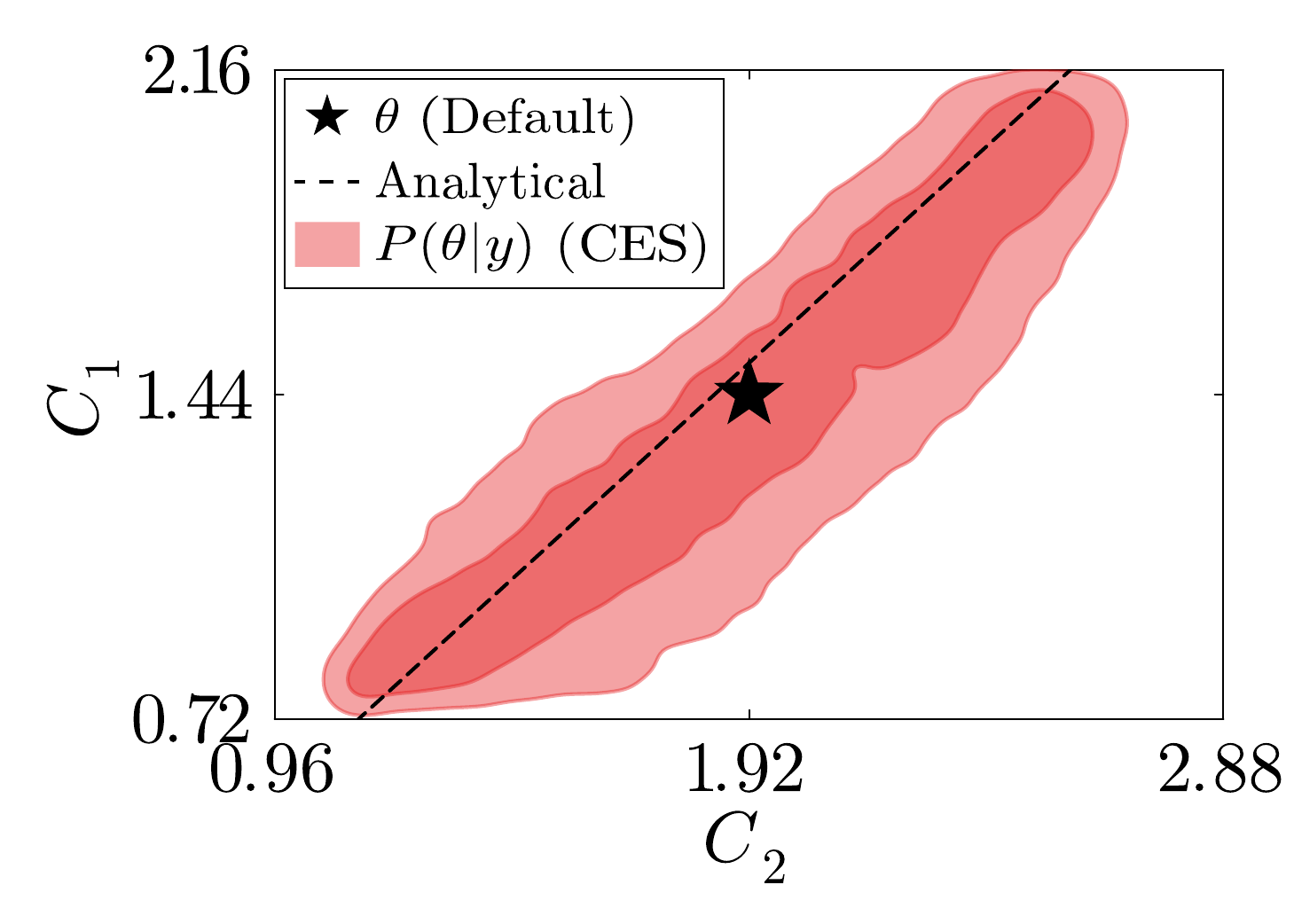} &
        \subfigimgthree[width=\linewidth,valign=t]
        {(b)}
        {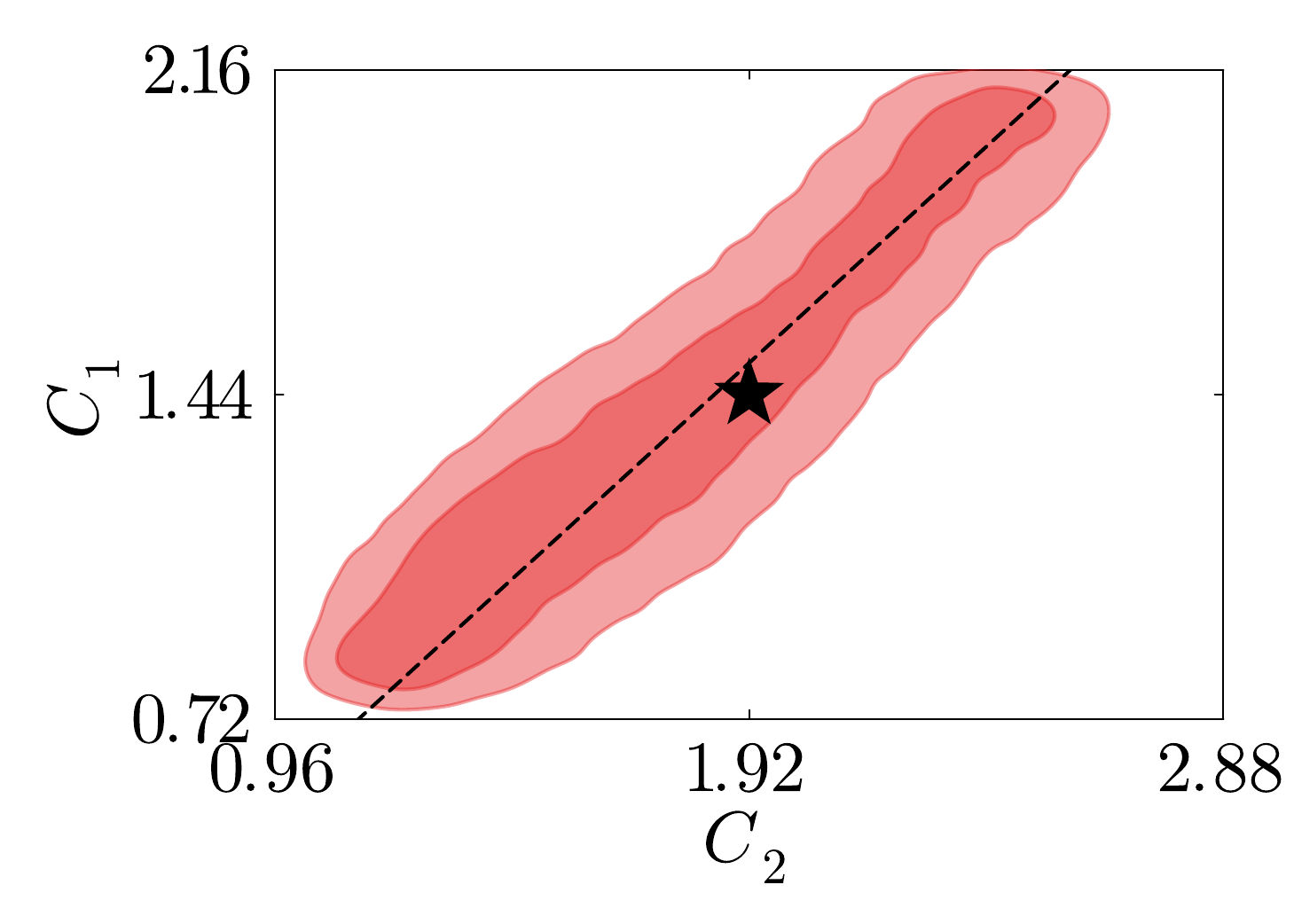} 
    \end{tabular}
    \caption{Joint posterior distribution of $C_1$ and $C_2$ in the STD $k$-$\epsilon$ model compared to the theoretical parameter relationship (dashed line) and the default parameters (star) for (a) neutral ASL and (b) TNBL. 68\% and 95\% HPDR are shown in dark and light shading, respectively. (The default parameters ($C_1,C_2=\{1.44,1.92\}$) are recommended by \cite{LAUNDER1974269} and do not exactly align with the analytical line.)}
    \label{fig:C1_C2_tnbl}
\end{figure}

\begin{figure}
    \centering
    \begin{tabular}
    {@{}p{0.45\linewidth}@{\quad}p{0.45\linewidth}}
        \subfigimgthree[width=\linewidth,valign=t]
        {(a)}
        {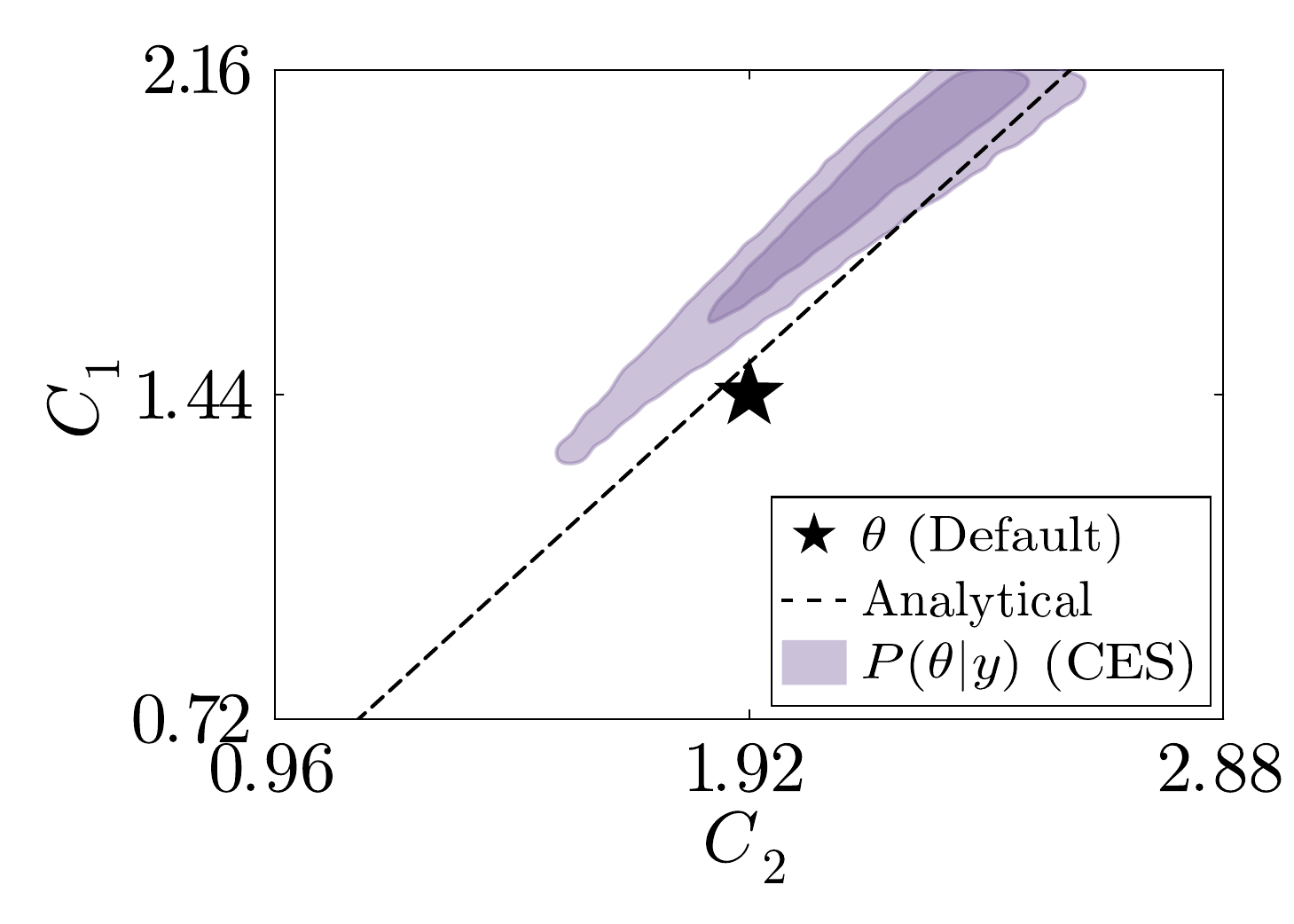} &
        \subfigimgthree[width=\linewidth,valign=t]
        {(b)}
        {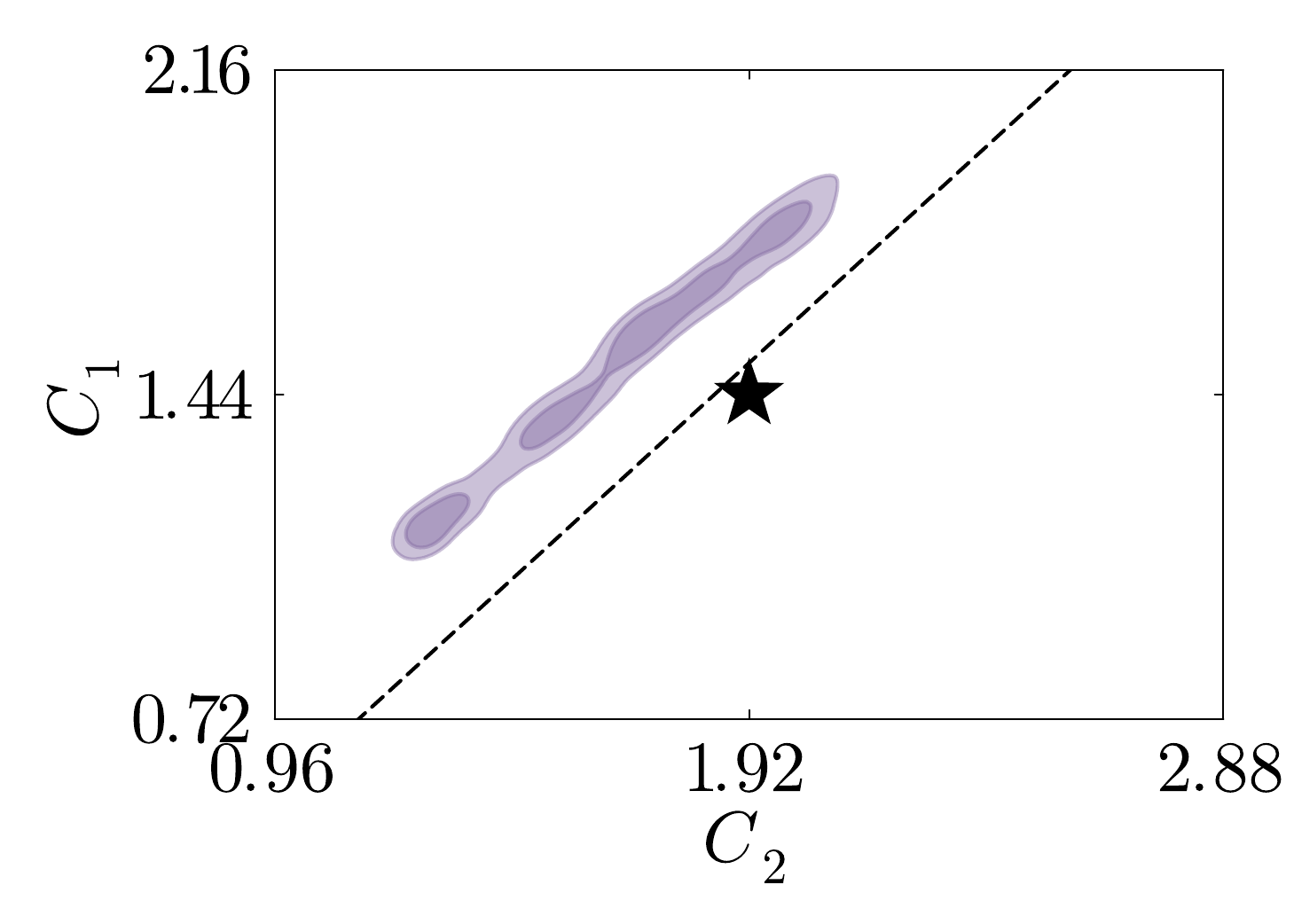} 
    \end{tabular}
    \caption{Joint posterior distribution of $C_1$ and $C_2$ in the STD $k$-$\epsilon$ model compared to the theoretical parameter relationship (dashed line) and the default parameters (star) for (a) moderately stable ASL and (b) SBL(M). 68\% and 95\% HPDR are shown in dark and light shading, respectively.}
    \label{fig:C1_C2_sbl}
\end{figure}

In this investigation, we infer the full set of model parameters in the baseline STD $k$-$\epsilon$ model from LES wind speed data from two representative vertical regions, the atmospheric surface layer (ASL) region and entire ABL region, and from two representative stability regimes, the truly neutral (TNBL) and moderately stable (SBL(M)) regimes.
The ASL is defined as the region in which the vertical momentum flux deviates from its surface value by less than 10\%.
This breakdown into four cases allows for the identification of stability- and height-dependent effects.
Since all parameters in the STD $k$-$\epsilon$ model are inferred, the resulting joint posterior of $C_1$ and $C_2$ provides a purely data-driven representation of their relationship and its uncertainty, which is compared to the theoretical expectation.

Figure~\ref{fig:C1_C2_tnbl} presents a comparison of the inferred joint posterior distributions of $C_1$ and $C_2$ for (a) neutral ASL and (b) entire ABL region for the TNBL against the analytical parameter relationship (dashed line).
In both cases, the 68\% HPDR contains the theoretical line, which implies that the modeling assumptions made in the analytical derivations are adequate approximations for neutral stratification given the observed wind speed data.
The strong variability in the data, explained by the high turbulence under neutral stratification, may explain how the relationship of the parameters remains valid even above the surface layer for the TNBL.
In Fig.~\ref{fig:C1_C2_sbl}, the comparison of the joint posterior distributions of $C_1$ and $C_2$ against the theoretical parameter relationship shows that the analytical line is contained in the 95\% HPDR for the stable ASL case but falls outside the 95\% HPDR when conditioned on the data across the entire ABL region of the SBL(M).
Although an analytical derivation is not straightforward with the inclusion of buoyancy effects, this deviation can be attributed to the combined effect of (1) sampling beyond the surface layer and (2) buoyancy effects induced by stable stratification, which are not taken into account in the derivation of Eq.~\ref{eq:C1C2}.
With an increase in buoyancy suppression of turbulence, the contribution of destruction by dissipation is predicted to decrease, causing the observed leftward shift in the $C_1$-$C_2$ balance.

These results confirm the stability-dependent and interdependent nature of model parameters and show how a data-informed approach using turbulence-resolving data could be utilized to verify other existing theoretical relationships for these model parameters.

\subsection{Learning from LES data of multiple stability regimes}
\label{sec:uq_multiple}

\begin{figure}
    \centering
        \includegraphics[width=0.9\textwidth]{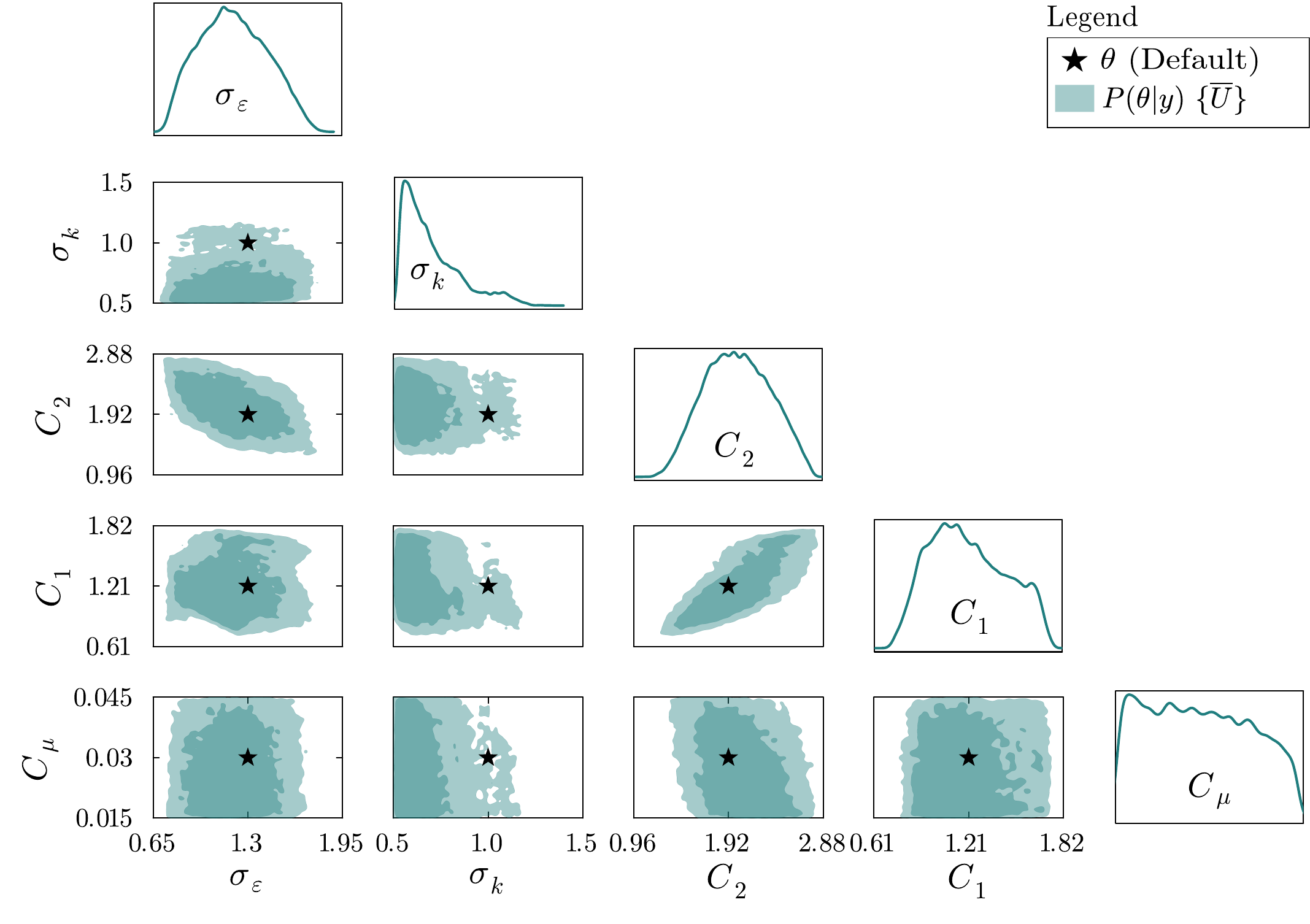}
    \caption{Marginal and joint posterior distributions of MOST $k$-$\epsilon$ parameters inferred from observing wind speed from the TNBL and SBL(M) stability regimes jointly. 68\% and 95\% HPDR are shown in dark and light shading, respectively.}
    \label{fig:combined_posterior_U}
\end{figure}

In the Bayesian framework, assimilating a larger and more diverse dataset into the model inversion allows the model to quantify uncertainties across a wider range of observations.
This facilitates a more comprehensive exploration of the parameter space and prevents the potential overfitting of parameters.
The STD $k$-$\epsilon$ model generally performs well in engineering flows because its original parameters have been calibrated, although deterministically, across a wide range of engineering flows \citep{LAUNDER1974269}.
This also explains why its default parameters struggle in predictions of ABL flows in which unobserved phenomena such as Coriolis and buoyancy effects emerge.
A larger and more diverse dataset increases the likelihood of capturing the physical processes observed in complex ABL flows \citep{baklanov2011nature}.

We build upon the previous inversion problem formulation based on a single stability regime and infer parameters from two representative stability regimes (TNBL and SBL(M)) jointly, by reformulating the loss function to minimize the data-model misfit for wind speed across both regimes.
We omit the SBL(W) case as an unobserved stability regime for testing the inferred posteriors in Sec.~\ref{sec:outofsampleflows}.
Here, we use the MOST $k$-$\epsilon$ model due to its robust performance in the different stability regimes.
To address the ill-conditioning of the sample covariance matrix caused by the disparity in the magnitude of noise between the TNBL and SBL(M) data, we utilize the Ledoit-Wolf shrinkage estimator to regularize the sample covariance matrix.
Additional details on this regularization approach are provided in \ref{sec:appendix4}.

Figure~\ref{fig:combined_posterior_U} presents the joint (68\% and 95\% HPDR) and marginal posterior distributions of the five model parameters for the MOST $k$-$\epsilon$ model, inferred jointly from the two stability regimes.
The posterior distributions of the five model parameters are analyzed in relation to the default values reported in \citet{van2017new} and those documented in previous studies.
First, the default values of all five parameters are within the statistically significant 95\% HPDR.
This indicates that, conditioned on the 10-minute averaged wind speed data of both stability regimes, MOST $k$-$\epsilon$ model's default parameters predict within the 95\% uncertainty bounds of the observational data.
Further analysis of the marginal posterior distributions shows a right-skewed distribution for $\sigma_k$ where the posterior peak is approximately 40\% below the deterministic default value.
$\sigma_k$ represents the turbulent Prandtl number for the transport of TKE.
Typically, a unitary value of $\sigma_k=1.0$ is assumed, based on traditional deterministic optimization \citep{hanjalic1972reynolds}.
However, different estimates of $\sigma_k$ exist in different models, as is apparent in the RNG $k$-$\epsilon$ model which uses $\sigma_k=0.72$.
Previous work by \cite{poroseva2001simulating} and \cite{yang2009new} indicate that the ratio of $\sigma_{\epsilon}/\sigma_k$ is more important than the exact values of the two parameters to accurately capture unbounded flows.
The variability in the uncertainty distribution for $\sigma_k$ is also observed in different flow configurations in \cite{edeling2014bayesian}.
Consequently, the posterior distribution of $\sigma_k$ observed here is deemed to be within the range of values recommended among different $k$-$\epsilon$ models.

The joint distribution of $C_1$ and $C_2$ shows a similar positive correlation to that observed in the STD $k$-$\epsilon$ model and is best explained by their contributions to the balance of shear production and dissipation of turbulence.
This data-driven result gives further support to previous efforts that have prescribed a dependency between these parameters \citep{richards1993appropriate, turgeon2001application, temel2017two}. 
The marginal posterior distribution of $C_2$ is centered around 1.92—a commonly used value that, while not exclusive, is frequently cited in the literature and has been associated with measurements from decaying turbulence \citep{detering1985application}.
As a final note on the relationship of $C_1$ and $C_2$, this study also observes the widely accepted constraint $C_2>C_1$, as discussed in \citet{xiao2019quantification} and supported by experimental evidence in \citet{ray2018learning}.
Further analysis of the posterior, including its comparison to those inferred from observing additional statistics, is presented in Sec.~\ref{sec:informative_statistics}.

\section{Results: Parameter uncertainty reduction}
\label{sec:results_uncertaintyreduction}

In this section, the CES methodology is used as a framework to reduce parameter uncertainty in the MOST $k$-$\epsilon$ model based on the degree of information contained within the observed data.
This addresses a key question of what data should be used in the calibration process to maximally reduce the uncertainty in model predictions.
We examine this framework in the context of selecting vertical sensing locations and fluid flow quantities to be assimilated into the Bayesian inversion.

\subsection{Targeted selection of vertical sensing locations}
\label{sec:height_experiment}

The role of observational sensing locations in the information gain for computational models has been extensively studied in optimal experimental design frameworks \citep{papadimitriou2015optimal,semaan2017optimal,chen2022optimal}.
Building on this foundation, we investigate whether targeted selection of vertical sensing locations can reduce parameter uncertainty in turbulence parameterizations and offer insight into optimal sensor placement for meteorological applications.
A central question we pose is whether assimilating measurements at higher elevations provides greater information for parameter inference. 
This has practical relevance for the design of meteorological (met) masts, which are typically 100–200 m tall, with only a few reaching up to 500 m \citep{ramon2020tall}. 
In practice, the height of met masts is often determined by engineering constraints and cost \citep{perrin2007effect, li2010boundary, munger2011measurement}. 
However, recent studies suggest that measurements between 200 and 600 m can enhance mesoscale wind speed forecasts \citep{sommerfeld_improving_2019}, motivating a deeper investigation into the value of higher-elevation data.

\begin{figure}
    \centering
        \includegraphics[width=0.45\textwidth]{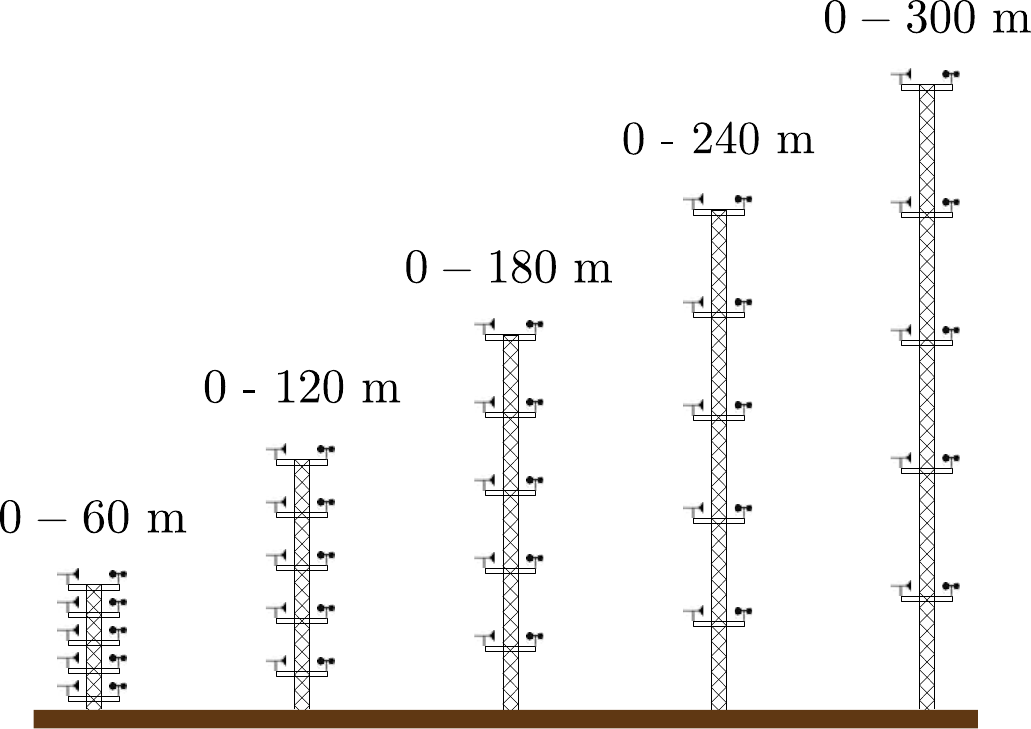}
    \caption{Visualization of virtual met masts spanning different vertical regions, increasing in 60 m increments to 300 m, each with five sensing locations.}
    \label{fig:met_mast_schematic}
\end{figure}

To assess this, we simulate virtual met masts with heights ranging up to 300 m, incremented in 60 m steps.
In other words, we assimilate wind speed data from increasingly taller vertical regions into the inverse problem.
Figure~\ref{fig:met_mast_schematic} illustrates this numerical experiment setup.
For each met mast, we fix the number of sensing locations to five—within the range commonly used in operational systems—to isolate the effect of height-dependent information gain from the effect of increased data volume.
This approach ensures that any observed improvements in parameter inference are due to the sensing height, not simply to more observations.
Therefore, each vertical region (or met mast) is spanned by five sensing locations.
The cumulative 60 m increment design reflects the fact that lower-elevation measurements are more financially and logistically feasible than those at higher altitudes. 

\begin{figure}
    \centering
    \begin{tabular}
    {@{}p{0.45\linewidth}@{\quad}p{0.45\linewidth}}
        \subfigimgthree[width=\linewidth,valign=t]
        {(a)}
        {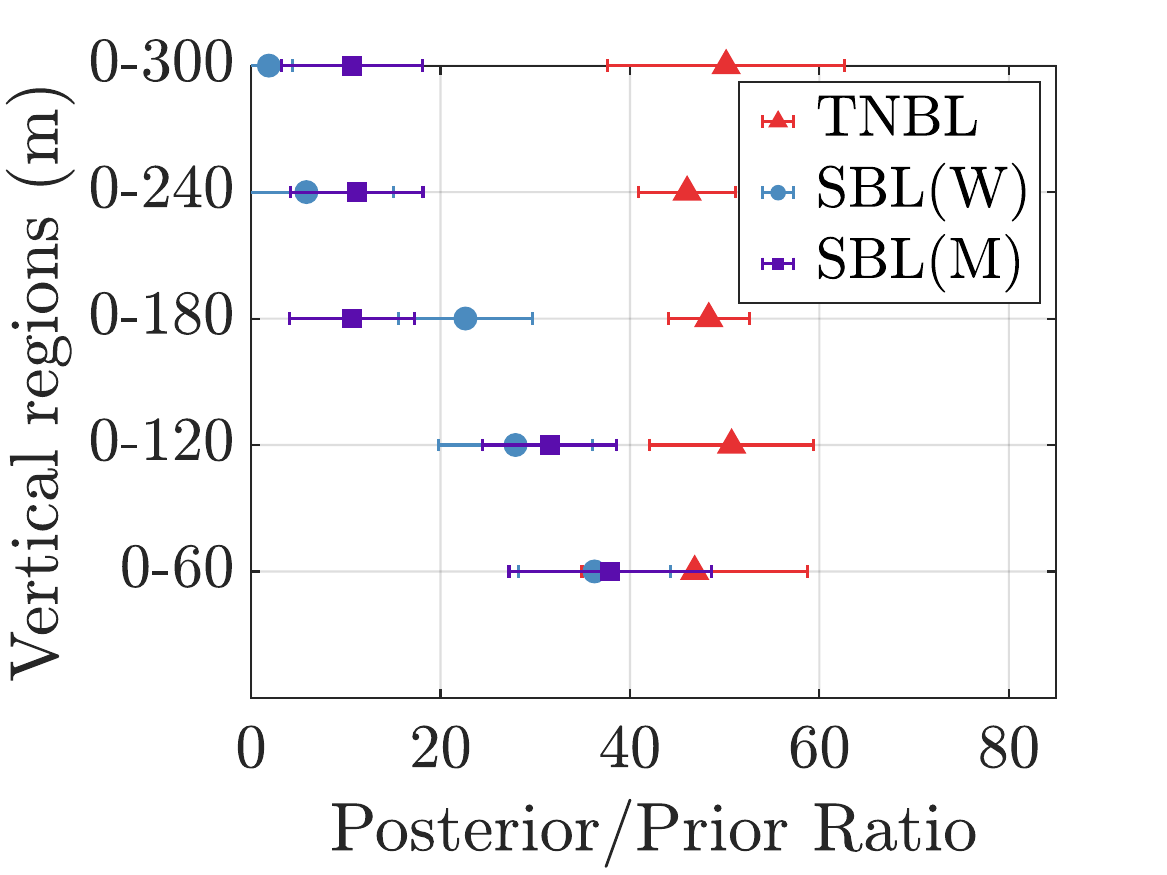} &
        \subfigimgthree[width=\linewidth,valign=t]
        {(b)}
        {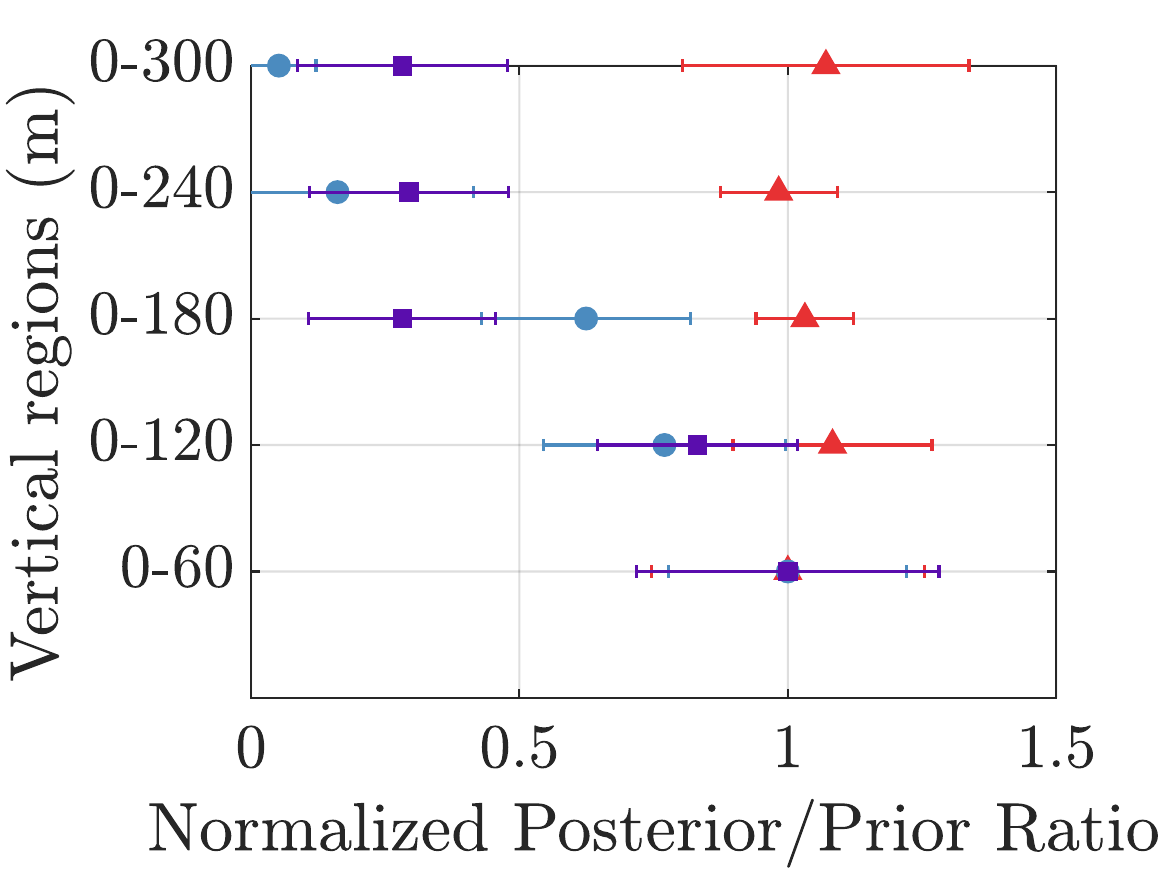} 
    \end{tabular}
    \caption{(a) Parameter uncertainty from sampling at different vertical regions for each stability regime, represented by the 68\%-posterior-prior ratio, (b) Same result as (a) but normalized by the 68\%-posterior-prior ratio at the first vertical region (0-60m).}
    \label{fig:ratio_vertical_regimes}
\end{figure}

Wind speed data from single stability regimes are used to drive the inverse problem, allowing for height- and stability-dependent effects to be isolated in the uncertainty analysis.
For each stability regime, a sample of the LES wind speed data from the specified vertical region is observed in the Bayesian inversion to infer turbulence model parameters in the MOST $k$-$\epsilon$ model.
Following a previous study \citep{howland2022parameter}, the reduction in parameter uncertainty, or information gain, in each case is quantified as the ratio of the 68\% area of the posterior to the 68\% area of the prior.
We quantify uncertainty reduction for two key parameters, $C_1$ and $C_2$, which are shown to dominantly influence the prediction of velocity profiles \citep{edeling2014bayesian}.
Figure~\ref{fig:ratio_vertical_regimes} shows the quantification of the parameter uncertainty reduction over the different vertical regions for each stability regime as 68\%-posterior-prior ratios.
The 95\% error bars are computed from five instantiations of parameter UQ observing a different LES sample each time.
First, no statistically significant trend in parameter uncertainty is observed for the TNBL case in sampling increasingly taller vertical regions up to 300 m.
The consistent posterior-prior ratio of approximately 50\% is caused by the larger internal variability in the observed data, due to the larger shear production of turbulence and the absence of buoyancy TKE destruction.
This implies that in neutral stratification, sampling wind speed measurement at higher elevations does not guarantee information gain.

In stable stratification, the prescribed cooling rate that drives stable stratification suppresses the TKE closer to the surface, which leads to smaller variability in the SBL truth data compared to the TNBL truth data.
This is likely the reason for the smaller posterior-prior ratio across all vertical regions in stable stratification in comparison to the TBNL case.
For both weakly (SBL(W)) and moderately stable (SBL(M)) stratification, we observe a clear trend of parameter uncertainty reduction with measurements from taller vertical regions, indicating an increase in information gain in the flow statistics from higher elevations.
Notably, the greatest reduction in the posterior-prior ratio occurs when sampling extends up to the respective ABL height of each SBL case.
The ABL height for the weakly stable case (SBL(W)) is estimated to be 234 m, which corresponds to a reduction of approximately 74\% in the posterior-prior ratio compared to the vertical region just below, and 84\% relative to the lowest vertical region.
A similar trend is observed in the moderately stable case (SBL(M)), where the ABL height is estimated at 180.9 m. 
In this case, the maximum parameter uncertainty reduction achieved by observing statistics up to the ABL height is around 66\% relative to the region just below and 71\% compared to the lowest region.
Sampling above these regions does not appear to provide a statistically significant additional information gain as further decreases/increases remain within the 95\% spread of multiple instantiations.
This is consistent with expectations that above the ABL height, the top boundary condition prescribes the geostrophic wind.

\begin{figure}
    \centering
    \begin{tabular}
    {@{}p{0.45\linewidth}@{\quad}p{0.45\linewidth}}
        \subfigimgthree[width=\linewidth,valign=t]
        {(a)}
        {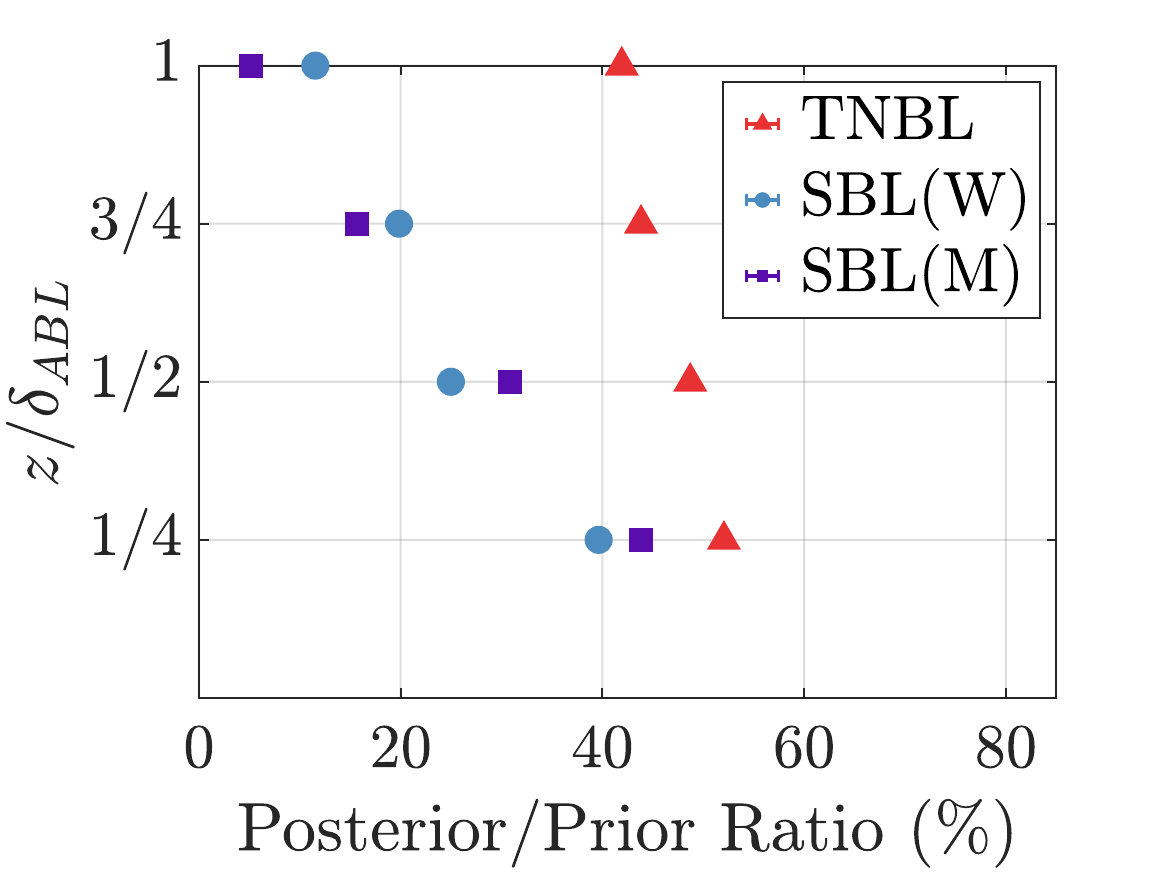} &
        \subfigimgthree[width=\linewidth,valign=t]
        {(b)}
        {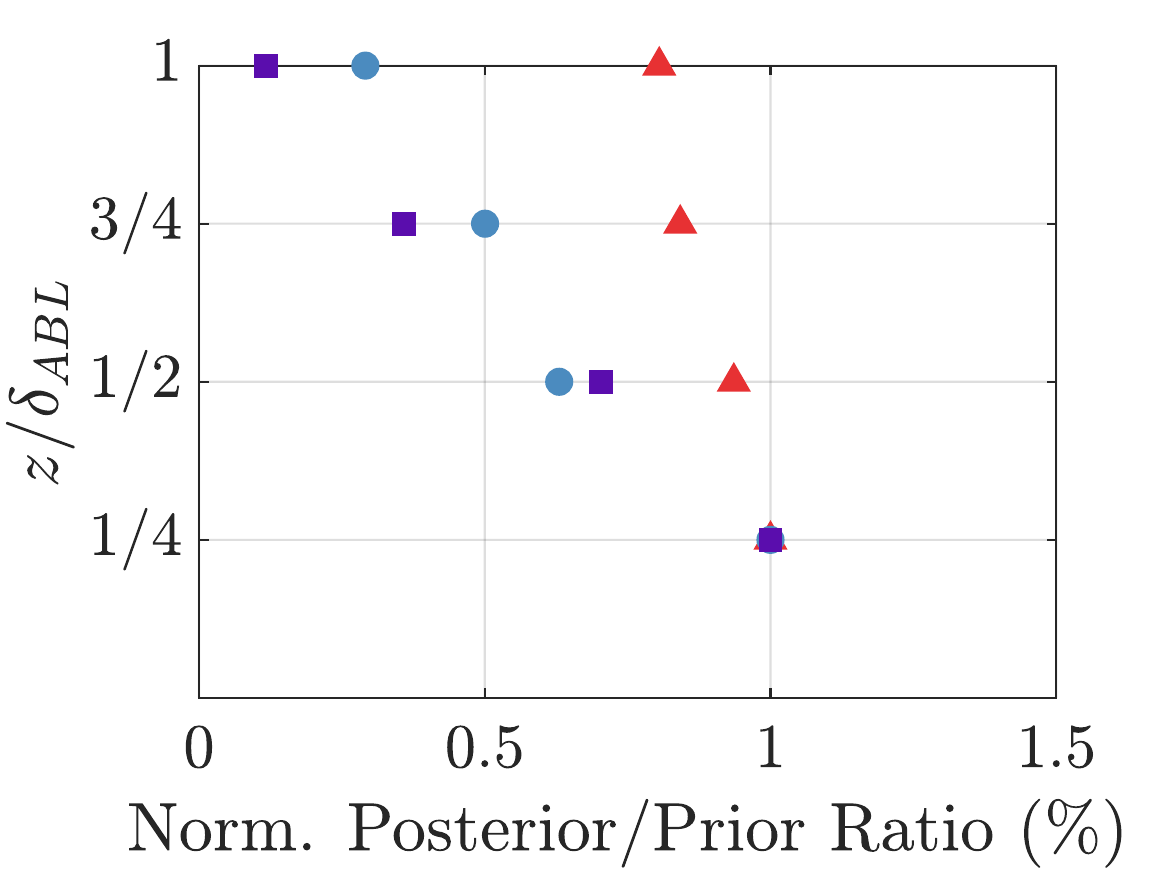} 
    \end{tabular}
    \caption{(a) Parameter uncertainty reduction from sampling at different nondimensional vertical regions up to the ABL height $\delta_{ABL}$ in each stability regime, represented by the 68\%-posterior-prior ratio, (b) Same result as (a) but normalized by the 68\%-posterior-prior ratio at the first vertical region (0-1/4 $z/\delta_{ABL}$).}
    \label{fig:ratio_nondim_vertical_regimes}
\end{figure}

We conduct a complementary test to better understand the physical mechanism for the observed reduction in parameter uncertainty.
The mechanisms for the ABL height are different for neutral and stable stratification: under stable stratification, the ABL height is primarily dictated by the buoyant suppression of turbulence, whereas under neutral stratification, where buoyancy effects are negligible, it is determined by the decay of shear-generated turbulence with height.
By sampling up to the respective ABL heights for each stability regime, we examine how buoyancy and shear production individually contribute to uncertainty reduction in parameter inference.

The vertical region up to the ABL height, $\delta_{ABL}$ is partitioned into four cumulative regions defined by normalized heights
\begin{equation}
    z/\delta_{ABL} \in \{ [0,1/4], [0,1/2], [0,3/4],[0,1] \}
\end{equation}
and number of observed data points is set to the maximum allowable to maintain a full-rank noise covariance matrix within the lowest vertical region, given the resolution of the LES data.
In each stability case, the sample of wind speed data is fixed but observed in the different vertical regions.
Figure~\ref{fig:ratio_nondim_vertical_regimes}a shows the results of the posterior-prior ratio calculated in the different vertical regions for the three stability regimes, while Fig.~\ref{fig:ratio_nondim_vertical_regimes}b shows the results normalized by the posterior-prior ratio of the first vertical region ($[0,1/4]\frac{z}{\delta_{ABL}}$).
Although a clear linear reduction in parameter uncertainty is observed with increasing height over all three stability regimes, the reduction is less pronounced for the neutral stratification case.
In stable cases, the posterior-to-prior ratio of approximately 40\%, inferred from wind speed data in the ($[0,1/4]\frac{z}{\delta_{ABL}}$) region, reduces to approximately 10\%, inferred from the ($[0,1]\frac{z}{\delta_{ABL}}$) region.
However, in neutral stratification (TNBL), the posterior-to-prior ratio at approximately 50\%, inferred from the ($[0,1/4]\frac{z}{\delta_{ABL}}$) region, reduces to 40\%, inferred from the ($[0,1]\frac{z}{\delta_{ABL}}$) region.
These results imply that buoyant suppression of turbulence plays a much more significant role than the reduction of shear production of turbulence in driving information gain for model parameters.
In other words, knowledge of atmospheric stratification is an important factor in reducing parameter uncertainty.

From a practical point of view, these two numerical experiments on vertical sensing locations offer insight into the added value of measurements taken at higher elevations. 
The information gain is especially pronounced under stable stratification, where the ABL height is determined to be a critical height at which parameter uncertainty is maximally reduced.
Future studies will validate this trend using observational data over a broader range of stability regimes, after which practical met mast height recommendations can be made.

\subsection{Targeted selection of fluid flow quantities}
\label{sec:informative_statistics}

\begin{table}
\caption{Combinations of Flow Quantities Tested for Information Gain}
\label{tab:informativestatistics}
\centering
\begin{tabular}{lcccc}
\hline\noalign{\smallskip}
 & Baseline & Wind direction-informed & TKE-informed & All-informed \\
\noalign{\smallskip}\hline\noalign{\smallskip}
Variables & $\overline{U}$ & $\overline{U},\overline{\phi}$ & $\overline{U},\overline{k}$ & $\overline{U},\overline{\phi},\overline{k}$ \\
\noalign{\smallskip}\hline
\end{tabular}
\end{table}

\begin{figure}
    \centering
        \includegraphics[width=0.9\textwidth]{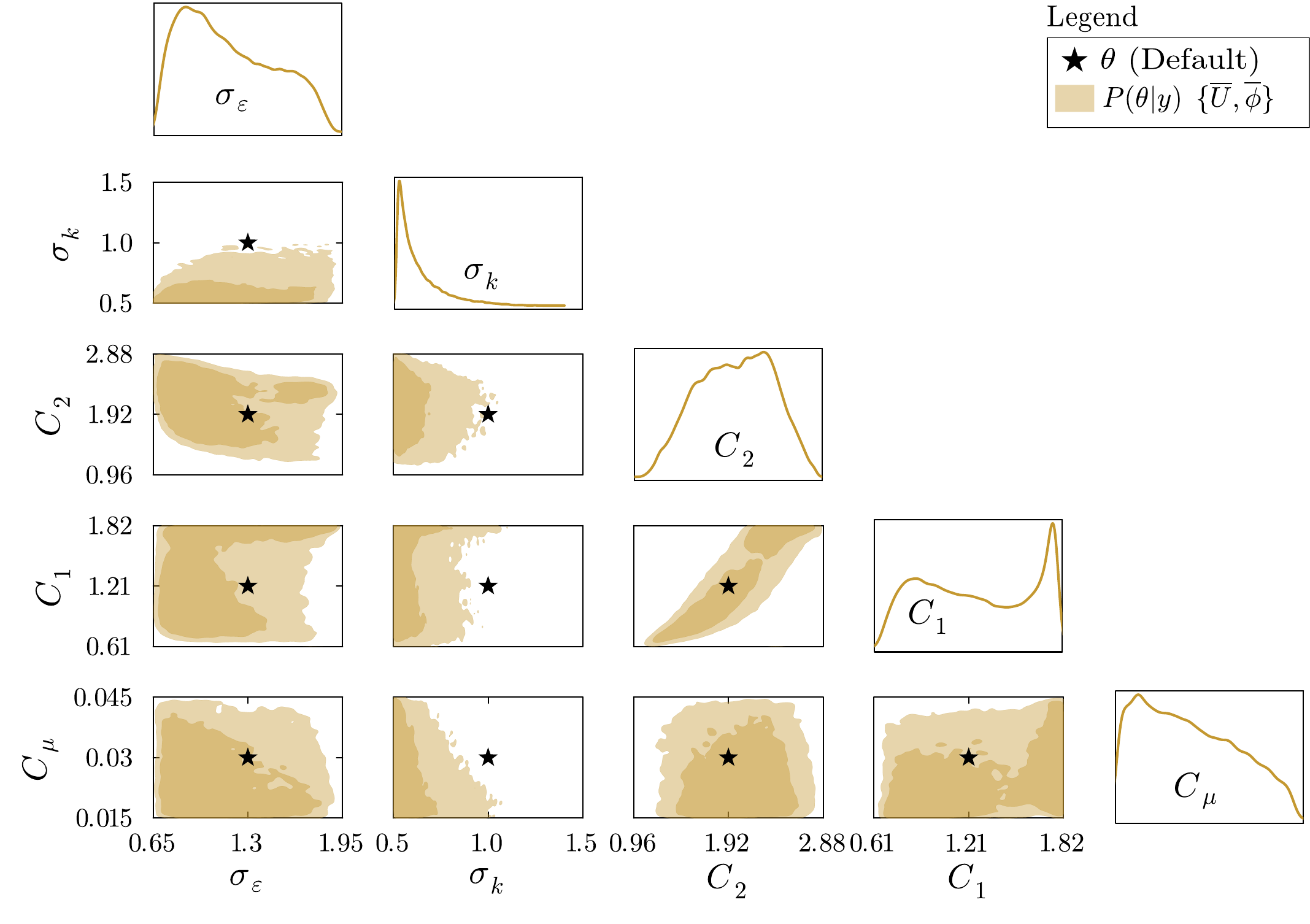}
    \caption{Marginal and joint posterior distributions of MOST $k$-$\epsilon$ parameters inferred from observing wind speed and wind direction from two stability regimes (TNBL, SBL(M)) jointly. 68\% and 95\% HPDR are shown in dark and light shading, respectively.}
    \label{fig:combined_posterior_Udir}
\end{figure}

\begin{figure}
    \centering
        \includegraphics[width=0.9\textwidth]{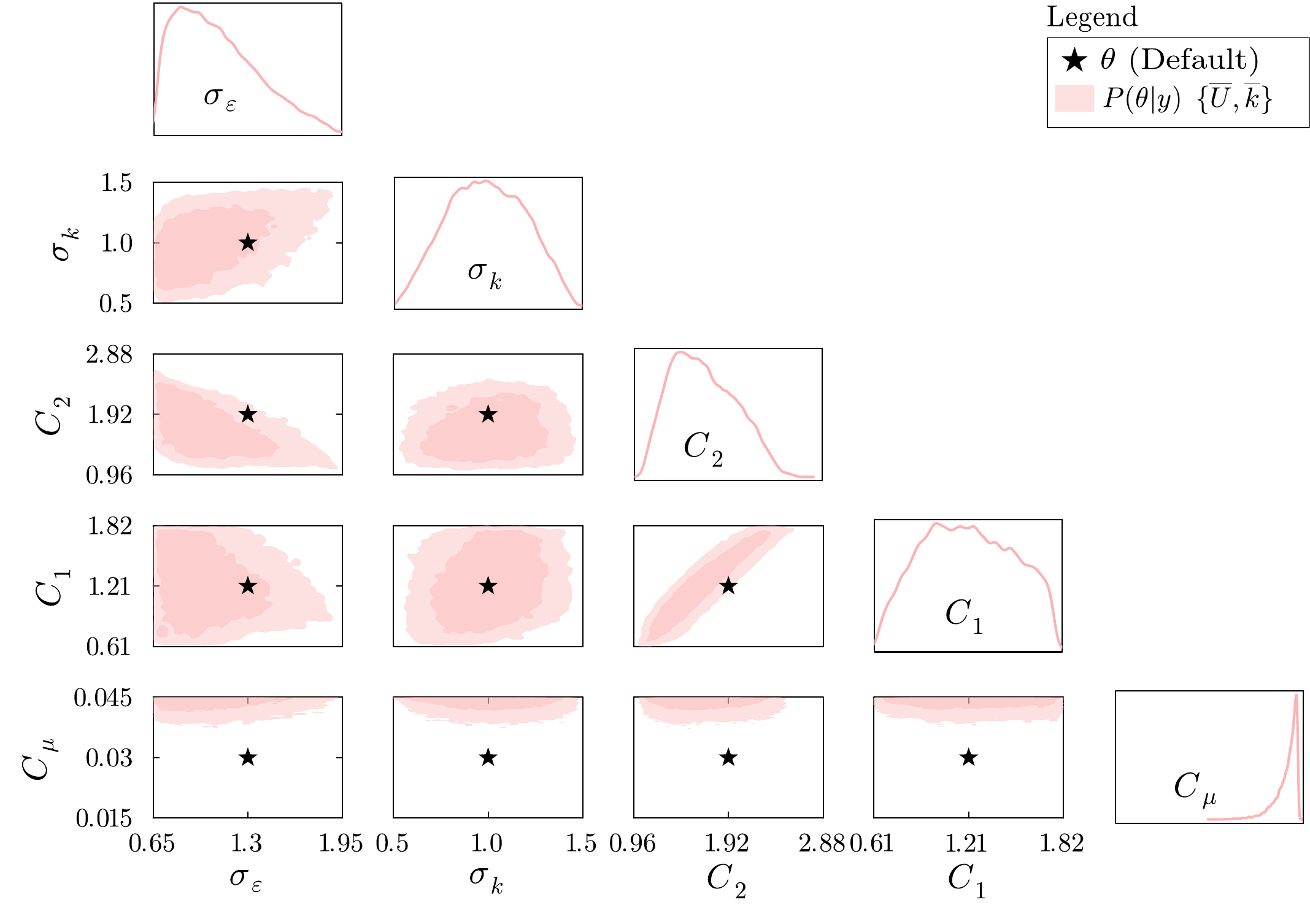}
    \caption{Marginal and joint posterior distributions of MOST $k$-$\epsilon$ parameters inferred from observing wind speed and TKE from two stability regimes (TNBL, SBL(M)) jointly. 68\% and 95\% HPDR are shown in dark and light shading, respectively.}
    \label{fig:combined_posterior_Uk}
\end{figure}

We investigate the ABL flow quantities beyond wind speed that are most effective in reducing uncertainty in turbulence model parameters.
CES-based UQ to two stability regimes (see details in Sec.~\ref{sec:uq_multiple}) is extended to facilitate exploration of the additional information value of the wind direction $\overline{\phi}$ and TKE $\overline{k}$ in the model inversion.
These two statistics are useful QOIs that depend significantly on the turbulence model used, and therefore show potential for additional information gain.
Table~\ref{tab:informativestatistics} presents the four different combinations of statistics that are used in the formulation of the inverse problem.
The baseline approach consists of calibration to wind speed profiles $\overline{U}$ from two stability regimes, TNBL and SBL(M).
The second combination, which consists of wind speed $\overline{U}$ and wind direction $\overline{\phi}$, is the wind direction-informed variant.
The third combination, which consists of wind speed $\overline{U}$ and TKE $\overline{k}$ is the TKE-informed variant.
The fourth combination consists of wind speed $\overline{U}$, wind direction $\overline{\phi}$, and TKE $\overline{k}$.
All four combinations include wind speed, as it is the most common variable measured during a measurement campaign.
Although wind speed and direction are commonly measured using standard instruments, TKE is a measurement derived from the three-dimensional wind vector sampled at high frequency using a fast-response sensor \citep{bonin2017evaluation}, making it a more difficult quantity to measure.

In each combination, the profiles of the relevant variables are concatenated into a single vector of the observed statistics.
To ensure a fair comparison and eliminate the potential confounding effect of differing noise covariance matrix dimensions, we use the same number of combined data points across the four combinations.
The Ledoit-Wolf shrinkage estimator is used to regularize the sample covariance matrix as in Sec.~\ref{sec:uq_multiple}.

\begin{figure}
    \centering
        \includegraphics[width=0.5\textwidth]{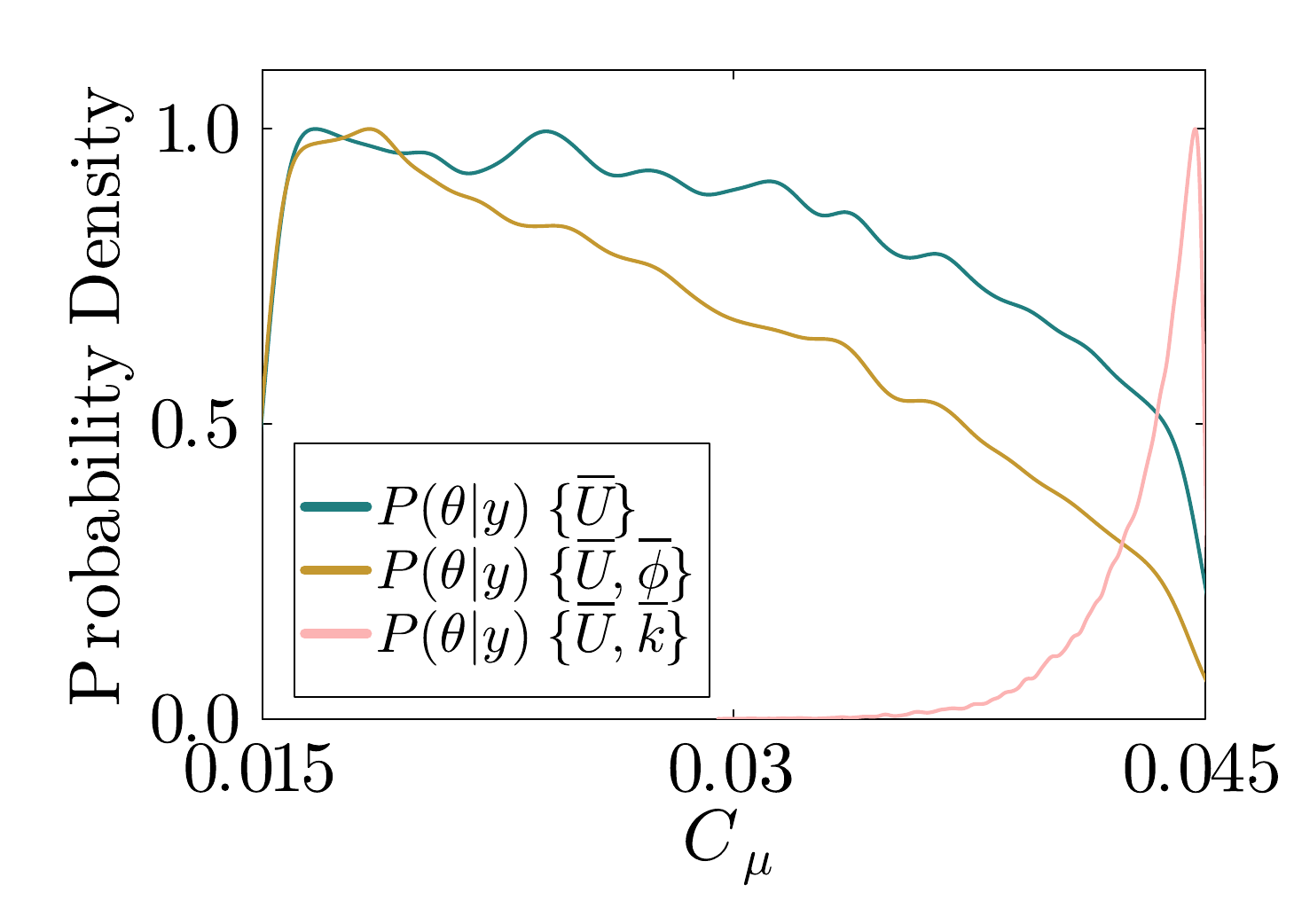}
    \caption{Marginal posterior distributions of $C_\mu$ inferred from observing combinations of wind speed $\{\overline{U}\}$ (turqoise), wind speed and wind direction $\{\overline{U},\overline{\phi}\}$ (brown), and wind speed and TKE $\{\overline{U},\overline{k}\}$ (pink) from two stability regimes (TNBL, SBL(M)).}
    \label{fig:marginal_Cmu}
\end{figure}

The information gain from the different flow quantities is evaluated from two perspectives: an inverse perspective that examines the parameter uncertainty and a forward perspective that examines the uncertainty propagated into the model predictions.
In the inverse perspective, the posteriors inferred from the different combinations of flow quantities are compared.
Figure~\ref{fig:combined_posterior_U} from Sec.~\ref{sec:uq_multiple} presents the joint and marginal posterior distributions of MOST $k$-$\epsilon$ parameters inferred from wind speed LES data from the TNBL and SBL(M) cases jointly.
Figure~\ref{fig:combined_posterior_Udir} presents the joint and marginal posterior distributions of parameters inferred from wind speed and wind direction LES data.
And finally, Fig.~\ref{fig:combined_posterior_Uk} presents the joint and marginal posterior distributions of parameters inferred from wind speed and resolved TKE LES data.
The posterior inferred from the fourth combination is qualitatively similar to the posterior inferred from wind speed and TKE (not shown).
Since the assimilated model output now includes more than just the wind speed, the joint posterior distribution of $C_1$ and $C_2$ is no longer the primary indicator of the parameter uncertainty.
Other parameters may exhibit greater sensitivity to different model outputs, and uncertainty quantification offers a systematic way to identify these sensitivities.
As an example of this phenomenon, Fig.~\ref{fig:marginal_Cmu} shows a comparison of the marginal distributions of the eddy viscosity coefficient $C_\mu$ inferred from the three combinations.
While the baseline and wind direction-informed combinations result in posteriors that are similar to the uniform prior, the TKE-informed combination exhibits a strongly left-skewed distribution near the upper prior bound.
The information gain in each case, quantified by the previously defined 68\% posterior-to-prior ratio, is 89\%, 78\%, and 11\% for the baseline, wind direction-informed, and TKE-informed combinations, respectively.
This result highlights the value of assimilating different model outputs, which reduces parameter uncertainty more effectively than simply increasing the quantity of observed wind speed data.

\begin{figure}
    \centering
        \includegraphics[width=0.6\textwidth]{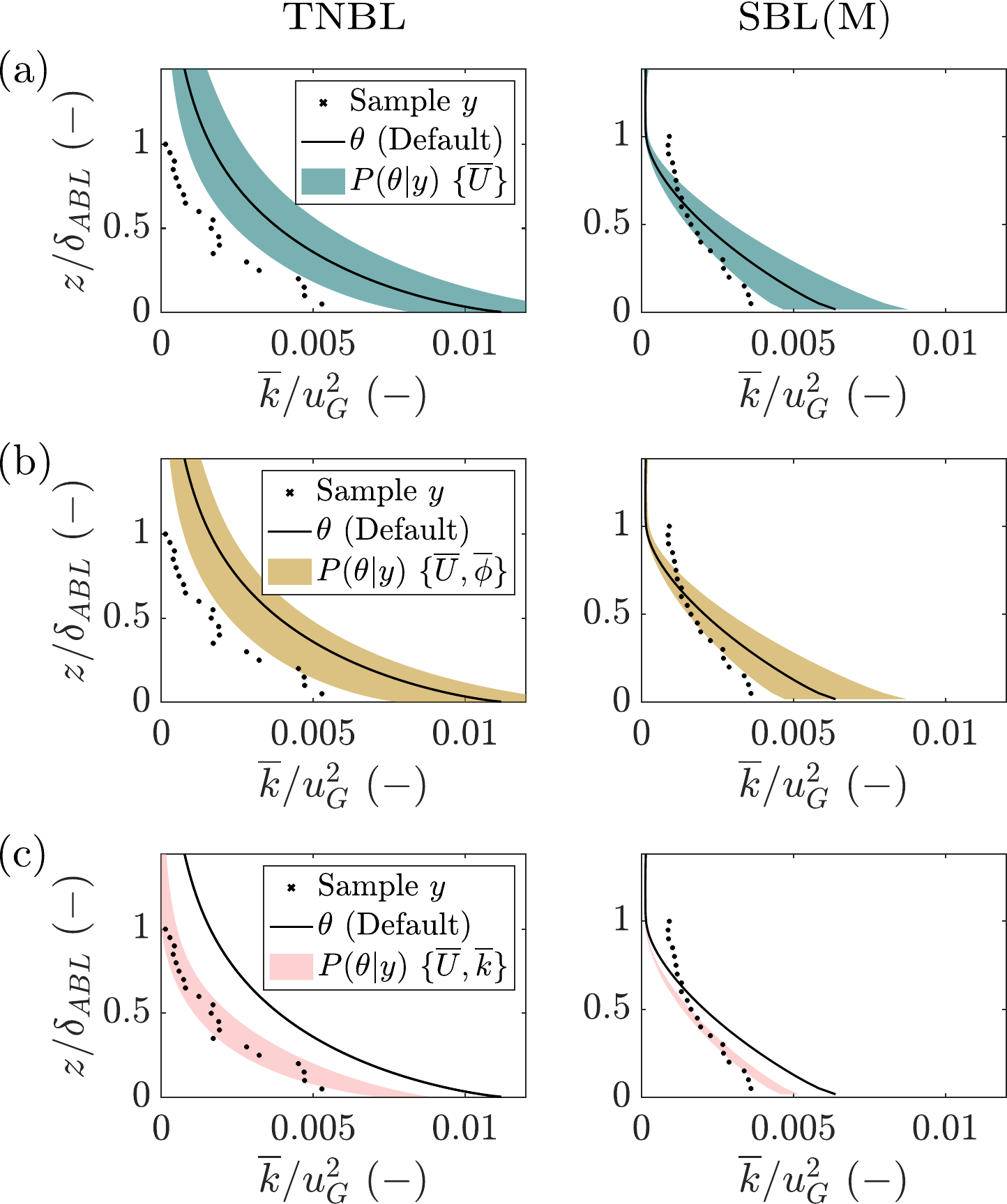}
    \caption{TKE $k$ predictions with 95\% credible intervals for the two stability regimes (TNBL, SBL(M)) generated from posteriors inferred from (a) baseline combination, (b) wind direction-informed combination, and (c) TKE-informed combination. Variables: $z$ is height, $\delta_{ABL}$ is ABL height, $\overline{k}$ is TKE, and $u_G$ is geostrophic wind speed.}
    \label{fig:tke_credibleintervals}
\end{figure}

The information gain from the choice of fluid flow quantities is further explored in the corresponding forward UQ predictions.
The posteriors inferred from the different combinations of flow quantities are forward propagated to predict flow quantities in the TNBL and SBL(M) regimes.
We focus on the credible intervals for TKE because the predictions of wind speed and wind direction are shown to be similar in all three combinations (see \ref{sec:appendix4}).
Figure~\ref{fig:tke_credibleintervals} shows the TKE predictions with 95\% credible intervals from each posterior, the default-parameter predictions, and the sample $\boldsymbol{y}$, which is observed only in the TKE-informed variant.
Posteriors inferred using only wind speed or wind speed and wind direction fail to capture the LES TKE sample within their credible intervals, and show a systematic overprediction of up to 50\%.
In contrast, the TKE-informed posterior yields significantly narrower credible intervals that fully contain the LES sample in the TNBL case and partially capture it in the SBL(M) case.
The latter is attributed to model-form discrepancy due to the nonstationary nature of higher-order moment statistics, especially in the stable boundary layer \citep{mahrt2009characteristics}.
This improvement in TKE prediction is explained by the information gain in $C_\mu$ (Fig.~\ref{fig:marginal_Cmu}), which formulates the surface boundary conditions for TKE and the TKE dissipation rate.
Here, we use $k_s=u_*^2/\sqrt{C_\mu}$ and $\epsilon_s=u_*^3/\kappa z$ where the subscript $s$ indicates the surface value, as commonly recommended in previous studies \citep{detering1985application, richards1993appropriate}.
As the formulation for the surface value of $k$ suggests, an increase in $C_\mu$ leads to a decrease in TKE, which results in the improvement in prediction shown in Fig.~\ref{fig:tke_credibleintervals}.
The overprediction is possibly related to the low default value $C_\mu=0.03$, as discussed previously in \citep{richards2019appropriate} and more recently in \cite{baungaard2024simulation}.

These results of reduced uncertainty in TKE are significant for two reasons.
First, they align with the hypothesis that model parameters inferred from first-order moment statistics can be sufficient to predict first-order moment statistics but may fall short when attempting to capture higher-order statistics.
This is critical as TKE predictions by operation models are important in applications such as dispersion modeling or energy meteorology.
Second, since the amount of observed data remains constant across combinations, the reduction in parameter and prediction uncertainty highlights the additional information gain provided by TKE in calibrating turbulence parameterizations.
Assimilating a quantity more directly linked to the model parameters, such as TKE, proves informative and contributes to reducing parameter uncertainty.
Whether this remains true with field measurement data remains an open question and is left for future investigation.

\subsubsection{Testing parameter posteriors in out-of-sample flows}
\label{sec:outofsampleflows}

\begin{figure}
    \centering
        \includegraphics[width=0.8\textwidth]{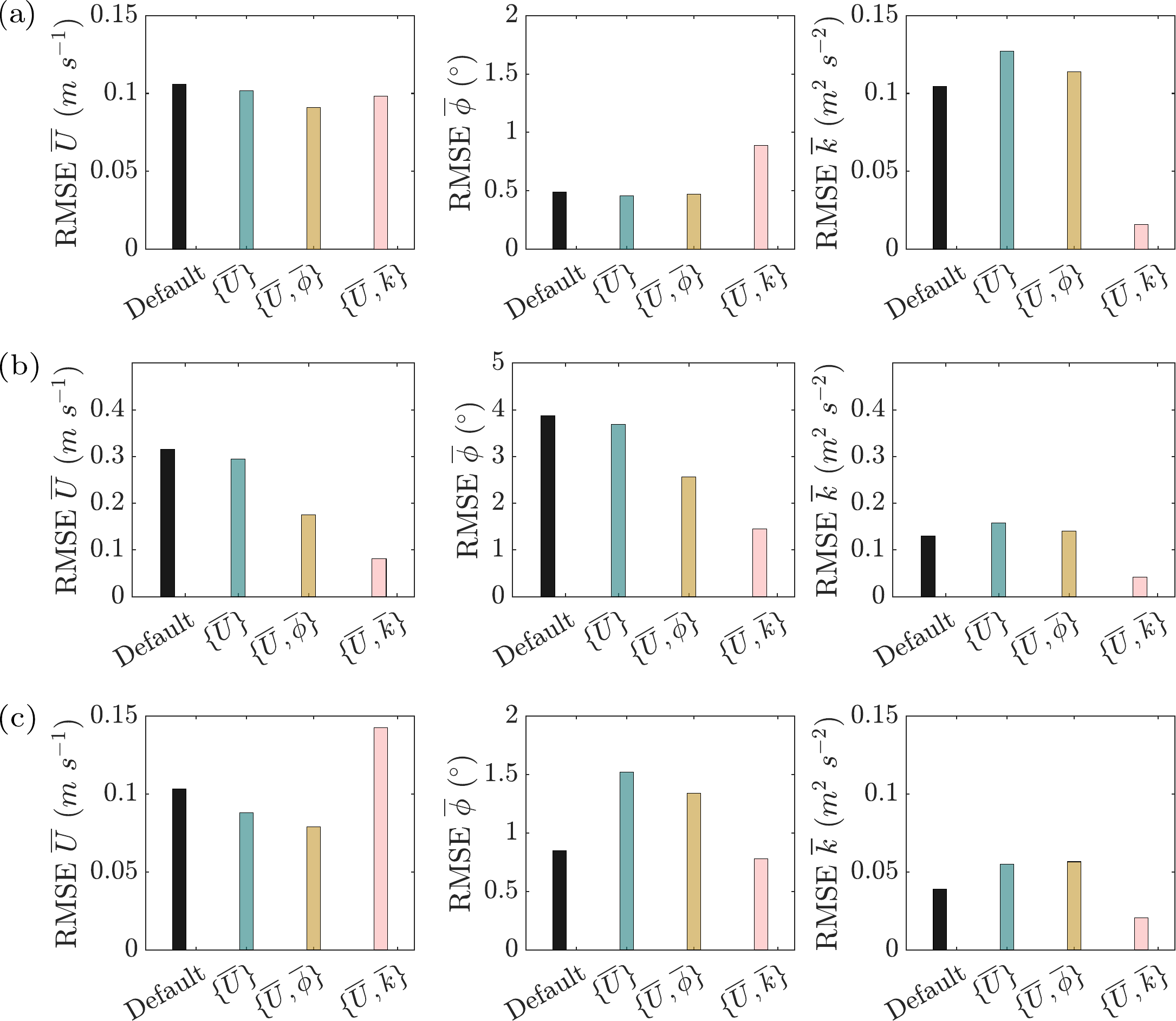}
    \caption{Root-mean-squared-error (RMSE) of dimensional wind speed $\overline{U}$, wind direction $\overline{\phi}$, and TKE $\overline{k}$ predictions by default parameters and the three posteriors against mean LES data for out-of-sample flows: (a) observed stability regime (TNBL) with unobserved surface roughness $z_0=0.0001~\mathrm{m}$ (b) unobserved stability regime (SBL(W)) with observed surface roughness $z_0=0.1~\mathrm{m}$ and (c) unobserved stability regime (SBL(W)) with unobserved surface roughness $z_0=0.0001~\mathrm{m}$.}
    \label{fig:outofsample_rmse}
\end{figure}

\begin{figure}
    \centering
        \includegraphics[width=0.8\textwidth]{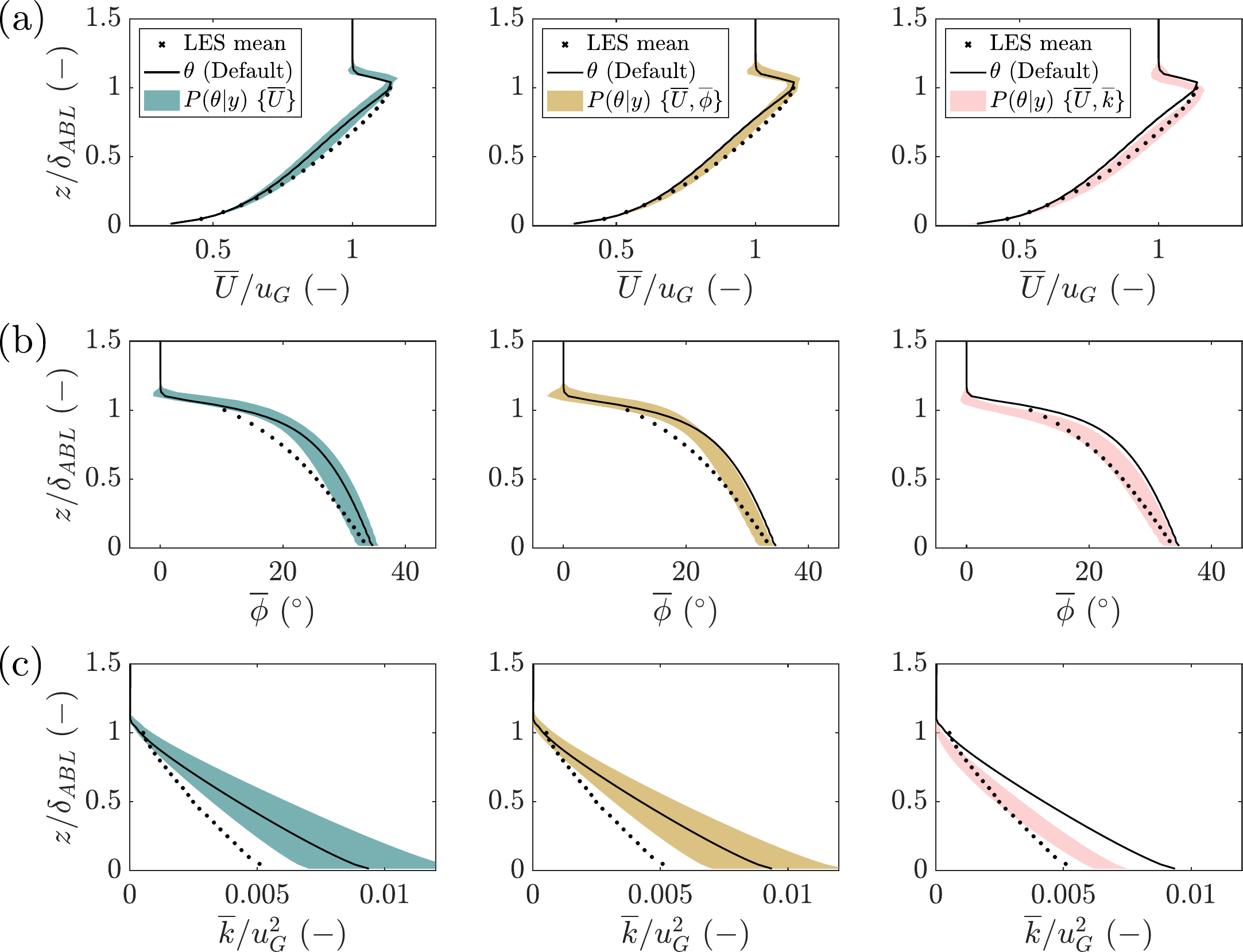}
    \caption{ Out-of-sample flow of unobserved stability regime (SBL(W)) and observed surface roughness ($z_0=0.1~\mathrm{m}$). Predictions by inferred posteriors compared against default-parameter predictions and LES data for (a) non-dimensional wind speed $\overline{U}/u_G$, (b) wind direction $\overline{\phi}$, and (c) non-dimensional TKE $\overline{k}/u^2_G$, for the three different posteriors (organized by column). Variables: $z$ is height, $\delta_{ABL}$ is ABL height, $\overline{U}$ is time-averaged wind speed, and $u_G$ is geostrophic wind speed.}
    \label{fig:outofsample_profiles}
\end{figure}

A major challenge in turbulence modeling is the generalizability of turbulence models to many different flows \citep{xiao2019quantification}.
In addressing both the predictive accuracy and generalizability of the parameters learned using UQ, we test the three posteriors inferred in Sec.~\ref{sec:informative_statistics} against unobserved data.
The observed (training) data are the model outputs from the calibration flow assimilated during the Bayesian inversion.
The unobserved (test) data may differ from the observed data in several aspects, including unseen model outputs, flow parameters, and geometric configurations \citep{wu2018physics}.

We consider changes in the stability and surface roughness $z_0$ for out-of-sample flows, since both of these ABL flow parameters significantly change the turbulence characteristics of the flow.
First, posteriors, which are calibrated to two stability regimes (TNBL, SBL(M)) at $z_0=0.1~\mathrm{m}$, are tested for an observed TNBL stability regime with an unobserved roughness length, $z_0=0.0001~\mathrm{m}$.
The three orders of magnitude difference between the training and test roughness length values, $z_0=[0.1, 0.0001]~\mathrm{m}$, ensures a fair evaluation of the calibrated parameters’ performance under unobserved turbulence conditions.
Second, posteriors are tested for the unobserved SBL(W) stability regime with an observed roughness length, $z_0=0.1~\mathrm{m}$.
Lastly, the posteriors are tested for the unobserved SBL(W) stability regime with an unobserved roughness length, $z_0=0.0001~\mathrm{m}$.

Figure~\ref{fig:outofsample_rmse} compares the root-mean-squared errors (RMSE) of wind speed, wind direction, and TKE predictions with respect to mean LES data for out-of-sample flows.
Predictions by default parameters are shown in black, while predictions by the three posteriors, representing different variables assimilated in the UQ process, are shown in colors consistent with Sec.~\ref{sec:informative_statistics}.
For the first out-of-sample flow test of the observed TNBL stability regime with unobserved surface roughness, $z_0=0.0001~\mathrm{m}$, shown in Fig.~\ref{fig:outofsample_rmse}a, all three posteriors yield lower wind speed prediction error than the default parameters, with reductions of up to 13\%.
Although the additional assimilation of wind direction to the posterior inference minimally impacts the wind direction predictions, the assimilation of TKE significantly reduces the mean TKE prediction error by approximately 85\%, underscoring the importance of TKE as detailed in Sec.~\ref{sec:informative_statistics}.
For the second out-of-sample flow test of the unobserved SBL(W) stability regime with observed surface roughness, $z_0=0.1~\mathrm{m}$, shown in Fig.~\ref{fig:outofsample_rmse}b, the RMSE for wind speed and wind direction demonstrate consistent improvements across all three posteriors.
With the additional assimilation of wind direction in the model inversion, the RMSE of wind direction reduces by approximately 34\% and 30\% from error by the default-parameters and baseline wind speed-assimilated posterior, respectively.
With the additional assimilation of TKE, a 63\% reduction in error is observed for TKE predictions with respect to the error in the default-parameter prediction.
We note that the assimilation of TKE also results in the lowest error for wind direction prediction.
This is explored further in Fig.~\ref{fig:outofsample_profiles}, which presents the 95\% spread of the forward UQ predictions from each of the three posteriors, predictions by the default parameters, and mean LES predictions, for wind speed, wind direction, and TKE.
The posterior calibrated jointly to wind speed and TKE ${\overline{U}, \overline{k}}$ (pink) improves TKE predictions with respect to mean LES data, aligning with observations made in Sec.~\ref{sec:informative_statistics}.
For the weakly stable stability regime, this improvement in the modeled TKE is accompanied by a reduced error in the predictions of wind speed and wind direction.
However, since this correlation is not observed in the results of the first out-of-sample test for the TNBL, this observed improvement does not generalize universally.

Lastly, in the third out-of-sample test for unobserved SBL(W) stability regime with unobserved surface roughness, $z_0=0.0001~\mathrm{m}$, shown in Fig.~\ref{fig:outofsample_rmse}c, varying both stability and surface roughness results in inconsistent trends of improvement across posteriors and predicted flow quantities.
In general, the out-of-sample tests show error reduction for the variables that are assimilated, even in unseen ABL conditions. 
However, these improvements in the predictions of variables seen during UQ can yield lower accuracy for the variables that were not seen by the UQ (e.g., higher error in TKE predictions when assimilating wind speed and direction).
This highlights the extent to which the inferred posteriors can predict out-of-sample and suggests the need to assimilate a broader range of variables and flow regimes-made feasible by the CES methodology-to further improve model generalizability.

\FloatBarrier

\section{Conclusions}
\label{sec:conclusions}

The current study demonstrates an efficient approach to Bayesian calibration and UQ for RANS-based atmospheric models, leveraging (1) scale-resolving datasets for broad parameter exploration and (2) a machine learning-based solution for computationally efficient sampling.
The methodology is applied to turbulence model parameters to show specifically how the combination of data and machine learning can be leveraged to drive physics-based insights. 
The key findings of this study are summarized as follows:
\begin{enumerate}

    \item Bayesian model calibration and UQ to data of stability-specific flow regimes enables an analysis of the stability-dependent modeling of turbulence parameterizations based on higher-fidelity data, and enables improvements in modeling not only in the variables used in the inversion but also in unseen terms in the mean momentum balance.
    The rapid inverse UQ method enables the generation of predictions with credible intervals for probabilistic predictions for RANS-based ABL models.
    
    \item Parameter UQ analysis using scale-resolving data enables an evaluation of the validity of theoretically derived parameter relationships for the regimes for which they are derived, and their potential deficiency for regimes out-of-sample from their derivation context, which can be useful for data-informed improvements of physics-based parameterizations.
    
    \item The investigation into the impact of vertical sensing locations on parameter uncertainty highlights a dependence of the information available in the data on the ABL height.
    The assimilation of data up to the ABL height is shown to reduce the parameter uncertainty by up to 84\%, demonstrating the value of ABL statistics beyond the surface layer.

    \item The sensitivity of parameters to different flow quantities can be leveraged to maximize information gain in the assimilation of additional relevant flow quantities.
    In this case, additionally assimilating TKE reduces eddy viscosity parameter uncertainty by up to 80\%, demonstrating the value of assimilating targeted flow quantities beyond wind speed.

    \item The parameters estimated via Bayesian inversion not only reduce the mean error in out‑of‑sample flow predictions but also provide uncertainty intervals.
    However, performance deteriorates under increasingly distant out‑of‑sample conditions, indicating that further work is required to improve generalizability.
    
\end{enumerate}
These findings (1) pave a way toward developing a Bayesian experimental design framework for optimal measurement acquisition in the ABL and (2) represent a step towards a more synergistic approach between model development and observation use for parameterizations of turbulence in the ABL.

Future work must broaden the application of Bayesian inversion by calibrating turbulence parameterizations on data sources beyond LES, which relies on the modeling of subgrid-scale processes. 
The use of experimental data and field observations is crucial for validating the CES methodology and assessing its performance in real-world measurement settings.
There is also a need to incorporate model-form discrepancy into the inverse problem framework to support a more comprehensive account of the uncertainty in numerical models.
In applications where multiple sources of uncertainties and multiple model fidelities are present, finding creative and computationally efficient methods to accurately quantify uncertainty becomes increasingly important \citep{peherstorfer2016multifidelity, peherstorfer2018survey}.
These extensions will enable us to leverage observational data in refining parameterizations and revealing interactions among coupled schemes in numerical weather prediction, ultimately to yield more accurate representations of the inherently chaotic atmosphere.

\section*{Acknowledgments}

E.Y.S. and M.F.H. gratefully acknowledge the support of Office of Naval Research, Young Investigator Program (YIP), grant no. N000142512045.
The authors acknowledge funding from the MIT UMRP program. 
E.Y.S. acknowledges gracious support from the MIT MathWorks fellowship. 
Simulations were performed on the Stampede3 supercomputers under the NSF ACCESS project ATM170028 and on MIT’s Engaging cluster.

\FloatBarrier

\appendix
\renewcommand{\thesection}{Appendix \arabic{section}}

\section{Details and verification of RANS turbulence models}
\label{sec:appendix1}

STD $k$-$\epsilon$ model \citep{LAUNDER1974269} is one of the most utilized and well-known RANS models. The target parameters with default values are $C_\mu = 0.09$, $C_1 = 1.44$, $C_2 = 1.92$, $C_3 = 1.0$, $\sigma_k = 1.0$, $\sigma_\epsilon = 1.3$.

Eddy viscosity formulation:

\begin{equation}
    \nu_t = C_\mu \frac{k^2}{\epsilon}
\end{equation}
with transport equations each for TKE $k$ and TKE dissipation rate $\epsilon$:

\begin{equation}
    \frac{\partial k}{\partial t} = -\overline{u'w'}\frac{\partial \overline{u}}{\partial z} - \overline{v'w'}\frac{\partial \overline{v}}{\partial z} + \frac{g}{\Theta_0} \overline{w'\theta'} + \frac{\partial}{\partial z} \left( \frac{\nu_t}{\sigma_k} \frac{\partial k}{\partial z} \right) - \epsilon 
\end{equation}

\begin{equation}
    \frac{\partial \epsilon}{\partial t} = C_{1} \frac{\epsilon}{k} \left( -\overline{u'w'}\frac{\partial \overline{u}}{\partial z} - \overline{v'w'}\frac{\partial \overline{v}}{\partial z} \right) + C_{3} \frac{\epsilon}{k } \frac{g}{\Theta_0} \overline{w'\theta'} + \frac{\partial}{\partial z} \left( \frac{\nu_t}{\sigma_\epsilon} \frac{\partial \epsilon}{\partial z} \right) - C_{2} \frac{\epsilon^2}{k}
\end{equation}

MOST $k$-$\epsilon$ model \citep{van2017new} is a $k$-$\epsilon$ variant that is motivated by the fact that the default STD $k$-$\epsilon$ model does not sustain MOST profiles across different stability conditions.
Therefore, an analytical source term is introduced in the $k$-equation and a stability function-based formulation of the parameter $C_3(\zeta)$ is used. The target parameters with default values are $C_\mu = 0.03$, $C_1 = 1.21$, $C_2 = 1.92$, $\sigma_k = 1.0$, $\sigma_\epsilon = 1.3$.

Eddy viscosity formulation:

\begin{equation}
    \nu_t = C_\mu \frac{k^2}{\epsilon}
\end{equation}
with transport equations each for TKE $k$ and TKE dissipation rate $\epsilon$:

\begin{equation}
    \frac{\partial k}{\partial t} = -\overline{u'w'}\frac{\partial \overline{u}}{\partial z} - \overline{v'w'}\frac{\partial \overline{v}}{\partial z} + \frac{g}{\Theta_0} \overline{w'\theta'} + \frac{\partial}{\partial z} \left( \frac{\nu_t}{\sigma_k} \frac{\partial k}{\partial z} \right) - \epsilon - S_k
\end{equation}

\begin{equation}
    \frac{\partial \epsilon}{\partial t} = C_{1} \frac{\epsilon}{k} \left( -\overline{u'w'}\frac{\partial \overline{u}}{\partial z} - \overline{v'w'}\frac{\partial \overline{v}}{\partial z} \right) + C_{3} \frac{\epsilon}{k } \frac{g}{\Theta_0} \overline{w'\theta'} + \frac{\partial}{\partial z} \left( \frac{\nu_t}{\sigma_\epsilon} \frac{\partial \epsilon}{\partial z} \right) - C_{2} \frac{\epsilon^2}{k}.
\end{equation}

Source term in the TKE equation:

\begin{equation}
    S_k = \frac{u_*^3}{\kappa L} \times \left\{ 1 - \frac{\Phi_h}{\sigma_\theta \Phi_m} - \frac{C_{k\mathcal{D}}}{4} \Phi_m^{-7/2} \Phi_\epsilon^{-3/2} f_{st} (\zeta, \beta) \right\}, \quad \zeta > 0
\end{equation}
where 
\begin{equation}
    C_{k\mathcal{D}} = \frac{\kappa}{\sigma_k \sqrt{C_\mu}}
\end{equation}

\begin{equation}
    f_{st}(\zeta, \beta) \equiv (2 - \zeta) - 2 \beta \zeta \left( 1 - 2 \zeta + 2 \beta \zeta \right)
\end{equation}

$C_3$ formulation: 

\begin{equation}
    C_3(\zeta) = \frac{\sigma_\theta}{\zeta} \frac{\Phi_m}{\Phi_h} \left( C_1 \Phi_m - C_2 \Phi_\epsilon + \left[ C_2 - C_1 \right] \Phi_\epsilon^{-1/2} f_\epsilon (\zeta) \right)
    \label{eq:C3}
\end{equation}
where $C_1$ and $C_2$ are model parameters, $\sigma_\theta$ is the turbulent Prandtl number, $\Phi_m$ and $\Phi_h$ are universal functions for momentum and heat fluxes respectively, and $f_\epsilon(\zeta)$ is a stability function:
    
\begin{equation}
f_\varepsilon(\zeta) \equiv
\begin{cases}
\displaystyle \Phi_m^{\frac{5}{2}} \!\bigl(1 - \tfrac{3}{4}\,\gamma_1 \zeta \bigr),
  & \zeta < 0, \\[6pt]
\displaystyle \Phi_m^{-\frac{5}{2}} \!\bigl(2\,\Phi_m - 1\bigr),
  & \zeta > 0.
\end{cases}
\label{eq:stabilityfunction}
\end{equation}  
based on the stability parameter $\zeta$ defined as
    
\begin{equation}
    \zeta=\frac{z}{L}=-\frac{\kappa z}{u_*^3} \frac{g}{\theta_0} \overline{w'\theta'}
\end{equation}
where $L$ is the Monin-Obukhov length, $\kappa$ is the von Kármán constant, $g$ is the gravitational acceleration, $u_*$ is the friction velocity, $\theta_0$ is the reference potential temperature, and $\gamma_1$ (in Eq.~\ref{eq:stabilityfunction}) is a constant.

The two models are verified for the stability conditions (TNBL, SBL) considered in this study against existing literature.

\begin{figure}[ht]
    \centering
        \includegraphics[width=0.5\textwidth]{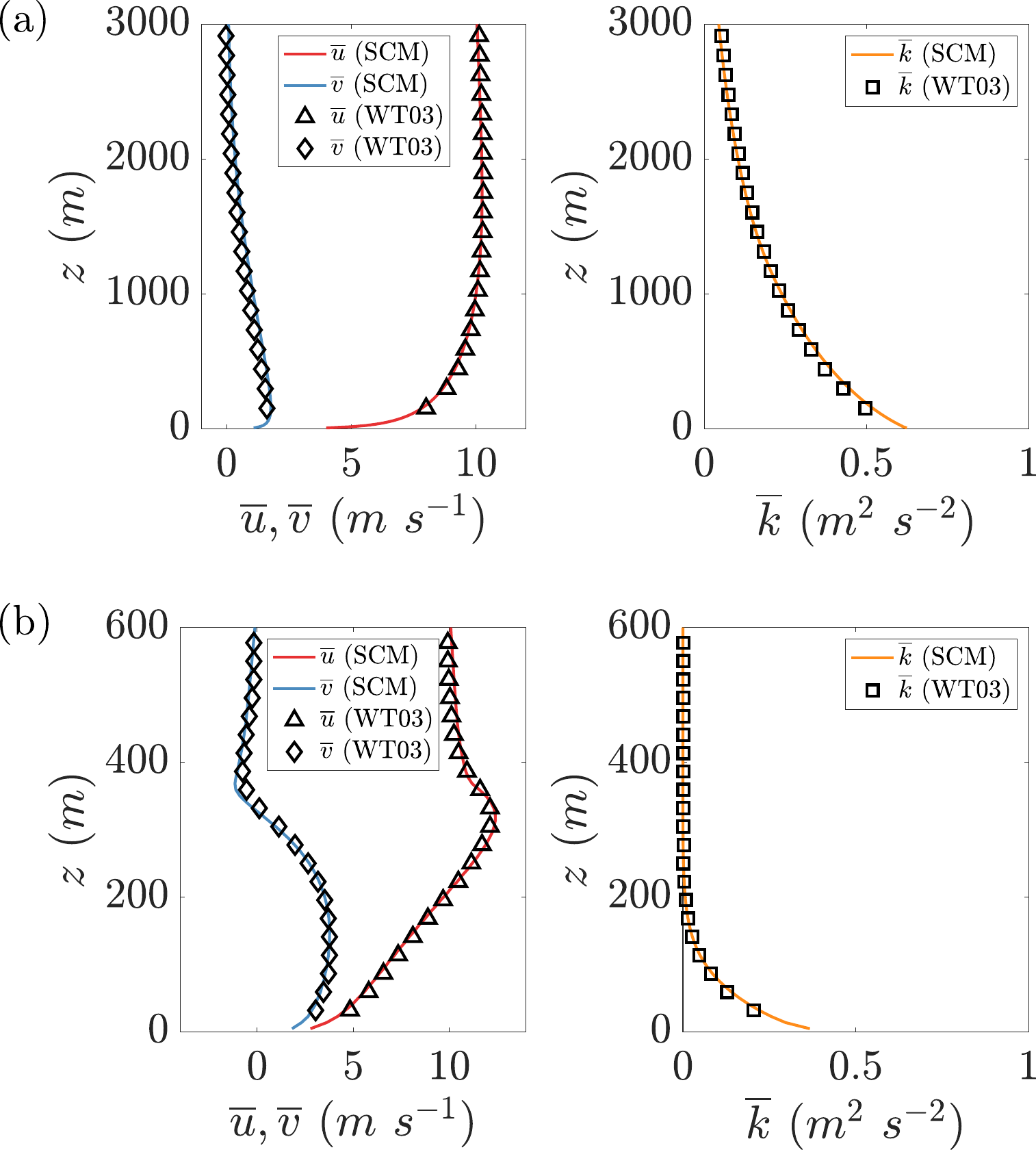}
    \caption{Verification of the STD $k$-$\epsilon$ model for wind velocity components ($\overline{u},\overline{v}$) and TKE ($\overline{k}$) in two stability regimes: (a) TNBL and (b) SBL. Reference data from \cite{weng2003modelling} are labeled as WT03.} 
    \label{fig:stdke_verification}
\end{figure}

\begin{figure}[ht]
    \centering
        \includegraphics[width=0.5\textwidth]{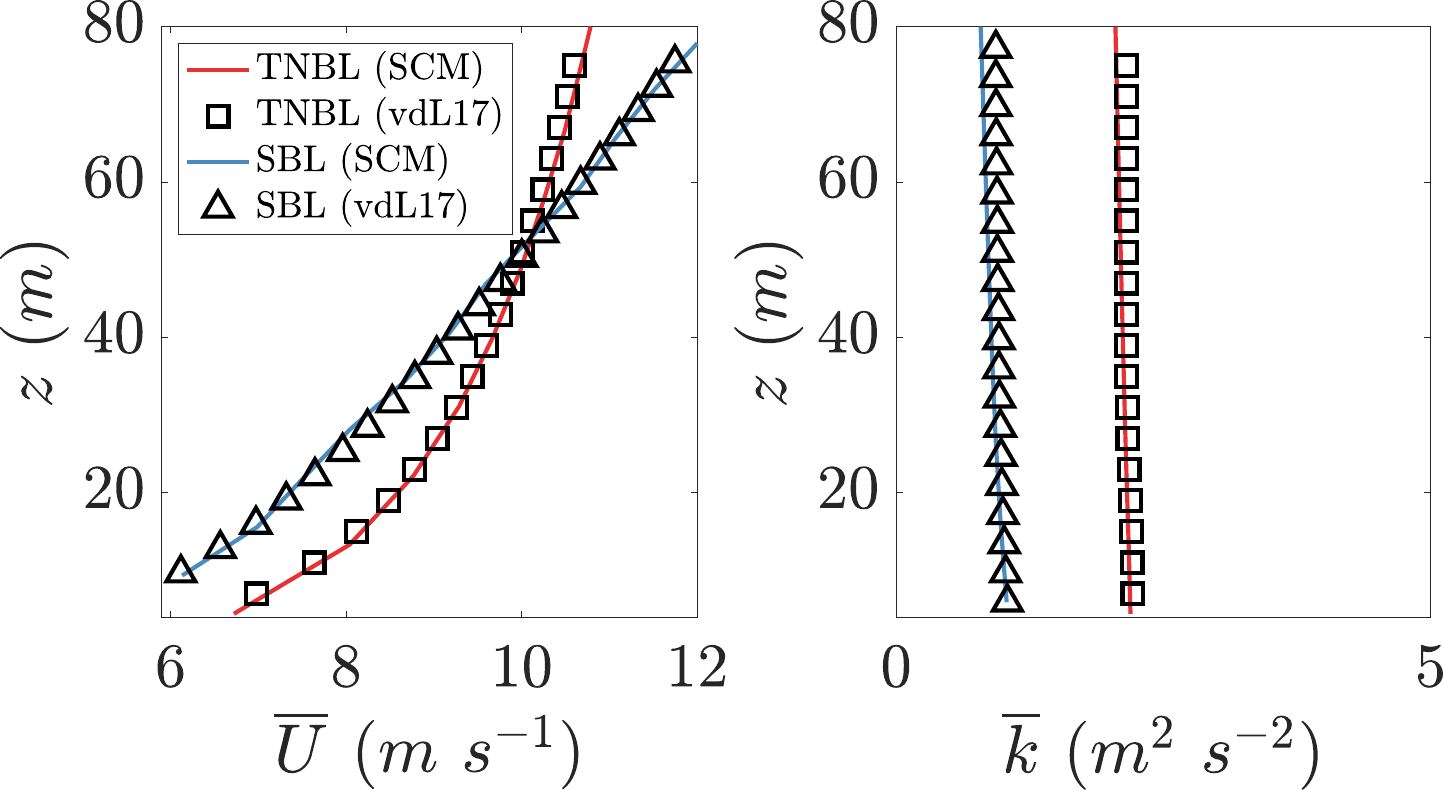}
    \caption{Verification of the MOST $k$-$\epsilon$ model for wind speed ($\overline{U}$) and TKE ($\overline{k}$) in two stability regimes: TNBL and SBL. Reference data from \cite{van2017new} are labeled as vdL17.} 
    \label{fig:mostke_verification}
\end{figure}

\FloatBarrier

\section{Gaussianity approximation for observational noise}
\label{sec:appendix2}

\FloatBarrier

\begin{figure}[ht]
    \centering
    \begin{tabular}
    {@{}p{0.33\linewidth}@{\quad}p{0.33\linewidth}@{\quad}p{0.33\linewidth}@{}}
        \subfigimgthree[width=\linewidth,valign=t]
        {(a)}
        {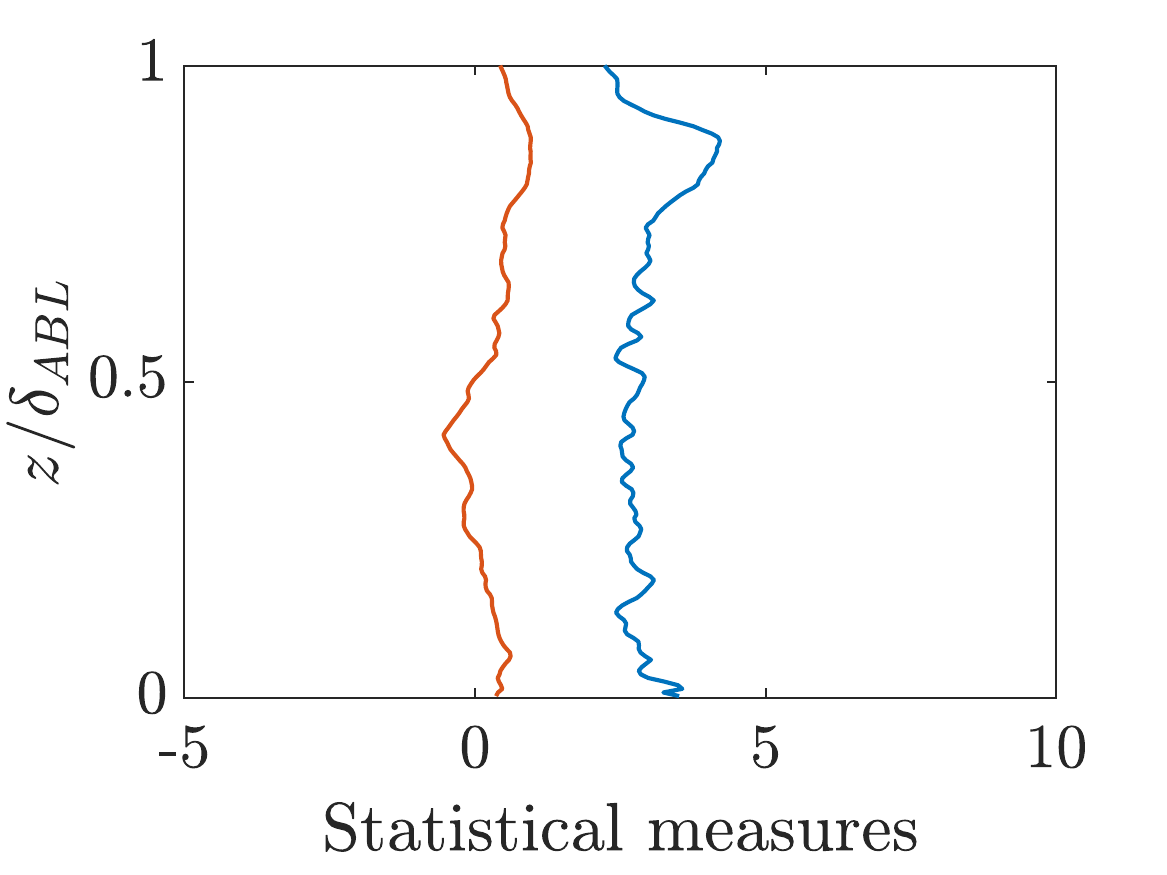} &
        \subfigimgthree[width=\linewidth,valign=t]
        {(b)}
        {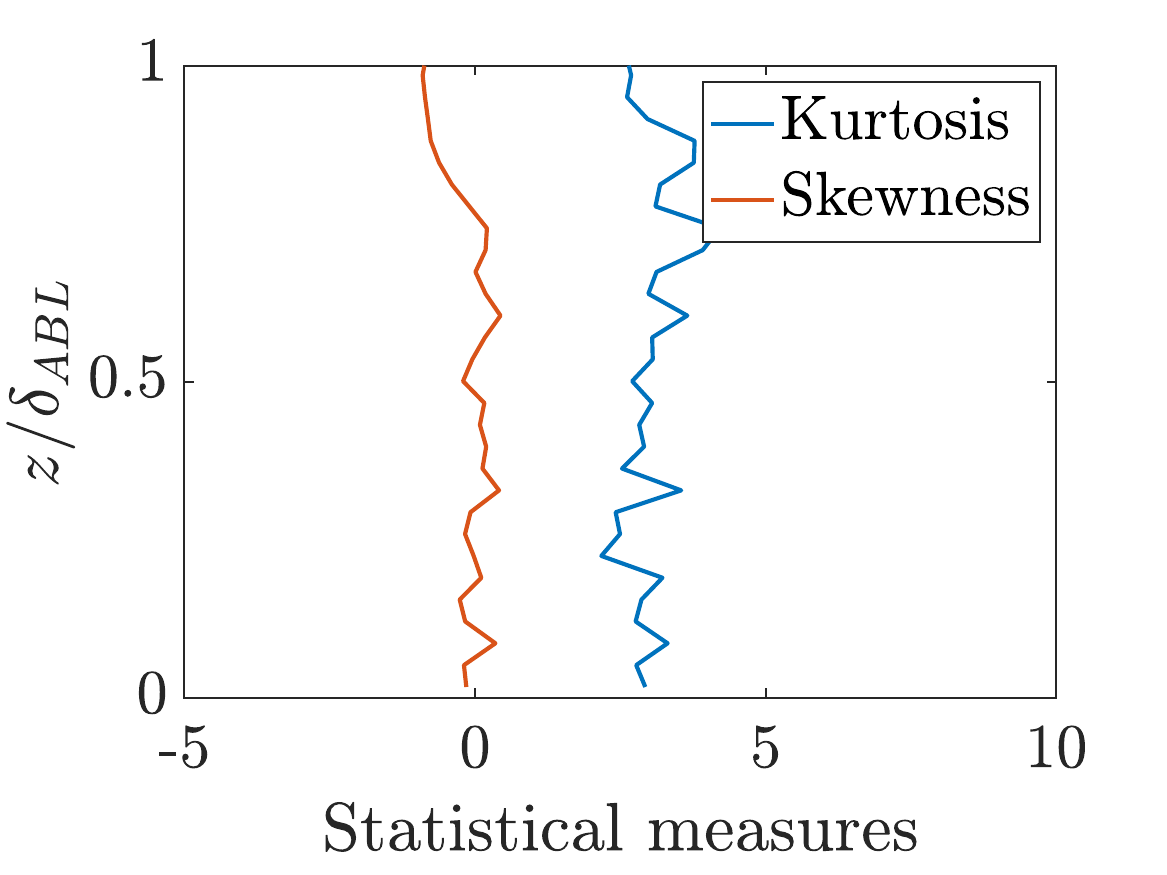} &
        \subfigimgthree[width=\linewidth,valign=t]
        {(c)}
        {figures/appendix/sblm_U_gaussianity.pdf} 
    \end{tabular}
    \caption{Kurtosis and skewness of the distribution of wind speed samples at each height for the three stability regimes: (a) TNBL, (b) SBL(W), and (c) SBL(M)}
    \label{fig:U_gaussianity}
\end{figure}

\begin{figure}[ht]
    \centering
    \begin{tabular}
    {@{}p{0.33\linewidth}@{\quad}p{0.33\linewidth}@{\quad}p{0.33\linewidth}@{}}
        \subfigimgthree[width=\linewidth,valign=t]
        {(a)}
        {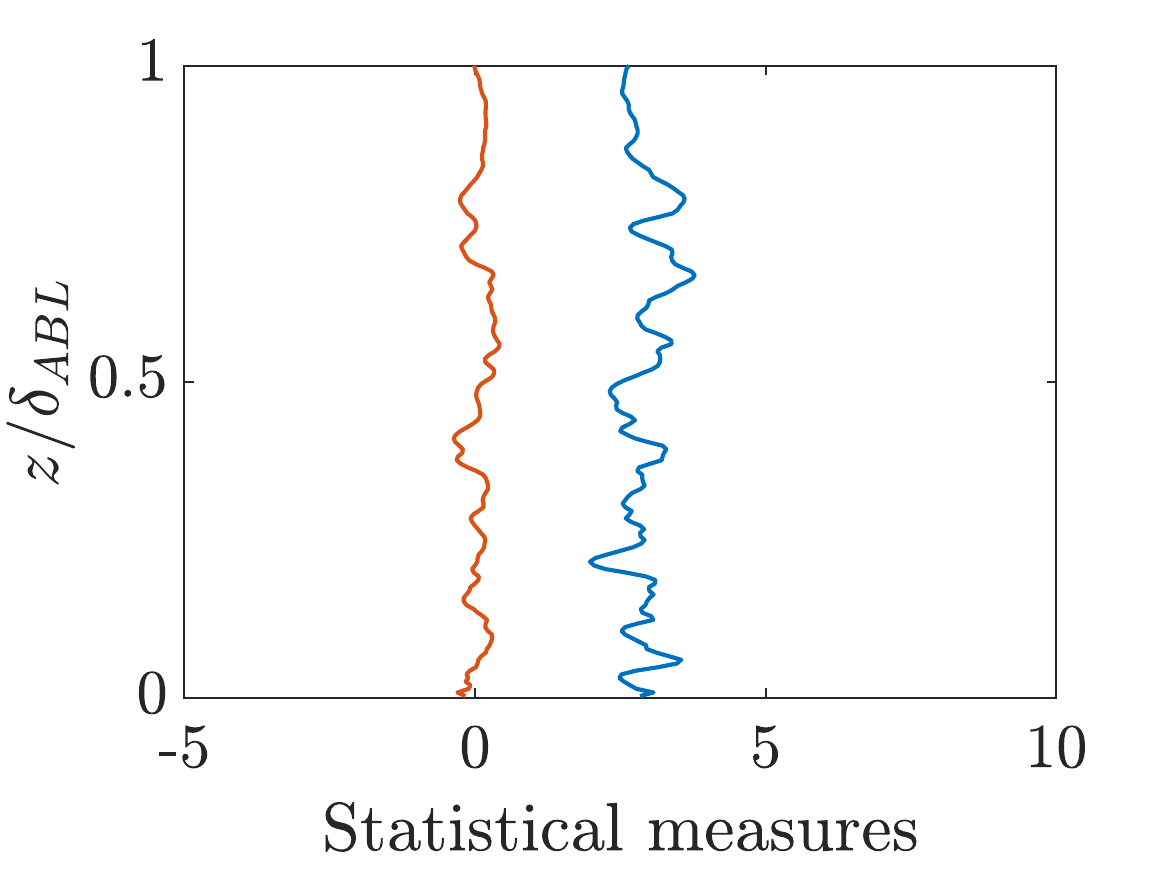} &
        \subfigimgthree[width=\linewidth,valign=t]
        {(b)}
        {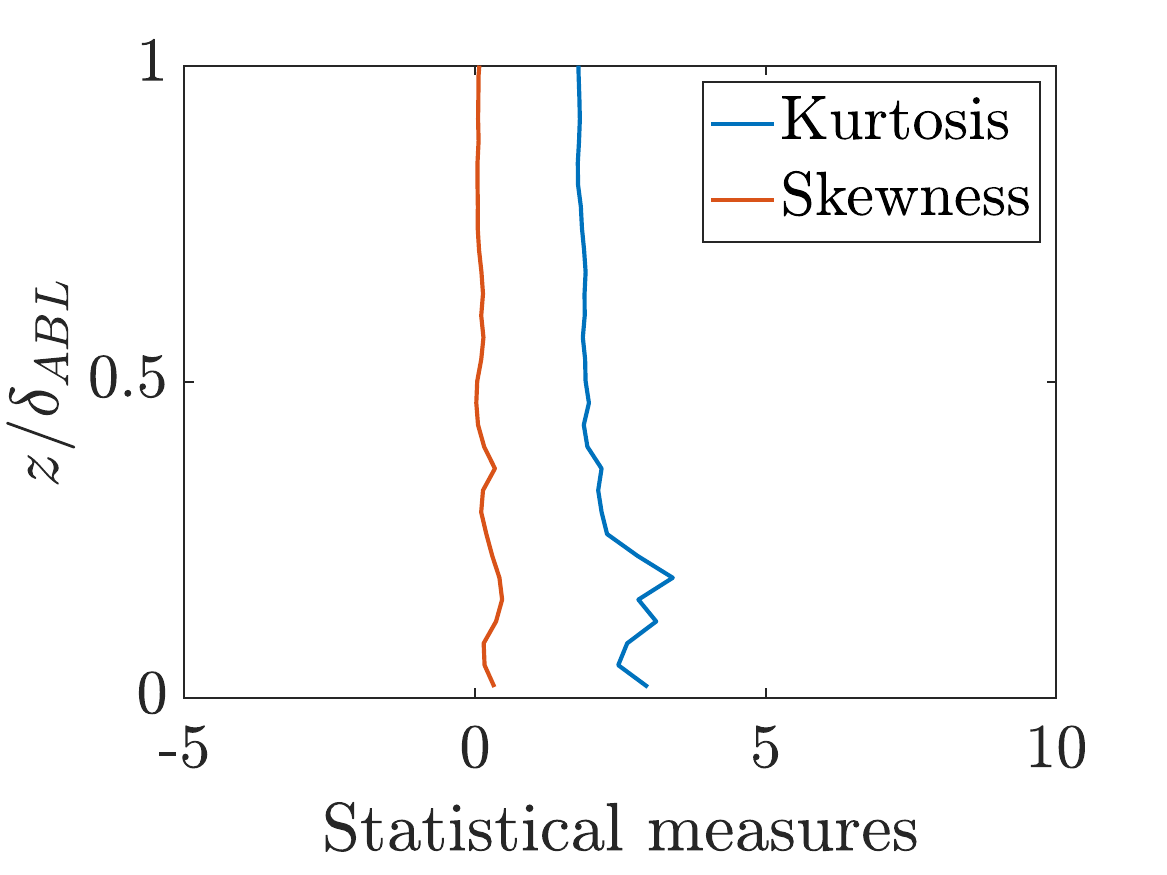} &
        \subfigimgthree[width=\linewidth,valign=t]
        {(c)}
        {figures/appendix/sblm_dir_gaussianity.pdf}
    \end{tabular}
    \caption{Kurtosis and skewness of the distribution of wind direction samples at each height for the three stability regimes: (a) TNBL, (b) SBL(W), and (c) SBL(M)}
    \label{fig:dir_gaussianity}
\end{figure}

\begin{figure}[ht]
    \centering
    \begin{tabular}
    {@{}p{0.33\linewidth}@{\quad}p{0.33\linewidth}@{\quad}p{0.33\linewidth}@{}}
        \subfigimgthree[width=\linewidth,valign=t]
        {(a)}
        {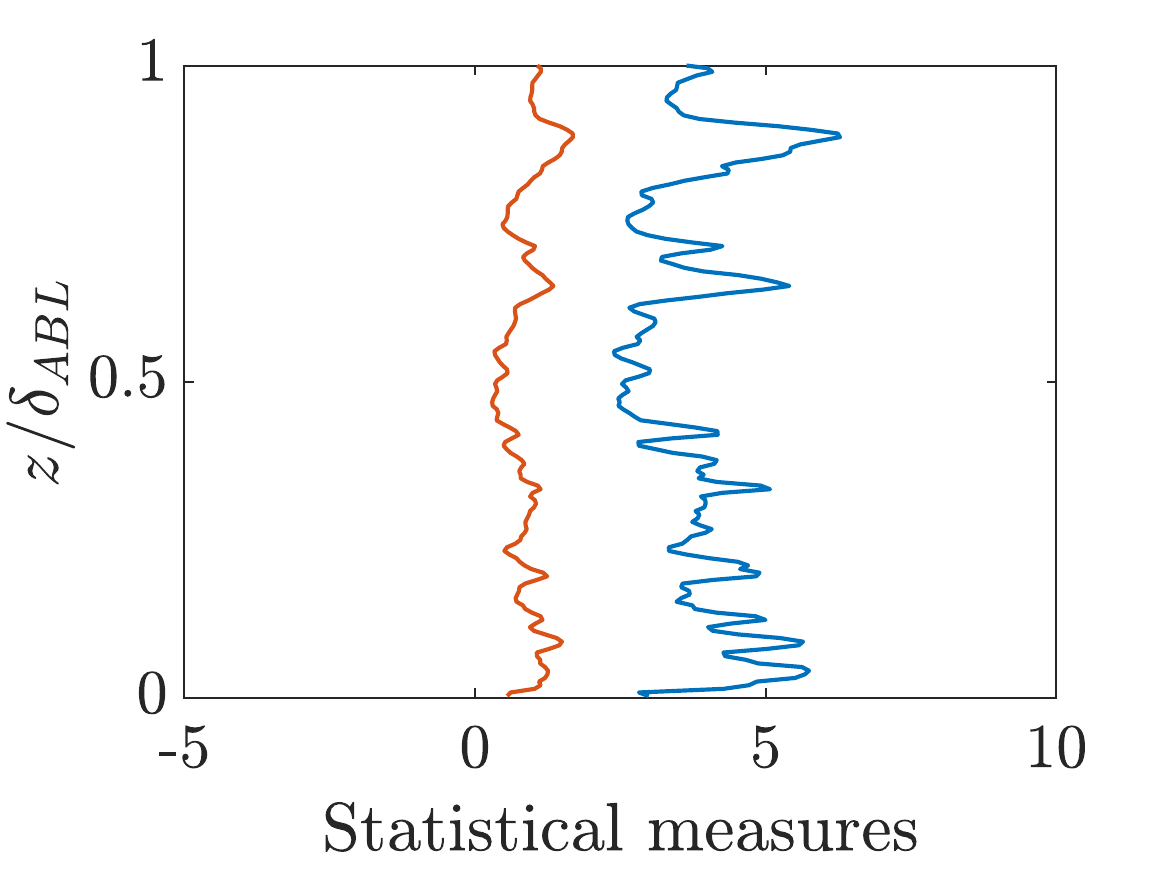} &
        \subfigimgthree[width=\linewidth,valign=t]
        {(b)}
        {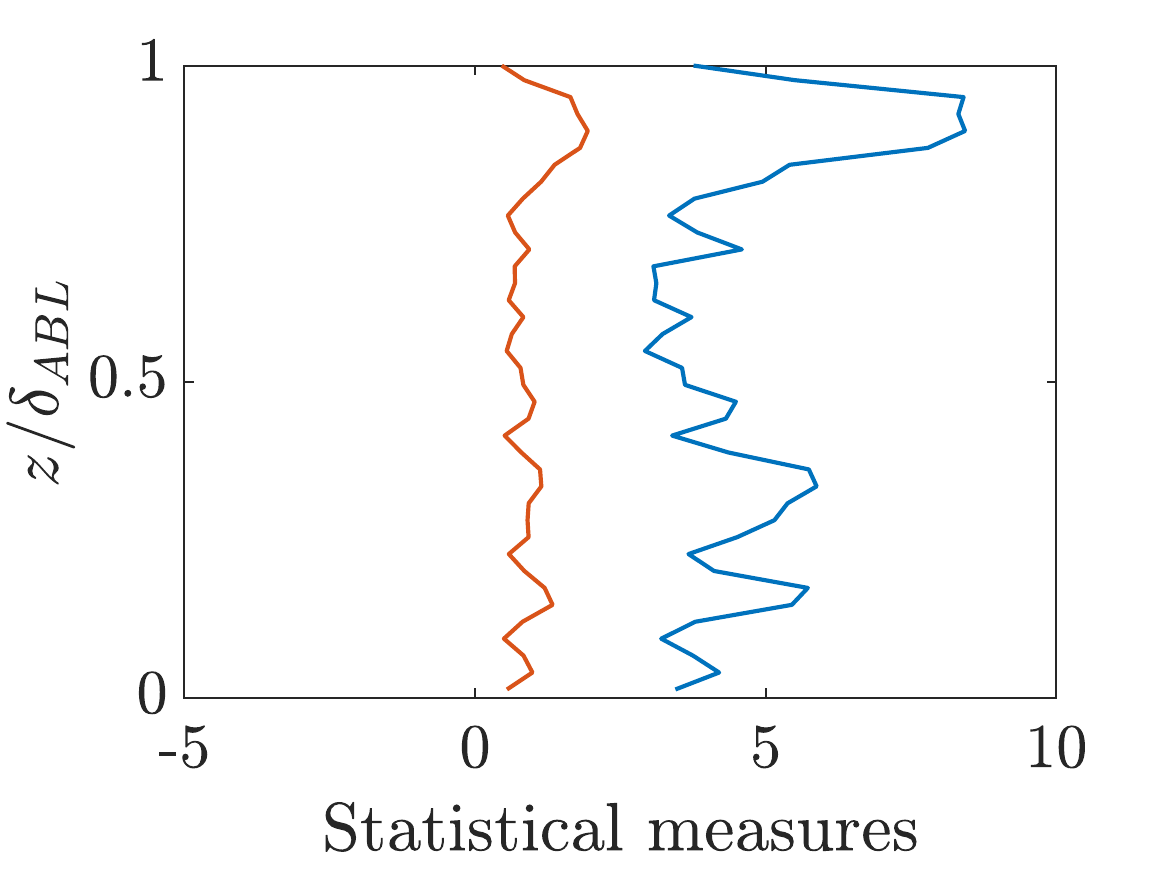} &
        \subfigimgthree[width=\linewidth,valign=t]
        {(c)}
        {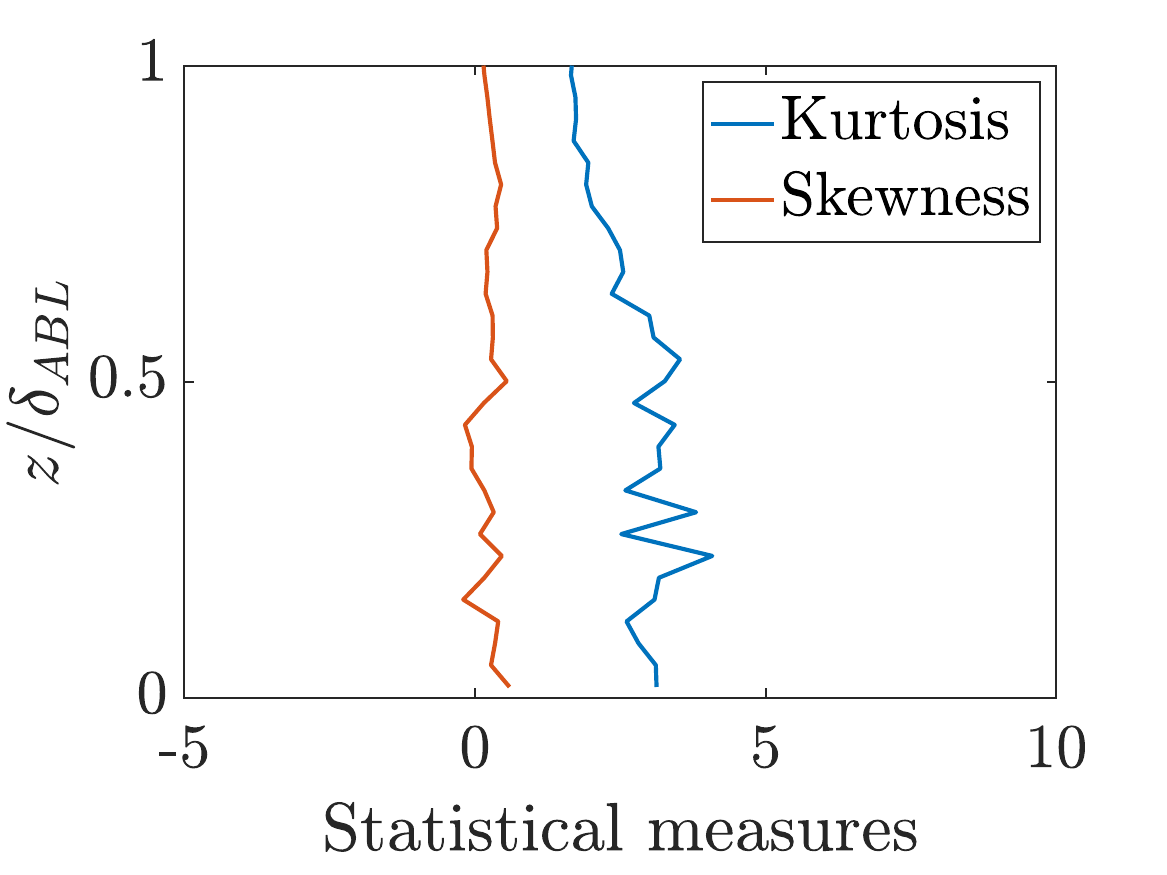}
    \end{tabular}
    \caption{Kurtosis and skewness of the distribution of TKE samples at each height for the three stability regimes: (a) TNBL, (b) SBL(W), and (c) SBL(M)}
    \label{fig:k_gaussianity}
\end{figure}

To check the validity of the Gaussian distribution for the observational noise, we compute two statistical measures, skewness and kurtosis, that quantify the divergence from normality for the distribution of 90 samples of 10-minute averaged profiles for wind speed, wind direction, and TKE.

The skewness measures the assymetry of the distribution and is defined as 
\begin{equation}
    s = \frac{E\big((x - \mu)^3\big)}{\sigma^3},
    \label{eq:skewness}
\end{equation}
where $x$ is the variable of interest, $\mu$ is the mean, $\sigma$ is the standard deviation, and $E(x)$ is the expected value.
The sample variant uses the sample mean and standard deviation.

The kurtosis measures the extent to which the distribution's tails differ from that of a Gaussian distribution and is defined as
\begin{equation}
    k = \frac{E\big((x - \mu)^4\big)}{\sigma^4},
\end{equation}
with the same variables used in Eq.~\ref{eq:skewness}.
Here, the sample kurtosis is computed at each height, and $k=3$ is the value for the Gaussian distribution.

Vertical profiles of the sample skewness and kurtosis of the sample distributions for the TNBL, SBL(W), and SBL(M) cases are shown for wind speed, wind direction, and TKE data (Fig.~\ref{fig:U_gaussianity}, Fig.~\ref{fig:dir_gaussianity}, Fig.~\ref{fig:k_gaussianity}).
For wind speed and wind direction of all three stability cases, the sample skewness is close to zero and sample kurtosis is centered around $k=3$, indicating approximately Gaussian distributions for the samples.
However, for TKE data of the TNBL and SBL(W) cases, sample kurtosis deviates from the Gaussian value with a positive $\mathcal{O}(1)$ bias, indicating moderately longer tails in the sample distributions.
Despite the higher kurtosis, the sample skewness—critical to the formulation of the ensemble Kalman gain—remains near zero.
Hence, we anticipate that LES data of all three flow quantities adequately satisfy the assumptions for approximate Bayesian inference via the CES methodology, which has proven effective even with noisy observations.

\FloatBarrier

\section{Regularization of the sample covariance matrix}
\label{sec:appendix3}

\FloatBarrier

The sample covariance matrix $\Gamma$ can be noisy or ill-conditioned, especially when the number of samples $n$ is small relative to the dimension $p$ of the covariance matrix \citep{ledoit_well-conditioned_2004}.
Increasing the dimension $p$ of the covariance matrix provides more information of the data, but in many practical applications of model inversion, the number of samples is limited and often comparable to the dimension $p$, making the ratio $p/n$ effectively non-negligible.
This may be addressed using two different regularization methods: (1) singular value decomposition (SVD) truncation and (2) Ledoit-Wolf shrinkage.

SVD truncation is an established method of regularizing ill-posed linear least squares problems by accounting for the ill-conditioning of the sample covariance matrix \citep{hansen1987truncated}.
This has also previously been utilized with the CES methodology for rank-deficient or ill-conditioned covariance matrices \citep{howland2022parameter}.
The decomposition and truncation up to a specified singular value $k$ may be defined as
\begin{equation}
    \Sigma^* \approx \Sigma_k^* = V_k D_k^{*2} V_k^T 
\end{equation}
where $D_k^{*2}=\mathrm{diag}(\sigma_1^*, ...,\sigma_k^*)$ is the diagonal matrix of eigenvalues and $V_k=[v_1,...v_k]$ are corresponding eigenvectors.
The resulting truncated SVD space for the covariance matrix $\tilde{\Sigma}_k$ and data $\tilde{{y}}_k$ used in the emulation stage of CES are given as
\begin{equation}
\tilde{\Sigma}_k = D_k^{-1} V_k^\top \Sigma^* V_k D_k^{-1}, 
\end{equation}
\begin{equation}
    \tilde{{y}}_k = D_k^{-1} V_k^\top {y}^* .
\end{equation}
In this application, the number of singular values retained $k$ is automatically chosen such that 99\% of the singular value energy is retained.

Ledoit-Wolf shrinkage is an estimation method for an asymptotically optimal linear combination of the sample covariance matrix with the identity matrix \citep{ledoit_well-conditioned_2004}.
The estimator is defined as
\begin{equation}
    \Sigma^*= \frac{\beta^2}{\delta^2} \mu I + \frac{\alpha^2}{\delta^2} \Gamma
\end{equation}
where $\Gamma$ is the sample covariance matrix, $I$ is the identity matrix, and $\beta^2, \alpha^2, \delta^2, \mu$ are scalar functions defined as
\begin{equation}
    \mu=\langle \Sigma, I \rangle
\end{equation}
\begin{equation}
    \beta^2 = E[||\Gamma-\Sigma||^2]
\end{equation}
\begin{equation}
    \alpha^2 = ||\Sigma-\mu I||^2
\end{equation}
\begin{equation}
    \delta^2 = E[||\Gamma - \mu I||^2]
\end{equation}
where $\langle A_1, A_2 \rangle = \mathrm{tr}(A_1, A_2)/p$ and $||\cdot||^2$ is the squared Frobenius norm.
In computing the measures of the true covariance matrix $\Sigma$, sample counterparts are used to formulate a consistent estimator of $\Sigma^*$.

\begin{figure}[ht]
    \centering
    \begin{tabular}
    {@{}p{0.33\linewidth}@{\quad}p{0.33\linewidth}@{\quad}p{0.33\linewidth}@{}}
        \subfigimgthree[width=\linewidth,valign=t]
        {(a)}
        {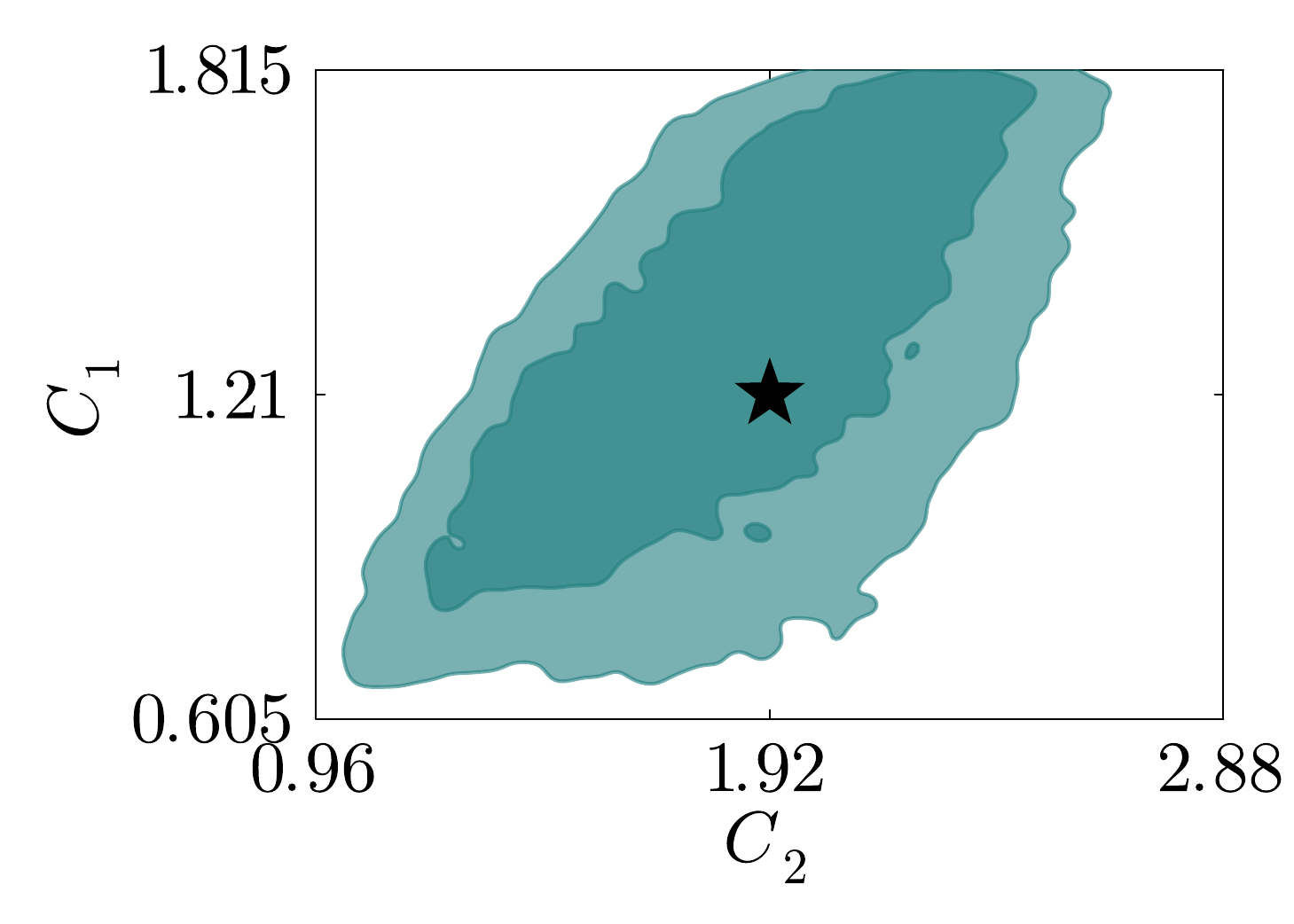} &
        \subfigimgthree[width=\linewidth,valign=t]
        {(b)}
        {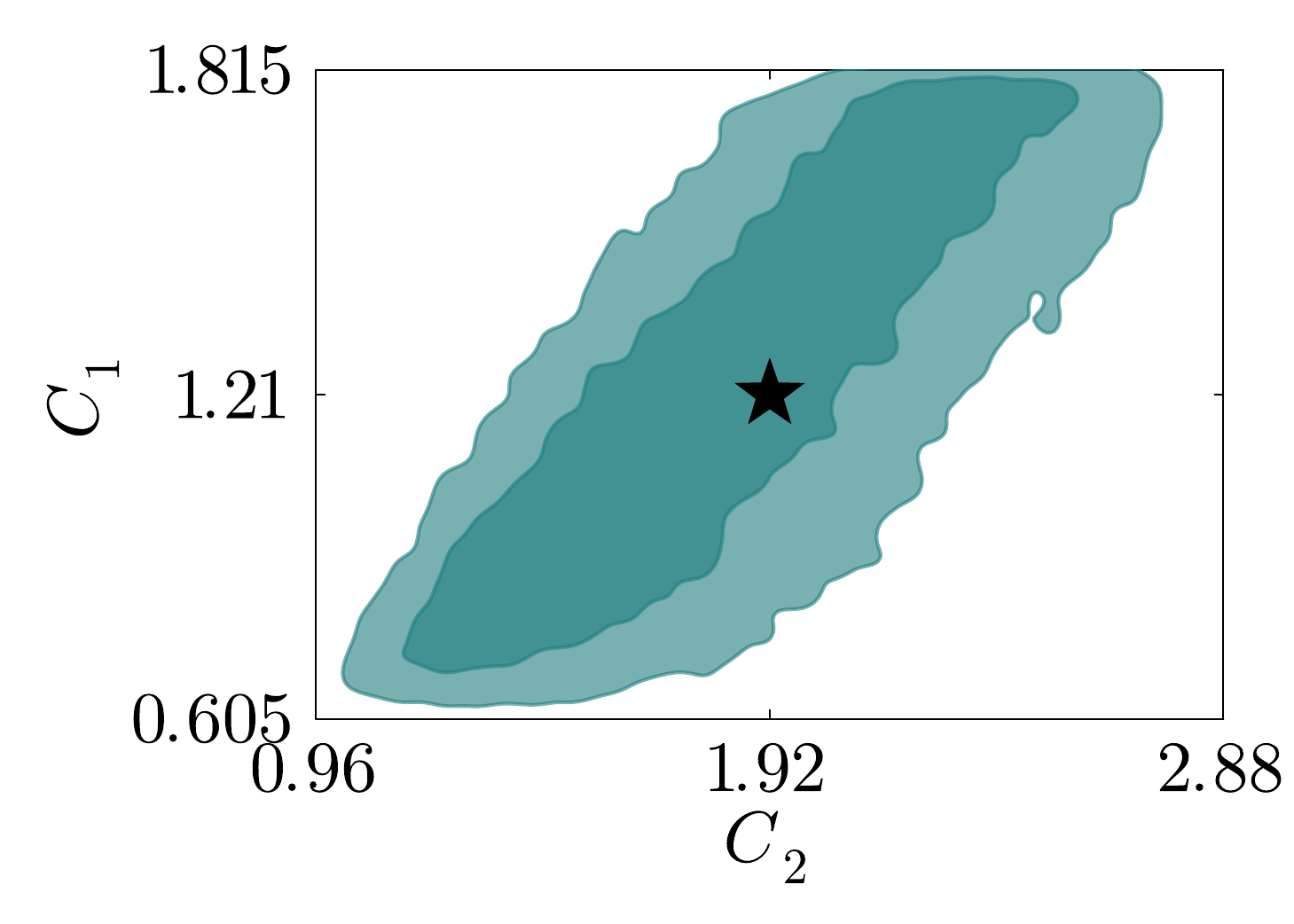} &
        \subfigimgthree[width=\linewidth,valign=t]
        {(c)}
        {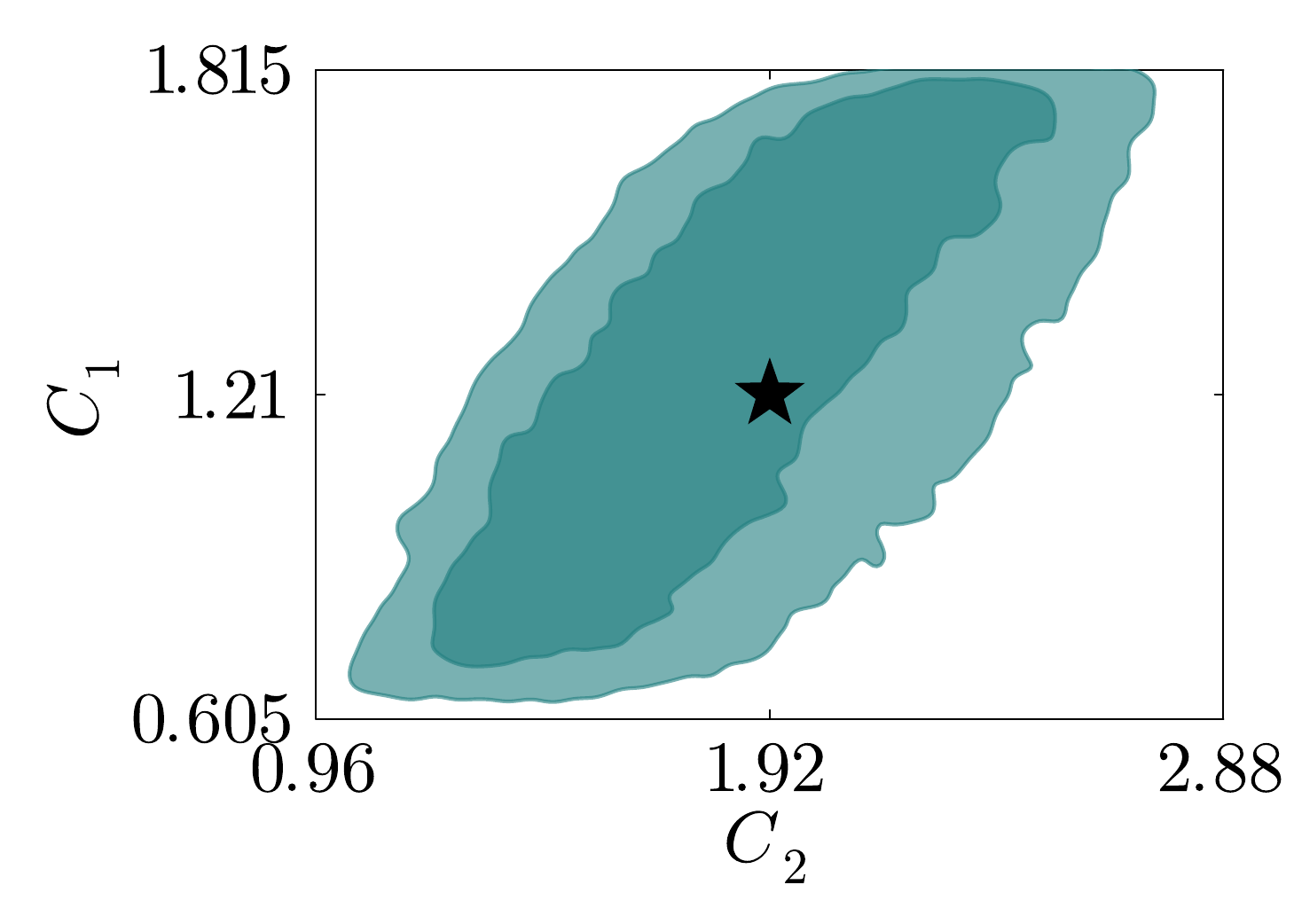}
    \end{tabular}
    \caption{Joint posterior distribution of $C_1$ and $C_2$ inferred from TNBL wind speed data using (a) no regularization, (b) SVD truncation at 99\%, and (c) Ledoit-Wolf shrinkage. 68\% and 95\% HPDR are shown in dark and light shading, respectively.}
    \label{fig:tnbl_regularization}
\end{figure}

\begin{figure}[ht]
    \centering
    \begin{tabular}
    {@{}p{0.33\linewidth}@{\quad}p{0.33\linewidth}@{\quad}p{0.33\linewidth}@{}}
        \subfigimgthree[width=\linewidth,valign=t]
        {(a)}
        {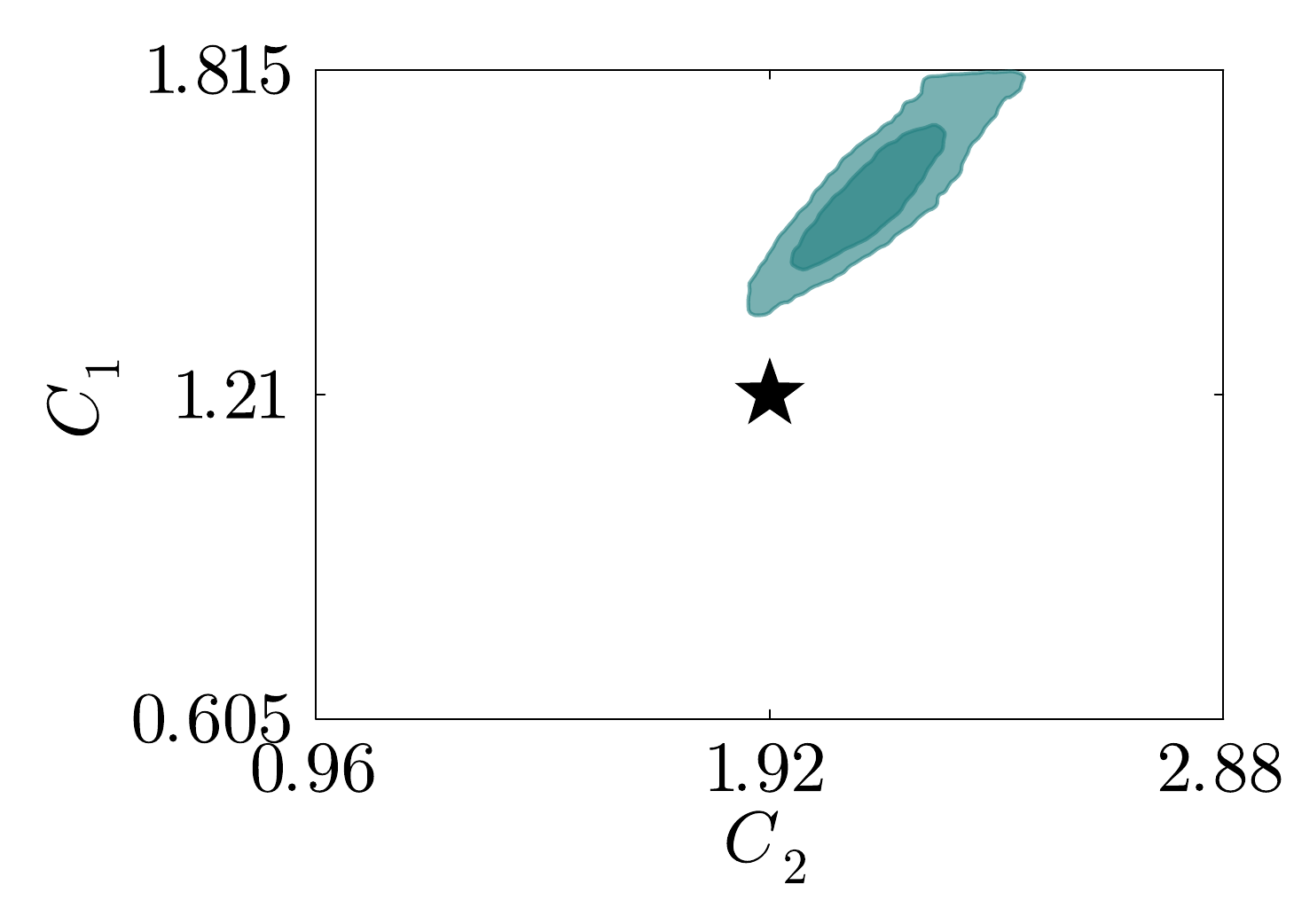} &
        \subfigimgthree[width=\linewidth,valign=t]
        {(b)}
        {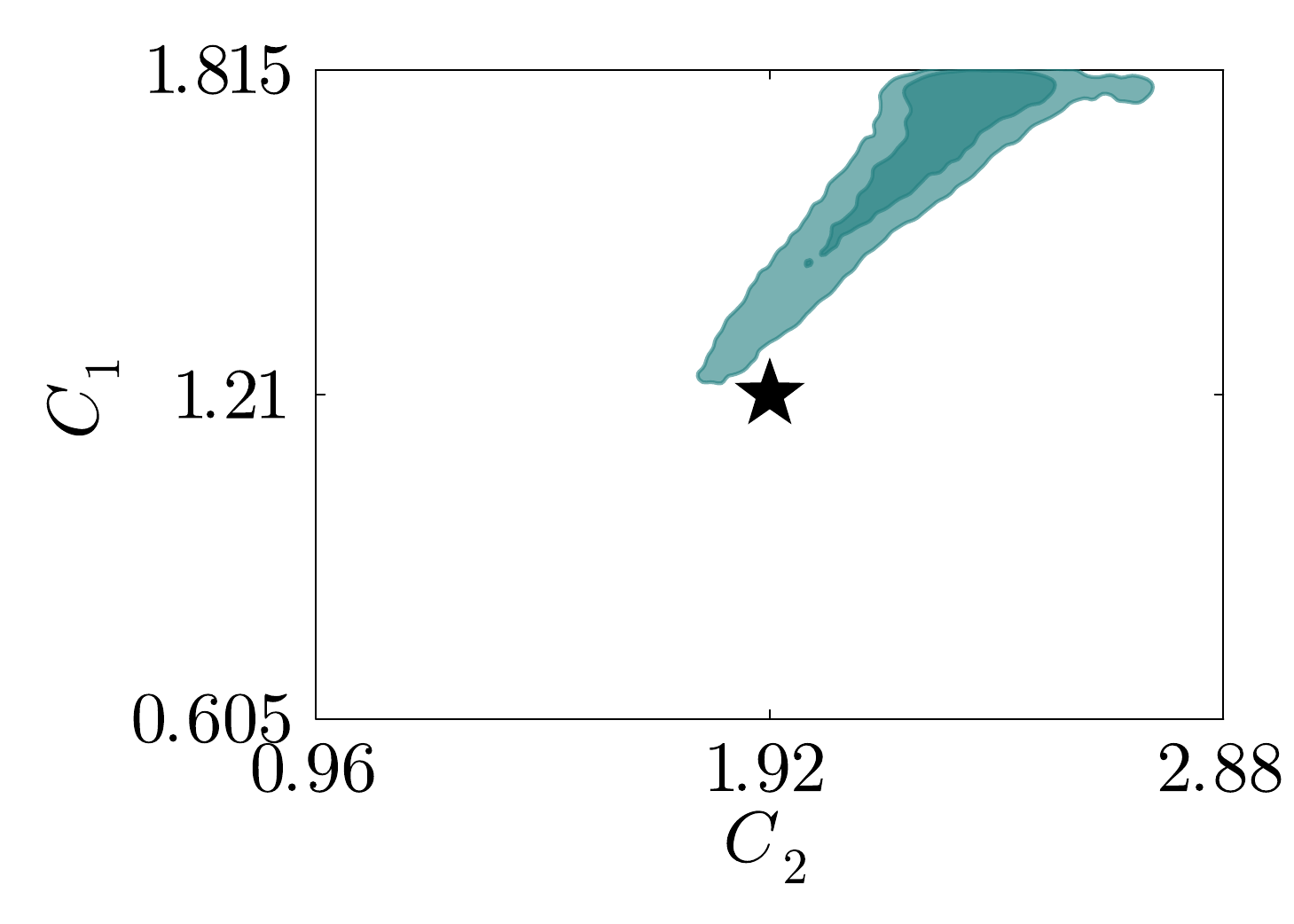} &
        \subfigimgthree[width=\linewidth,valign=t]
        {(c)}
        {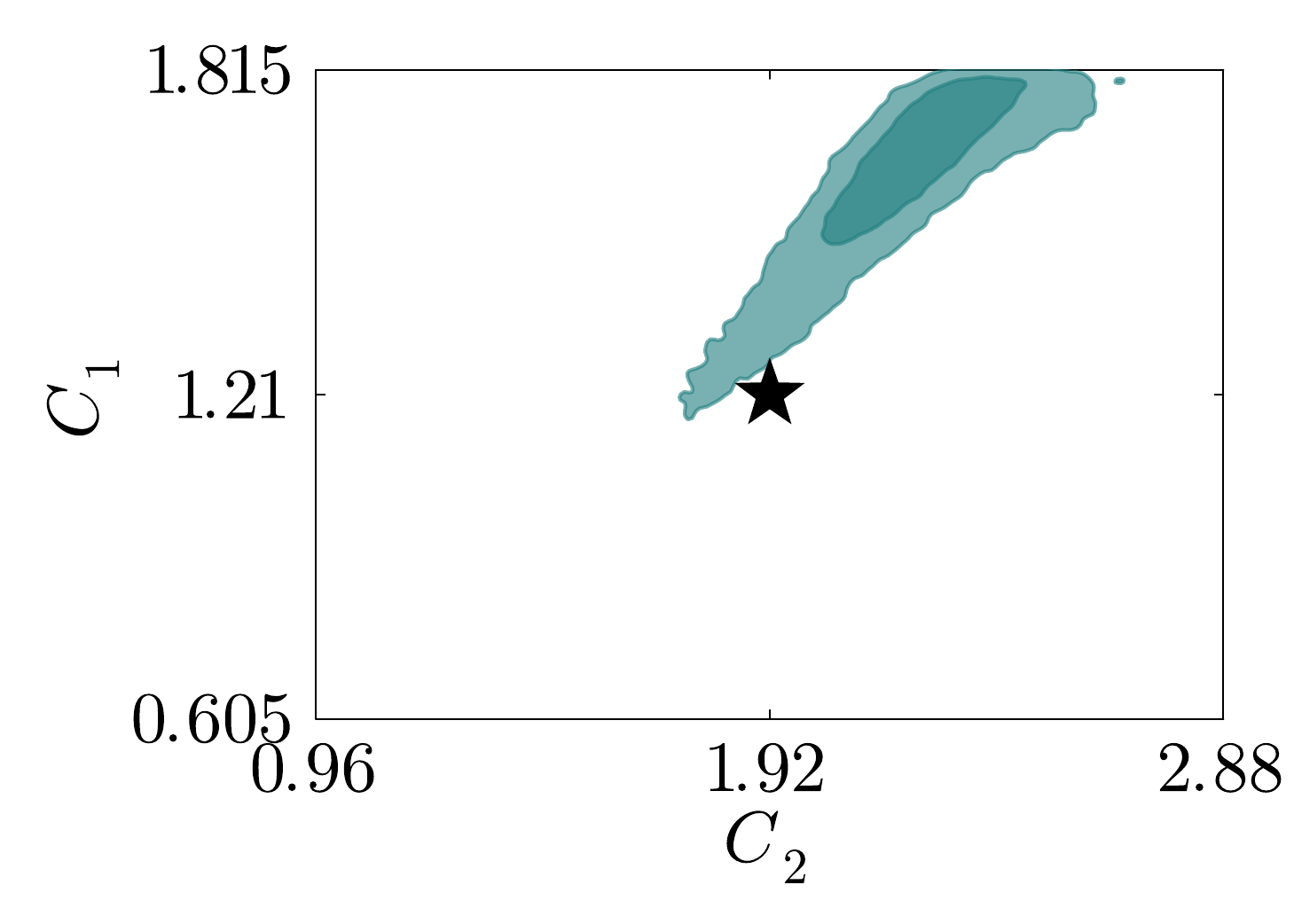}
    \end{tabular}
    \caption{Joint posterior distribution of $C_1$ and $C_2$ inferred from SBL(M) wind speed data using (a) no regularization, (b) SVD truncation at 99\%, and (c) Ledoit-Wolf shrinkage. 68\% and 95\% HPDR are shown in dark and light shading, respectively.}
    \label{fig:sblm_regularization}
\end{figure}

We study the effect of the regularization methods for three different cases (1) TNBL, (2) SBL(M), and (3) the combination of the two stabilities (TNBL, SBL(M)).
We assimilate $p_{\mathrm{TNBL}}=20, p_{\mathrm{SBL(M)}}=20$ points for each of the single stability regimes and $p_{\mathrm{total}}=p_{\mathrm{TNBL}}+p_{\mathrm{SBL(M)}}=40$ for the combined case, such that the same eigenvalues contribute to the estimation of the sample covariance matrix.
Figure~\ref{fig:tnbl_regularization} presents a comparison of the joint posterior distribution of $C_1$ and $C_2$ inferred from TNBL wind speed data using no regularization, SVD truncation at 99\% singular value energy, or the Ledoit-Wolf shrinkage method.
Figure~\ref{fig:sblm_regularization} presents a comparison counterpart inferred from SBL(M) wind speed data.
In the TNBL case (Fig.~\ref{fig:tnbl_regularization}), the posterior area is unaffected by the two regularization methods.
In the SBL(M) case (Fig.~\ref{fig:sblm_regularization}), the posterior area slightly increases for both methods of regularization, similar to results found in \cite{howland2022parameter}.

\begin{figure}[ht]
    \centering
    \begin{tabular}
    {@{}p{0.45\linewidth}@{\quad}p{0.45\linewidth}}
        \subfigimgthree[width=\linewidth,valign=t]
        {(a)}
        {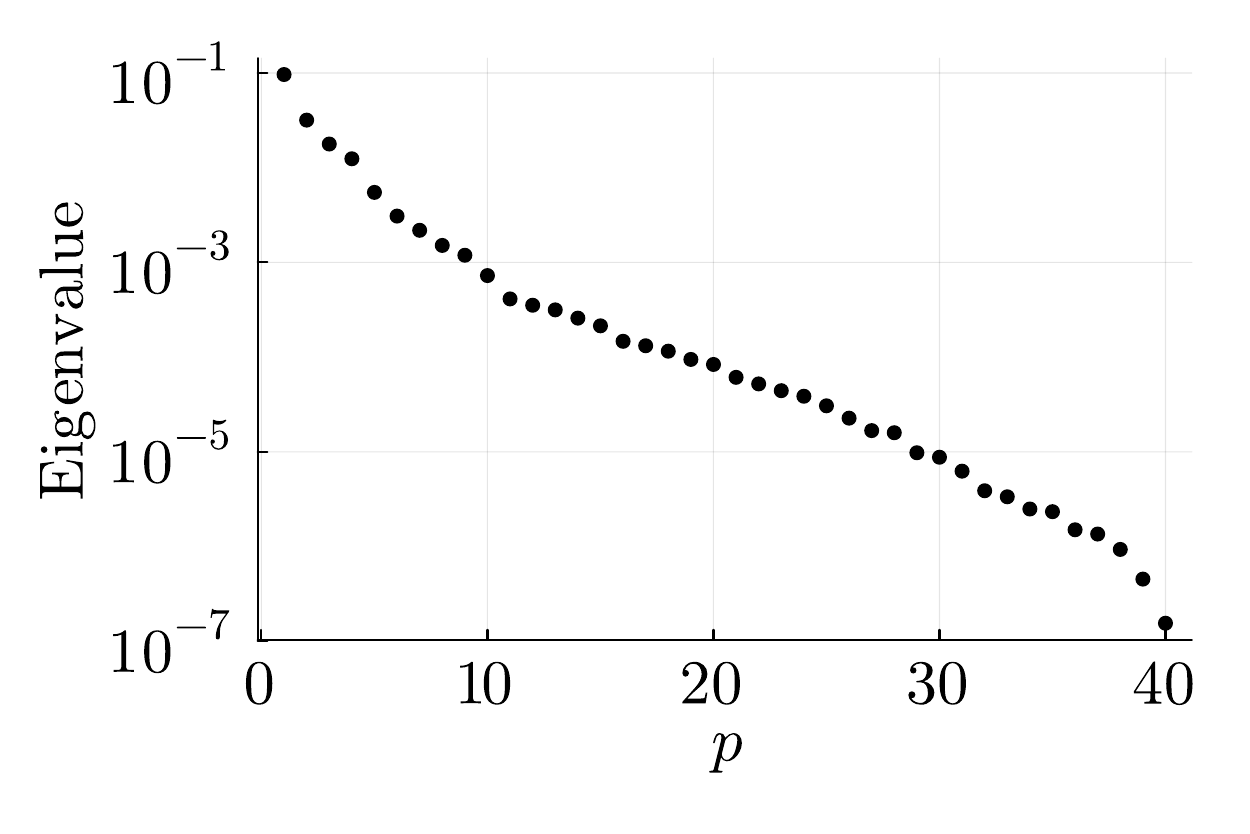} &
        \subfigimgthree[width=\linewidth,valign=t]
        {(b)}
        {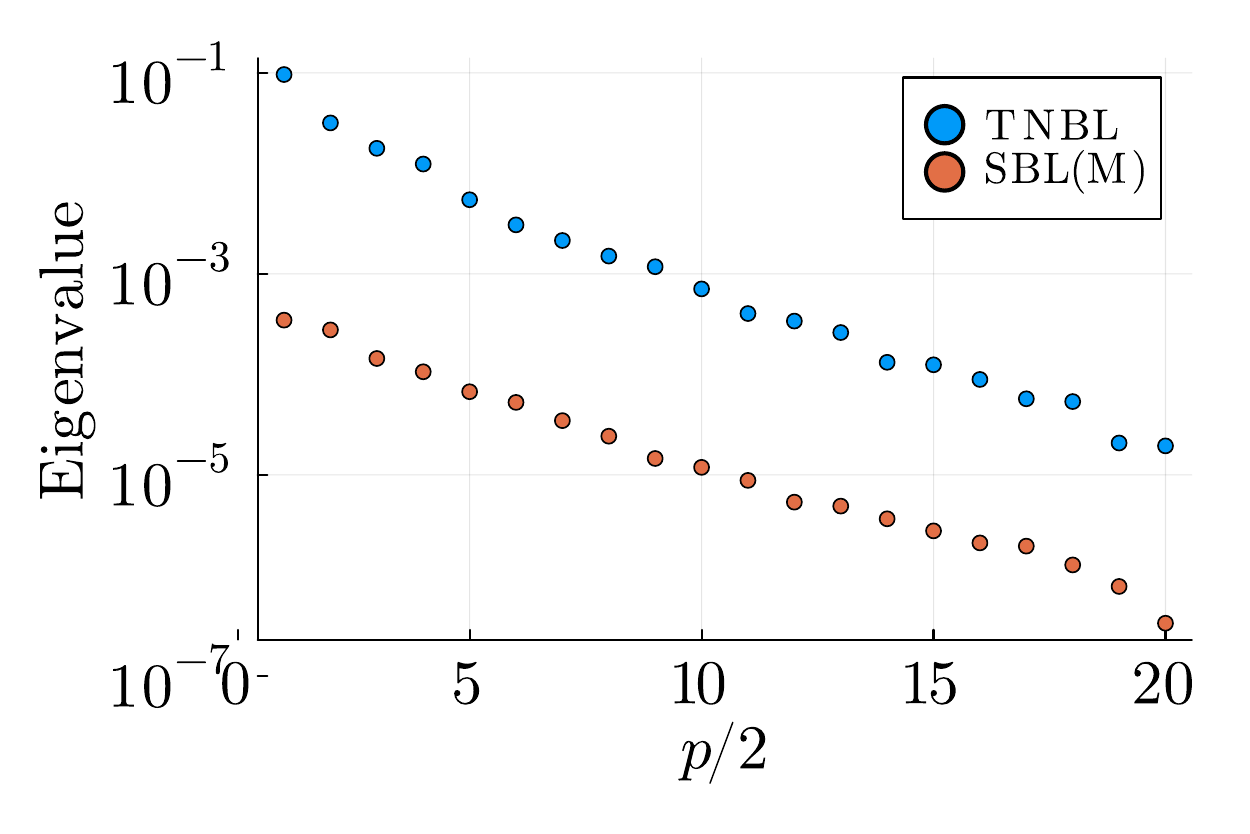} 
    \end{tabular}
    \caption{(a) Eigenvalues of the sample covariance matrix using wind speed data from two stability regimes (TNBL, SBL(M)), (b) Same result as (a) but separated to highlight the relevant eigenvalues for each stability regime.}
    \label{fig:eigenvalues}
\end{figure}

\begin{figure}[ht]
    \centering
    \begin{tabular}
    {@{}p{0.33\linewidth}@{\quad}p{0.33\linewidth}@{\quad}p{0.33\linewidth}@{}}
        \subfigimgthree[width=\linewidth,valign=t]
        {(a)}
        {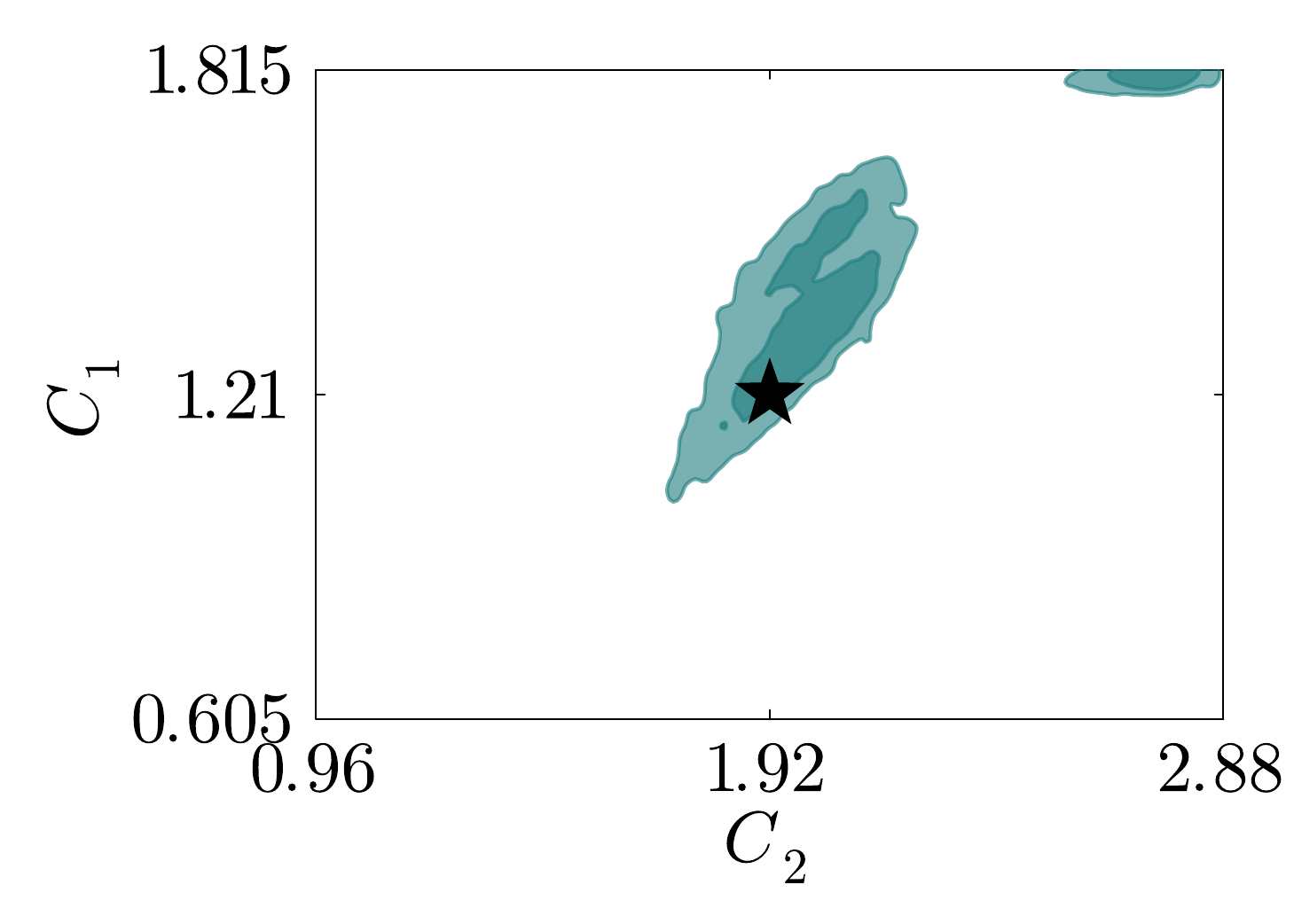} &
        \subfigimgthree[width=\linewidth,valign=t]
        {(b)}
        {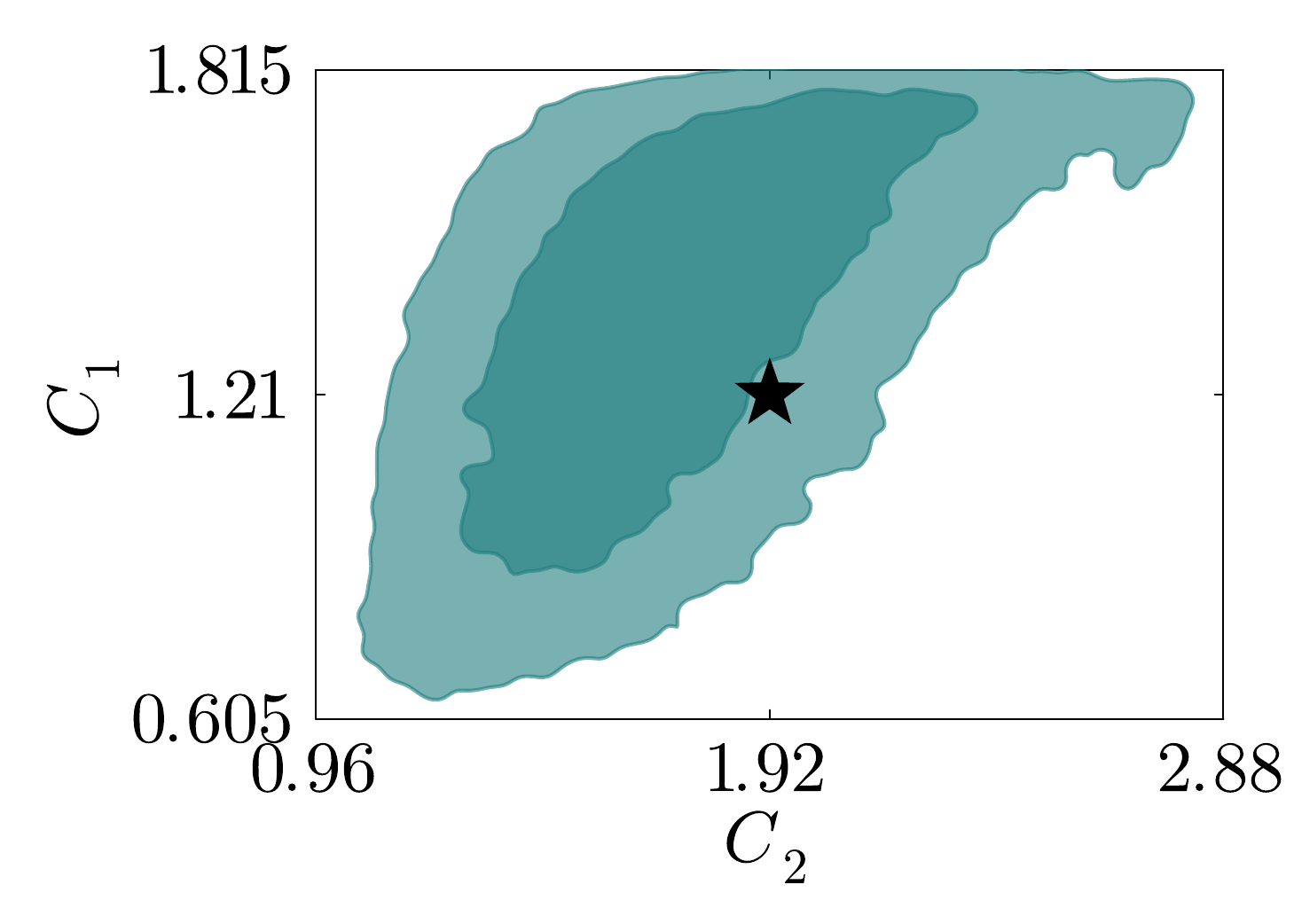} &
        \subfigimgthree[width=\linewidth,valign=t]
        {(c)}
        {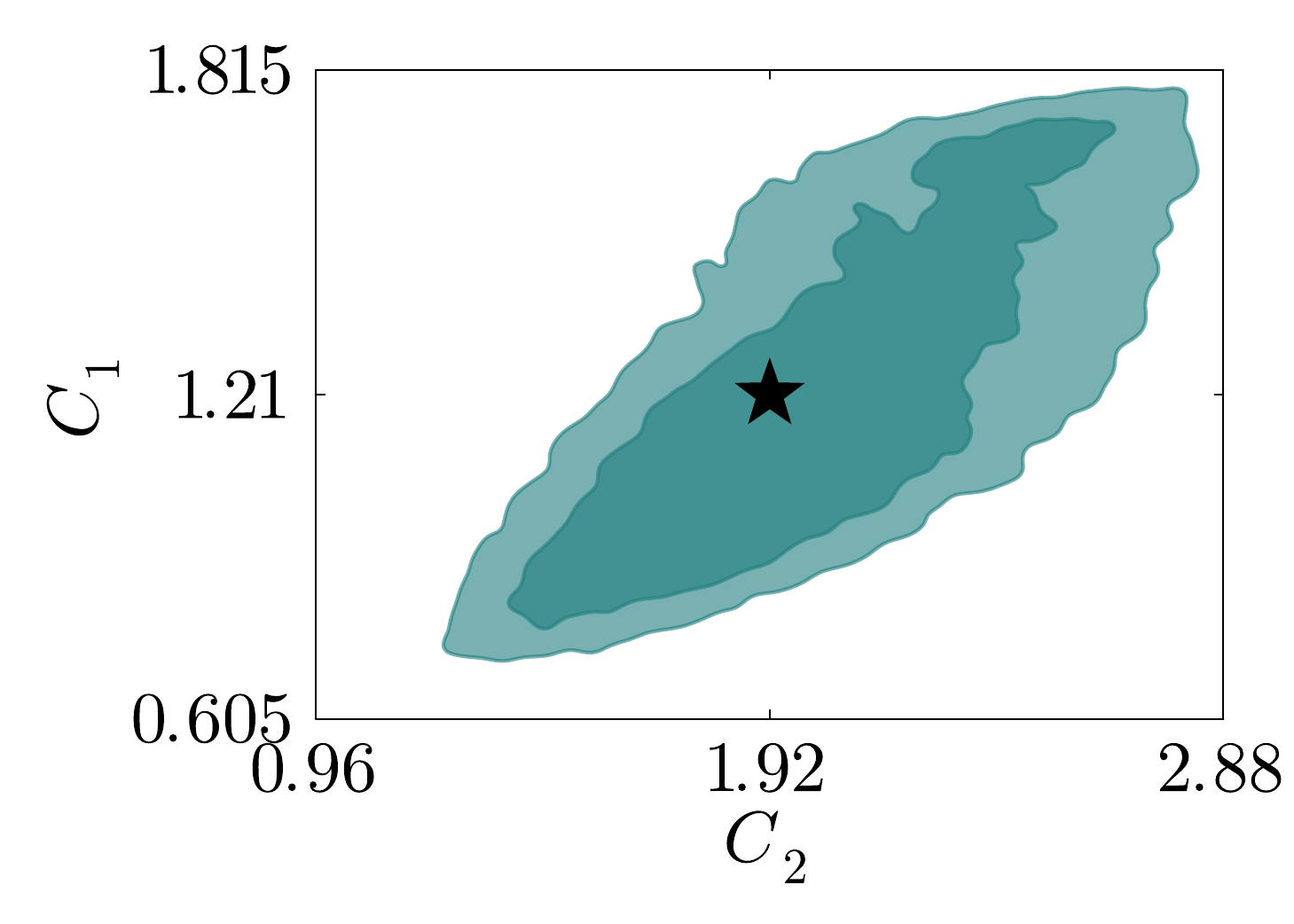}
    \end{tabular}
    \caption{Joint posterior distribution of $C_1$ and $C_2$ inferred from wind speed data in two stability regimes (TNBL, SBL(M)) combined using (a) no regularization, (b) SVD truncation at 99\%, and (c) Ledoit-Wolf shrinkage. 68\% and 95\% HPDR are shown in dark and light shading, respectively.}
    \label{fig:both_regularization}
\end{figure}

The third case of combining data from the two stability regimes presents a unique challenge because of the large spectrum of eigenvalues that span the combined (TNBL, SBL(M)) wind speed data.
Figure~\ref{fig:eigenvalues}a shows the eigenvalues of the sample covariance matrix, while Fig.~\ref{fig:eigenvalues}b separates them by stability regime to highlight the relevant eigenvalues for each case.
Arising from the differing turbulence characteristics, the wide gap in relevant eigenvalues is physically meaningful but leads to a highly ill-conditioned sample covariance matrix ($\sim \mathcal{O}(10^5)$).
Figure~\ref{fig:both_regularization} presents the inferred $C_1$-$C_2$ joint posteriors for the three different regularization scenarios.
First, the posterior with no regularization displays a bimodal property, which is identified as a result of poor mixing of the MCMC sampling chain. 
On the other hand, SVD truncation at 99\% captures approximately 30\% of the singular value energy of the SBL(M) while the Ledoit-Wolf shrinkage method captures approximately 60\%.
Therefore, in using regularization, the inferred posteriors are influenced by the retained dominant eigenvalues in the combined variance.

\begin{figure}[ht]
    \centering
        \includegraphics[width=0.5\textwidth]{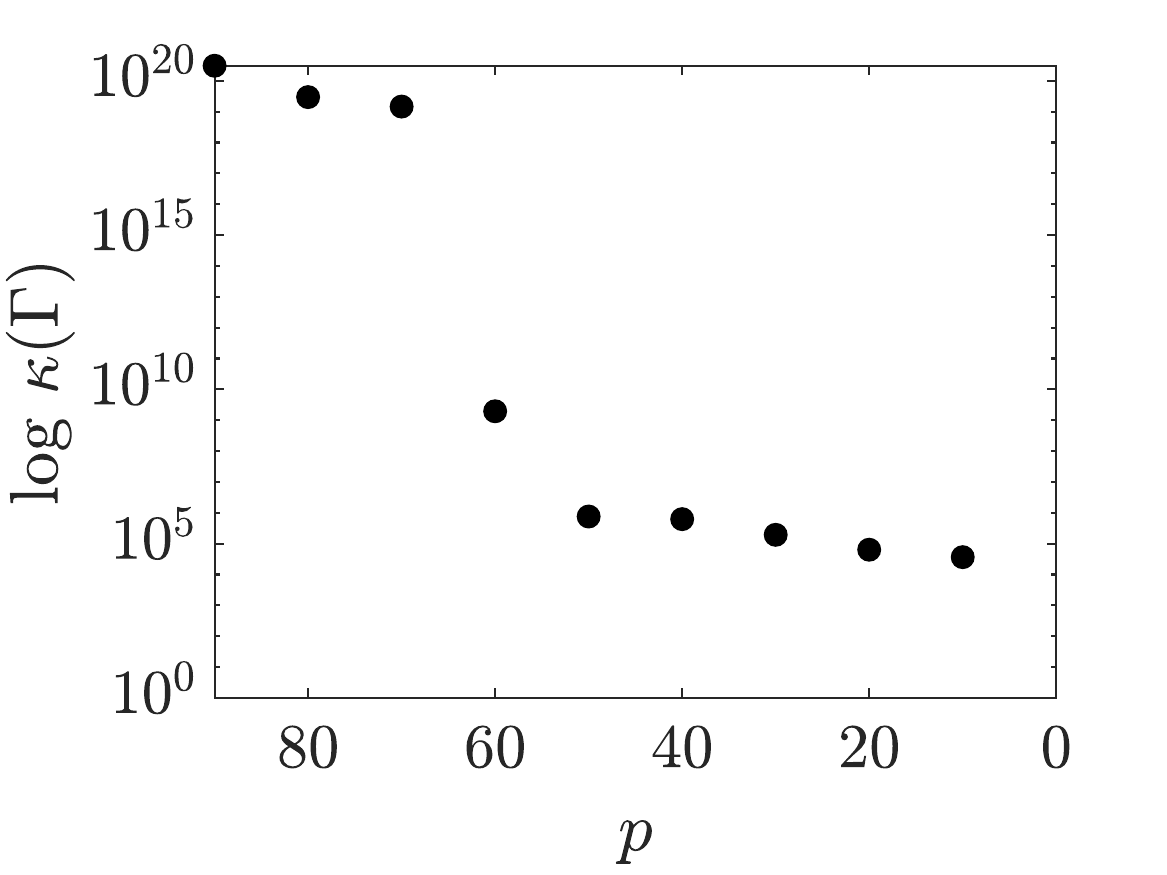}
    \caption{Condition number of the sample covariance matrix $\Gamma$ as a function of covariance matrix dimension $p$ for the two stability regimes (TNBL, SBL(M)) combined.} 
    \label{fig:condition_number}
\end{figure}

A central question is whether regularization is necessary for assimilating data across multiple stability regimes. 
This can be examined through the condition number of the sample covariance matrix.
Figure~\ref{fig:condition_number} shows the condition number as a function of the covariance matrix dimension $p$ for the combined stability regimes.
Although the condition number decreases with reduced $p$, the sharp difference in eigenvalue magnitudes implies that achieving a well-conditioned matrix requires reducing $p \sim \mathcal{O}(1)$ to the point where little information remains.
This highlights the need for regularization to retain a practical dimensionality of the sample covariance matrix while ensuring numerical stability.

\FloatBarrier

\section{Predictions from parameter posteriors conditioned on different fluid flow quantities}
\label{sec:appendix4}

Forward UQ predictions of wind speed, wind direction, and TKE by the posteriors inferred in Sec.~\ref{sec:informative_statistics} are shown.

\begin{figure}[ht]
    \centering
        \includegraphics[width=0.6\textwidth]{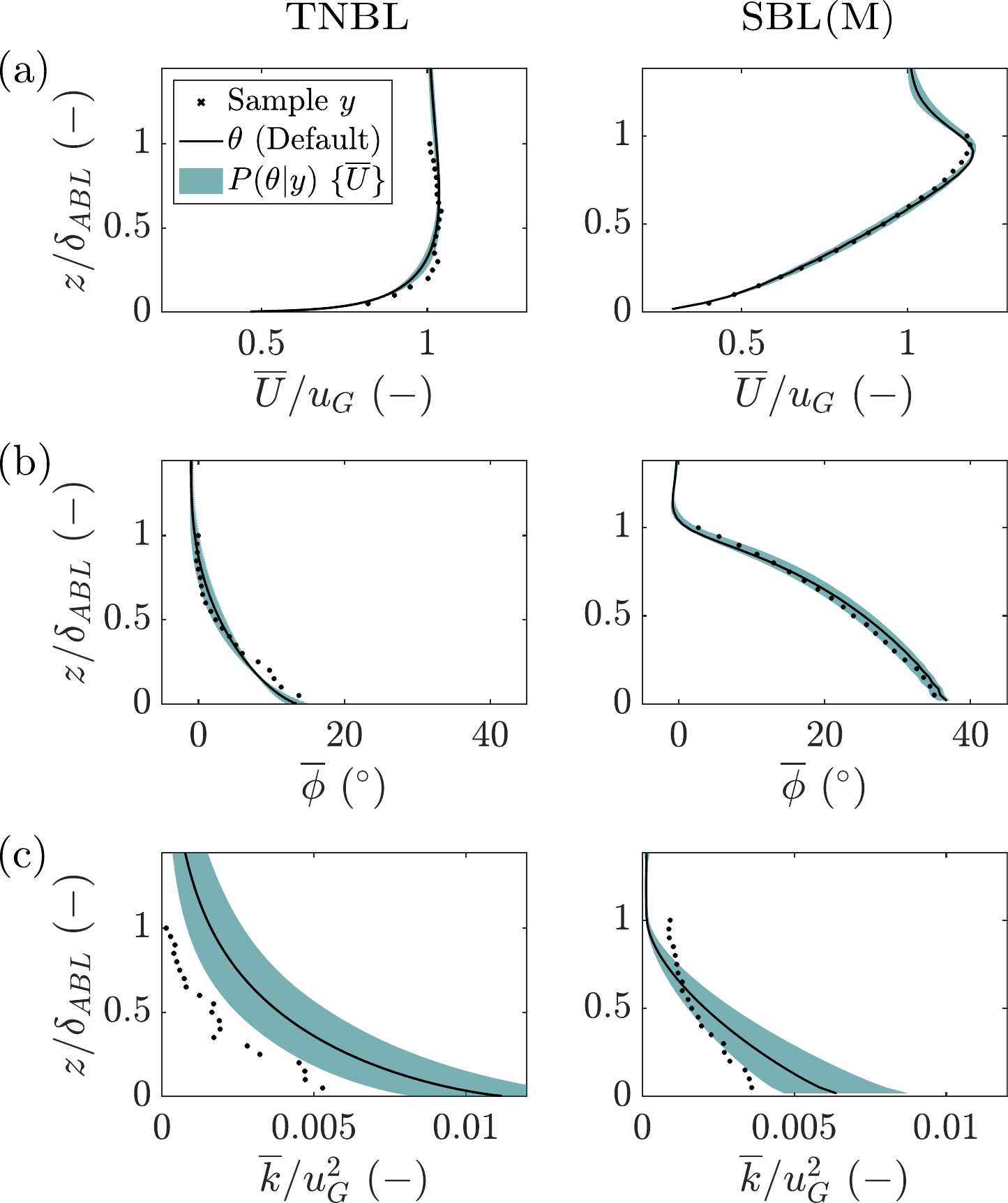}
    \caption{Forward UQ predictions of (a) wind speed, (b) wind direction, and (c) TKE for two regimes, generated by a posterior inferred from LES wind speed data of two stability regimes  (TNBL, SBL(M)) combined.} 
    \label{fig:credible_intervals_U}
\end{figure}

\begin{figure}[ht]
    \centering
        \includegraphics[width=0.6\textwidth]{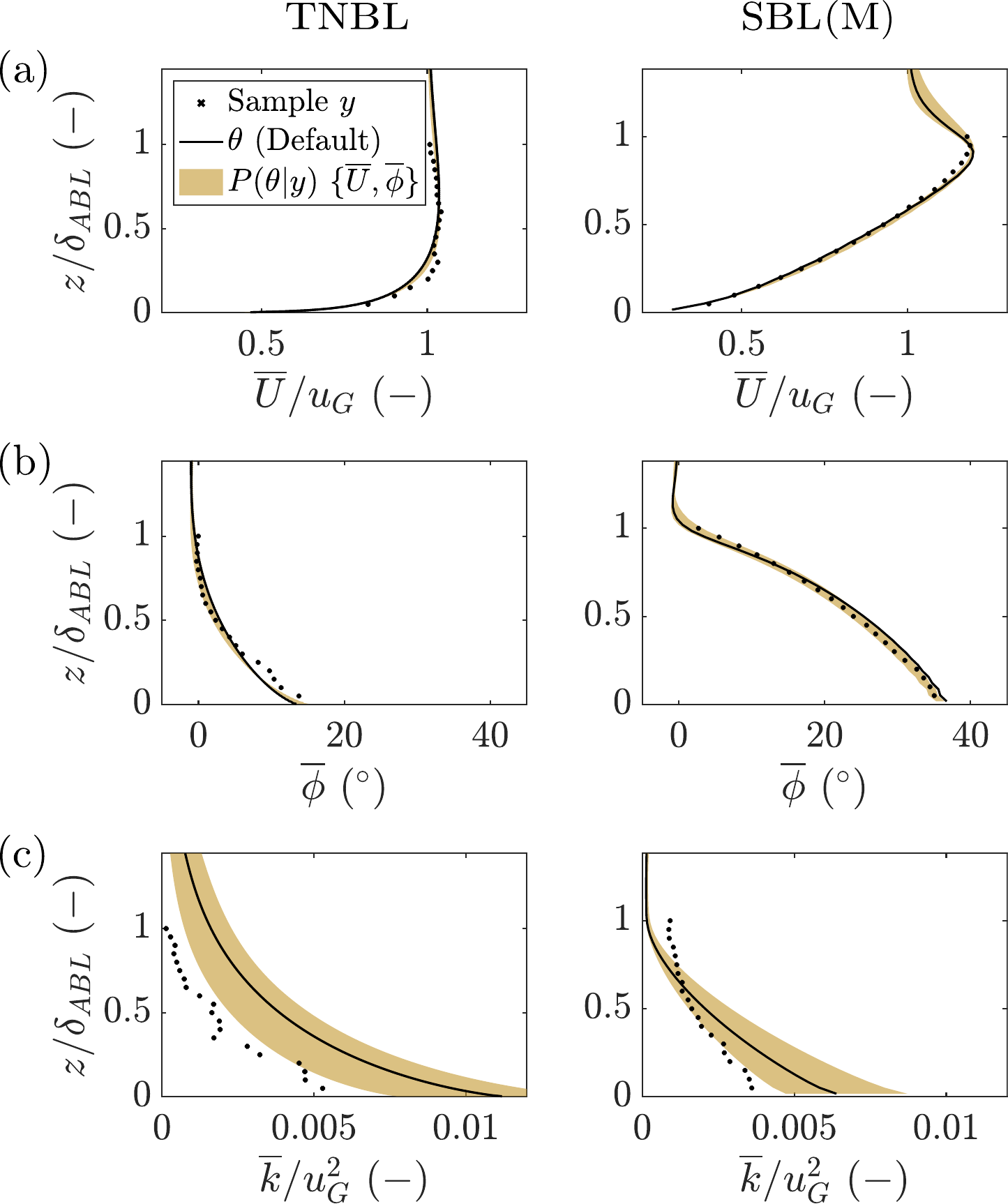}
    \caption{Forward UQ predictions of (a) wind speed, (b) wind direction, and (c) TKE for two regimes, generated by a posterior inferred from LES wind speed and wind direction data of two stability regimes  (TNBL, SBL(M)) combined.} 
    \label{fig:credible_intervals_Udir}
\end{figure}

\begin{figure}[ht]
    \centering
        \includegraphics[width=0.6\textwidth]{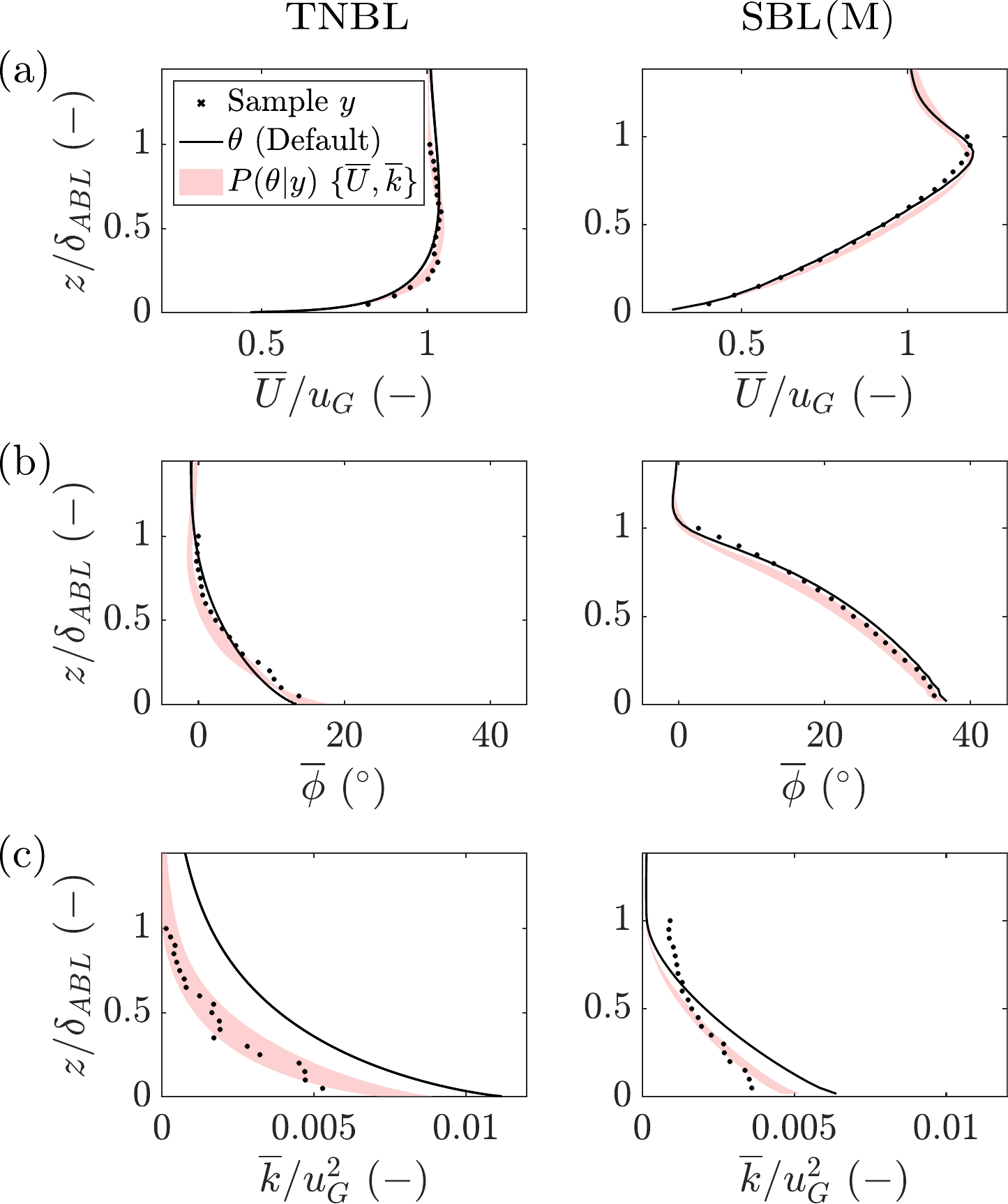}
    \caption{Forward UQ predictions of (a) wind speed, (b) wind direction, and (c) TKE for two regimes, generated by a posterior inferred from LES wind speed and TKE data of two stability regimes  (TNBL, SBL(M)) combined.} 
    \label{fig:credible_intervals_Uk}
\end{figure}

\FloatBarrier

\bibliographystyle{apalike}
\bibliography{references} 

\begin{thebibliography}{}

\bibitem[Andrieu et~al., 2010]{andrieu2010particle}
Andrieu, C., Doucet, A., and Holenstein, R. (2010).
\newblock Particle markov chain monte carlo methods.
\newblock {\em Journal of the Royal Statistical Society Series B: Statistical Methodology}, 72(3):269--342.

\bibitem[Baklanov et~al., 2011]{baklanov2011nature}
Baklanov, A.~A., Grisogono, B., Bornstein, R., Mahrt, L., Zilitinkevich, S.~S., Taylor, P., Larsen, S.~E., Rotach, M.~W., and Fernando, H. (2011).
\newblock The nature, theory, and modeling of atmospheric planetary boundary layers.
\newblock {\em Bulletin of the American Meteorological Society}, 92(2):123--128.

\bibitem[Basu et~al., 2008]{basu2008inconvenient}
Basu, S., Holtslag, A.~A., Van De~Wiel, B.~J., Moene, A.~F., and Steeneveld, G.-J. (2008).
\newblock An inconvenient “truth” about using sensible heat flux as a surface boundary condition in models under stably stratified regimes.
\newblock {\em Acta Geophysica}, 56(1):88--99.

\bibitem[Baungaard et~al., 2024]{baungaard2024simulation}
Baungaard, M., Van Der~Laan, M., Kelly, M., and Hodgson, E. (2024).
\newblock Simulation of a conventionally neutral boundary layer with two-equation urans.
\newblock In {\em Journal of Physics: Conference Series}, volume 2767, page 052013. IOP Publishing.

\bibitem[Beare et~al., 2006]{beare2006intercomparison}
Beare, R.~J., Macvean, M.~K., Holtslag, A.~A., Cuxart, J., Esau, I., Golaz, J.-C., Jimenez, M.~A., Khairoutdinov, M., Kosovic, B., Lewellen, D., et~al. (2006).
\newblock An intercomparison of large-eddy simulations of the stable boundary layer.
\newblock {\em Boundary-Layer Meteorology}, 118(2):247--272.

\bibitem[Biegler et~al., 2011]{biegler2011large}
Biegler, L., Biros, G., Ghattas, O., Heinkenschloss, M., Keyes, D., Mallick, B., Marzouk, Y., Tenorio, L., van Bloemen~Waanders, B., and Willcox, K. (2011).
\newblock {\em Large-Scale Inverse Problems and Quantification of Uncertainty}.
\newblock Wiley Online Library.

\bibitem[Blocken et~al., 2007]{blocken2007cfd}
Blocken, B., Carmeliet, J., and Stathopoulos, T. (2007).
\newblock Cfd evaluation of wind speed conditions in passages between parallel buildings—effect of wall-function roughness modifications for the atmospheric boundary layer flow.
\newblock {\em Journal of Wind Engineering and Industrial Aerodynamics}, 95(9-11):941--962.

\bibitem[Bolstad and Curran, 2016]{bolstad2016introduction}
Bolstad, W.~M. and Curran, J.~M. (2016).
\newblock {\em Introduction to Bayesian statistics}.
\newblock John Wiley \& Sons.

\bibitem[Bonin et~al., 2017]{bonin2017evaluation}
Bonin, T.~A., Choukulkar, A., Brewer, W.~A., Sandberg, S.~P., Weickmann, A.~M., Pichugina, Y.~L., Banta, R.~M., Oncley, S.~P., and Wolfe, D.~E. (2017).
\newblock Evaluation of turbulence measurement techniques from a single doppler lidar.
\newblock {\em Atmospheric Measurement Techniques}, 10(8):3021--3039.

\bibitem[Brooks et~al., 2011]{brooks2011handbook}
Brooks, S., Gelman, A., Jones, G., and Meng, X.-L. (2011).
\newblock {\em Handbook of markov chain monte carlo}.

\bibitem[Calaf et~al., 2010]{calaf2010large}
Calaf, M., Meneveau, C., and Meyers, J. (2010).
\newblock Large eddy simulation study of fully developed wind-turbine array boundary layers.
\newblock {\em Physics of fluids}, 22(1).

\bibitem[Chen and Gorl{\'e}, 2022]{chen2022optimal}
Chen, C. and Gorl{\'e}, C. (2022).
\newblock Optimal temperature sensor placement in buildings with buoyancy-driven natural ventilation using computational fluid dynamics and uncertainty quantification.
\newblock {\em Building and Environment}, 207:108496.

\bibitem[Christensen et~al., 2018]{christensen2018forcing}
Christensen, H.~M., Dawson, A., and Holloway, C.~E. (2018).
\newblock Forcing single-column models using high-resolution model simulations.
\newblock {\em Journal of Advances in Modeling Earth Systems}, 10(8):1833--1857.

\bibitem[Cleary et~al., 2021]{cleary2021calibrate}
Cleary, E. et~al. (2021).
\newblock Calibrate, emulate, sample.
\newblock {\em Journal of Computational Physics}, 424:109716.

\bibitem[Cohen et~al., 2015]{cohen2015review}
Cohen, A.~E., Cavallo, S.~M., Coniglio, M.~C., and Brooks, H.~E. (2015).
\newblock A review of planetary boundary layer parameterization schemes and their sensitivity in simulating southeastern us cold season severe weather environments.
\newblock {\em Weather and forecasting}, 30(3):591--612.

\bibitem[Cotteleer et~al., 2024]{cotteleer2024flow}
Cotteleer, L., Longo, R., Debaste, F., and Parente, A. (2024).
\newblock Flow-based stress-blended eddy simulation: A local rans/les turbulence model for urban flow cfd simulations.
\newblock {\em Results in Engineering}, 21:101679.

\bibitem[Cui et~al., 2016]{cui2016dimension}
Cui, T., Law, K.~J., and Marzouk, Y.~M. (2016).
\newblock Dimension-independent likelihood-informed mcmc.
\newblock {\em Journal of Computational Physics}, 304:109--137.

\bibitem[Cuxart et~al., 2006]{cuxart2006single}
Cuxart, J., Holtslag, A.~A., Beare, R.~J., Bazile, E., Beljaars, A., Cheng, A., Conangla, L., Ek, M., Freedman, F., Hamdi, R., et~al. (2006).
\newblock Single-column model intercomparison for a stably stratified atmospheric boundary layer.
\newblock {\em Boundary-Layer Meteorology}, 118:273--303.

\bibitem[Deardorff, 1970]{deardorff1970numerical}
Deardorff, J.~W. (1970).
\newblock A numerical study of three-dimensional turbulent channel flow at large reynolds numbers.
\newblock {\em Journal of Fluid Mechanics}, 41(2):453--480.

\bibitem[Detering and Etling, 1985]{detering1985application}
Detering, H. and Etling, D. (1985).
\newblock Application of the e-$\varepsilon$ turbulence model to the atmospheric boundary layer.
\newblock {\em Boundary-Layer Meteorology}, 33:113--133.

\bibitem[Doubrawa and Mu{\~n}oz-Esparza, 2020]{doubrawa2020simulating}
Doubrawa, P. and Mu{\~n}oz-Esparza, D. (2020).
\newblock Simulating real atmospheric boundary layers at gray-zone resolutions: How do currently available turbulence parameterizations perform?
\newblock {\em Atmosphere}, 11(4):345.

\bibitem[Dunbar et~al., 2021]{dunbar2021calibration}
Dunbar, O.~R., Garbuno-Inigo, A., Schneider, T., and Stuart, A.~M. (2021).
\newblock Calibration and uncertainty quantification of convective parameters in an idealized gcm.
\newblock {\em Journal of Advances in Modeling Earth Systems}, 13(9):e2020MS002454.

\bibitem[Duraisamy et~al., 2019]{duraisamy2019turbulence}
Duraisamy, K., Iaccarino, G., and Xiao, H. (2019).
\newblock Turbulence modeling in the age of data.
\newblock {\em Annual Review of Fluid Mechanics}, 51:357--377.

\bibitem[Edeling et~al., 2014]{edeling2014bayesian}
Edeling, W.~N., Cinnella, P., Dwight, R.~P., and Bijl, H. (2014).
\newblock Bayesian estimates of parameter variability in the k--$\varepsilon$ turbulence model.
\newblock {\em Journal of Computational Physics}, 258:73--94.

\bibitem[Evans and Stark, 2002]{evans2002inverse}
Evans, S.~N. and Stark, P.~B. (2002).
\newblock Inverse problems as statistics.
\newblock {\em Inverse problems}, 18(4):R55.

\bibitem[Fisher, 1970]{fisher1970statistical}
Fisher, R.~A. (1970).
\newblock Statistical methods for research workers.
\newblock In {\em Breakthroughs in statistics: Methodology and distribution}, pages 66--70. Springer.

\bibitem[Flores and Riley, 2011]{flores2011analysis}
Flores, O. and Riley, J. (2011).
\newblock Analysis of turbulence collapse in the stably stratified surface layer using direct numerical simulation.
\newblock {\em Boundary-Layer Meteorology}, 139:241--259.

\bibitem[Freedman and Jacobson, 2002]{freedman2002transport}
Freedman, A. F.~R. and Jacobson, B. M.~Z. (2002).
\newblock Transport-dissipation analytical solutions to the e-$\epsilon$ turbulence model and their role in predictions of the neutral abl.
\newblock {\em Boundary-layer meteorology}, 102:117--138.

\bibitem[Freedman and Jacobson, 2003]{freedman2003modification}
Freedman, F.~R. and Jacobson, M.~Z. (2003).
\newblock Modification of the standard $\epsilon$-equation for the stable abl through enforced consistency with monin--obukhov similarity theory.
\newblock {\em Boundary-layer meteorology}, 106:383--410.

\bibitem[Garc{\'\i}a-S{\'a}nchez and Gorl{\'e}, 2018]{garcia2018uncertainty}
Garc{\'\i}a-S{\'a}nchez, C. and Gorl{\'e}, C. (2018).
\newblock Uncertainty quantification for microscale cfd simulations based on input from mesoscale codes.
\newblock {\em Journal of Wind Engineering and Industrial Aerodynamics}, 176:87--97.

\bibitem[Geyer, 2011]{geyer2011introduction}
Geyer, C. (2011).
\newblock Introduction to markov chain monte carlo.
\newblock {\em Handbook of markov chain monte carlo}.

\bibitem[Ghate and Lele, 2017]{ghate2017subfilter}
Ghate, A.~S. and Lele, S.~K. (2017).
\newblock Subfilter-scale enrichment of planetary boundary layer large eddy simulation using discrete fourier--gabor modes.
\newblock {\em Journal of Fluid Mechanics}, 819:494--539.

\bibitem[Giacomini and Giometto, 2024]{giacomini2024quantification}
Giacomini, B. and Giometto, M.~G. (2024).
\newblock Quantification of approaching wind uncertainty in flow over realistic plant canopies.
\newblock {\em Boundary-Layer Meteorology}, 190(2):8.

\bibitem[Guillas et~al., 2014]{guillas2014bayesian}
Guillas, S., Glover, N., and Malki-Epshtein, L. (2014).
\newblock Bayesian calibration of the constants of the k--$\varepsilon$ turbulence model for a cfd model of street canyon flow.
\newblock {\em Computer methods in applied mechanics and engineering}, 279:536--553.

\bibitem[Hagen et~al., 1981]{hagen1981simulation}
Hagen, L.~J., Skidmore, E., Miller, P., and Kipp, J. (1981).
\newblock Simulation of effect of wind barriers on airflow.
\newblock {\em Transactions of the ASAE}, 24(4):1002--1008.

\bibitem[Hanjali{\'c} and Launder, 1972]{hanjalic1972reynolds}
Hanjali{\'c}, K. and Launder, B.~E. (1972).
\newblock A reynolds stress model of turbulence and its application to thin shear flows.
\newblock {\em Journal of fluid Mechanics}, 52(4):609--638.

\bibitem[Hansen, 1987]{hansen1987truncated}
Hansen, P.~C. (1987).
\newblock The truncated svd as a method for regularization.
\newblock {\em BIT Numerical Mathematics}, 27:534--553.

\bibitem[Heck and Howland, 2025]{heck2025coriolis}
Heck, K.~S. and Howland, M.~F. (2025).
\newblock Coriolis effects on wind turbine wakes across neutral atmospheric boundary layer regimes.
\newblock {\em Journal of Fluid Mechanics}, 1008:A7.

\bibitem[Holtslag and Boville, 1993]{holtslag1993local}
Holtslag, A. and Boville, B. (1993).
\newblock Local versus nonlocal boundary-layer diffusion in a global climate model.
\newblock {\em Journal of climate}, 6(10):1825--1842.

\bibitem[Holtslag et~al., 2013]{holtslag2013stable}
Holtslag, A., Svensson, G., Baas, P., Basu, S., Beare, B., Beljaars, A., Bosveld, F., Cuxart, J., Lindvall, J., Steeneveld, G., et~al. (2013).
\newblock Stable atmospheric boundary layers and diurnal cycles: challenges for weather and climate models.
\newblock {\em Bulletin of the American Meteorological Society}, 94(11):1691--1706.

\bibitem[Hong et~al., 2006]{hong2006new}
Hong, S.-Y., Noh, Y., and Dudhia, J. (2006).
\newblock A new vertical diffusion package with an explicit treatment of entrainment processes.
\newblock {\em Monthly weather review}, 134(9):2318--2341.

\bibitem[Howland et~al., 2022]{howland2022parameter}
Howland, M.~F., Dunbar, O.~R., and Schneider, T. (2022).
\newblock Parameter uncertainty quantification in an idealized {GCM} with a seasonal cycle.
\newblock {\em JAMES}, 14(3):e2021MS002735.

\bibitem[Howland et~al., 2020]{howland2020influence}
Howland, M.~F., Gonz{\'a}lez, C.~M., Mart{\'\i}nez, J. J.~P., Quesada, J.~B., Larranaga, F.~P., Yadav, N.~K., Chawla, J.~S., and Dabiri, J.~O. (2020).
\newblock Influence of atmospheric conditions on the power production of utility-scale wind turbines in yaw misalignment.
\newblock {\em Journal of Renewable and Sustainable Energy}, 12(6).

\bibitem[Huan and Marzouk, 2013]{huan2013simulation}
Huan, X. and Marzouk, Y.~M. (2013).
\newblock Simulation-based optimal bayesian experimental design for nonlinear systems.
\newblock {\em Journal of Computational Physics}, 232(1):288--317.

\bibitem[Iglesias et~al., 2013]{iglesias2013ensemble}
Iglesias, M.~A., Law, K.~J., and Stuart, A.~M. (2013).
\newblock Ensemble kalman methods for inverse problems.
\newblock {\em Inverse Problems}, 29(4):045001.

\bibitem[Kennedy and O'Hagan, 2001]{kennedy2001bayesian}
Kennedy, M.~C. and O'Hagan, A. (2001).
\newblock Bayesian calibration of computer models.
\newblock {\em Journal of the Royal Statistical Society: Series B (Statistical Methodology)}, 63(3):425--464.

\bibitem[Kim et~al., 1987]{kim1987turbulence}
Kim, J., Moin, P., and Moser, R. (1987).
\newblock Turbulence statistics in fully developed channel flow at low reynolds number.
\newblock {\em Journal of fluid mechanics}, 177:133--166.

\bibitem[Kosovi{\'c} and Curry, 2000]{kosovic2000large}
Kosovi{\'c}, B. and Curry, J.~A. (2000).
\newblock A large eddy simulation study of a quasi-steady, stably stratified atmospheric boundary layer.
\newblock {\em Journal of the atmospheric sciences}, 57(8):1052--1068.

\bibitem[Kumar et~al., 2006a]{kumar2006large}
Kumar, V., Kleissl, J., Meneveau, C., and Parlange, M.~B. (2006a).
\newblock Large-eddy simulation of a diurnal cycle of the atmospheric boundary layer: Atmospheric stability and scaling issues.
\newblock {\em Water resources research}, 42(6).

\bibitem[Kumar et~al., 2006b]{kumar_largeeddy_2006}
Kumar, V., Kleissl, J., Meneveau, C., and Parlange, M.~B. (2006b).
\newblock Large‐eddy simulation of a diurnal cycle of the atmospheric boundary layer: {Atmospheric} stability and scaling issues.
\newblock {\em Water Resources Research}, 42(6):2005WR004651.

\bibitem[Lamberti and Gorl{\'e}, 2018]{lamberti2018uncertainty}
Lamberti, G. and Gorl{\'e}, C. (2018).
\newblock Uncertainty quantification for modeling night-time ventilation in stanford’s y2e2 building.
\newblock {\em Energy and buildings}, 168:319--330.

\bibitem[Lamberti and Gorl{\'e}, 2019]{lamberti2019uncertainty}
Lamberti, G. and Gorl{\'e}, C. (2019).
\newblock Uncertainty quantification for rans predictions of wind loads on buildings.
\newblock In {\em Proceedings of the XV Conference of the Italian Association for Wind Engineering: IN-VENTO 2018 25}, pages 402--412. Springer.

\bibitem[Launder and Spalding, 1974]{LAUNDER1974269}
Launder, B. and Spalding, D. (1974).
\newblock The numerical computation of turbulent flows.
\newblock {\em Computer Methods in Applied Mechanics and Engineering}, 3(2):269--289.

\bibitem[Launder and Spalding, 1983]{launder1983numerical}
Launder, B.~E. and Spalding, D.~B. (1983).
\newblock The numerical computation of turbulent flows.
\newblock In {\em Numerical prediction of flow, heat transfer, turbulence and combustion}, pages 96--116. Elsevier.

\bibitem[Ledoit and Wolf, 2004]{ledoit_well-conditioned_2004}
Ledoit, O. and Wolf, M. (2004).
\newblock A well-conditioned estimator for large-dimensional covariance matrices.
\newblock {\em Journal of Multivariate Analysis}, 88(2):365--411.

\bibitem[Lettau, 1962]{lettau1962theoretical}
Lettau, H. (1962).
\newblock Theoretical wind spirals in the boundary layer of a barotropic atmosphere.
\newblock {\em Beitr. Phys. Atmosph.}, 35:195--212.

\bibitem[Li et~al., 2010]{li2010boundary}
Li, Q., Zhi, L., and Hu, F. (2010).
\newblock Boundary layer wind structure from observations on a 325 m tower.
\newblock {\em Journal of wind engineering and industrial aerodynamics}, 98(12):818--832.

\bibitem[Lilly, 1967]{lilly1967representation}
Lilly, D.~K. (1967).
\newblock The representation of small-scale turbulence in numerical simulation experiments.
\newblock In {\em Proc. IBM Sci. Comput. Symp. on Environmental Science}, pages 195--210.

\bibitem[Liu et~al., 2021]{liu2021geostrophic}
Liu, L., Gadde, S.~N., and Stevens, R.~J. (2021).
\newblock Geostrophic drag law for conventionally neutral atmospheric boundary layers revisited.
\newblock {\em Quarterly journal of the royal meteorological society}, 147(735):847--857.

\bibitem[L{\'o}pez et~al., 2010]{lopez2010optimization}
L{\'o}pez, D., De~Bustos, I.~F., Angulo, C., and Avil{\'e}s, R. (2010).
\newblock Optimization of k-$\varepsilon$turbulence models for incompressible flow around airfoils.
\newblock In {\em 2nd International Conference on Engineering Optimization, Lisbon, Portugal}.

\bibitem[L{\"u}pkes and Schl{\"u}nzen, 1996]{lupkes1996modelling}
L{\"u}pkes, C. and Schl{\"u}nzen, K.~H. (1996).
\newblock Modelling the arctic convective boundary-layer with different turbulence parameterizations.
\newblock {\em Boundary-Layer Meteorology}, 79:107--130.

\bibitem[Machado and Chaboureau, 2015]{machado2015effect}
Machado, L.~A. and Chaboureau, J.-P. (2015).
\newblock Effect of turbulence parameterization on assessment of cloud organization.
\newblock {\em Monthly Weather Review}, 143(8):3246--3262.

\bibitem[Mahrt, 2009]{mahrt2009characteristics}
Mahrt, L. (2009).
\newblock Characteristics of submeso winds in the stable boundary layer.
\newblock {\em Boundary-layer meteorology}, 130:1--14.

\bibitem[Mahrt and Bou-Zeid, 2020]{mahrt2020non}
Mahrt, L. and Bou-Zeid, E. (2020).
\newblock Non-stationary boundary layers.
\newblock {\em Boundary-Layer Meteorology}, 177(2):189--204.

\bibitem[Marzouk and Xiu, 2009]{marzouk2009stochastic}
Marzouk, Y. and Xiu, D. (2009).
\newblock A stochastic collocation approach to bayesian inference in inverse problems.

\bibitem[Mellor and Yamada, 1982]{mellor1982development}
Mellor, G.~L. and Yamada, T. (1982).
\newblock Development of a turbulence closure model for geophysical fluid problems.
\newblock {\em Reviews of Geophysics}, 20(4):851--875.

\bibitem[Menter, 1994]{menter1994two}
Menter, F.~R. (1994).
\newblock Two-equation eddy-viscosity turbulence models for engineering applications.
\newblock {\em AIAA journal}, 32(8):1598--1605.

\bibitem[Munger et~al., 2011]{munger2011measurement}
Munger, J.~W., Loescher, H.~W., and Luo, H. (2011).
\newblock Measurement, tower, and site design considerations.
\newblock In {\em Eddy covariance: a practical guide to measurement and data analysis}, pages 21--58. Springer.

\bibitem[Nakanishi and Niino, 2009]{nakanishi2009development}
Nakanishi, M. and Niino, H. (2009).
\newblock Development of an improved turbulence closure model for the atmospheric boundary layer.
\newblock {\em Journal of the Meteorological Society of Japan. Ser. II}, 87(5):895--912.

\bibitem[Nicoud and Ducros, 1999]{nicoud1999subgrid}
Nicoud, F. and Ducros, F. (1999).
\newblock Subgrid-scale stress modelling based on the square of the velocity gradient tensor.
\newblock {\em Flow, turbulence and Combustion}, 62(3):183--200.

\bibitem[Nicoud et~al., 2011]{nicoud2011using}
Nicoud, F., Toda, H.~B., Cabrit, O., Bose, S., and Lee, J. (2011).
\newblock Using singular values to build a subgrid-scale model for large eddy simulations.
\newblock {\em Phys. Fluids}, 23(8):085106.

\bibitem[Oliver et~al., 2024]{oliver2024calibrateemulatesample}
Oliver, R., Bieli, M., Garbuno-I{\~n}igo, A., Howland, M., de~Souza, A.~N., Mansfield, L.~A., Wagner, G.~L., Efrat-Henrici, N., et~al. (2024).
\newblock Calibrateemulatesample. jl: Accelerated parametric uncertainty quantification.
\newblock {\em Journal of Open Source Software}, 9(97):6372.

\bibitem[Oliver and Moser, 2011]{oliver2011bayesian}
Oliver, T.~A. and Moser, R.~D. (2011).
\newblock Bayesian uncertainty quantification applied to rans turbulence models.
\newblock In {\em Journal of Physics: Conference Series}, volume 318, page 042032. IOP Publishing.

\bibitem[Papadimitriou and Papadimitriou, 2015]{papadimitriou2015optimal}
Papadimitriou, D.~I. and Papadimitriou, C. (2015).
\newblock Optimal sensor placement for the estimation of turbulence model parameters in cfd.
\newblock {\em International Journal for Uncertainty Quantification}, 5(6).

\bibitem[Parente et~al., 2011]{parente2011comprehensive}
Parente, A., Gorl{\'e}, C., van Beeck, J., and Benocci, C. (2011).
\newblock A comprehensive modelling approach for the neutral atmospheric boundary layer: consistent inflow conditions, wall function and turbulence model.
\newblock {\em Boundary-layer meteorology}, 140(3):411--428.

\bibitem[Peherstorfer et~al., 2016]{peherstorfer2016multifidelity}
Peherstorfer, B., Cui, T., Marzouk, Y., and Willcox, K. (2016).
\newblock Multifidelity importance sampling.
\newblock {\em Computer Methods in Applied Mechanics and Engineering}, 300:490--509.

\bibitem[Peherstorfer et~al., 2018]{peherstorfer2018survey}
Peherstorfer, B., Willcox, K., and Gunzburger, M. (2018).
\newblock Survey of multifidelity methods in uncertainty propagation, inference, and optimization.
\newblock {\em Siam Review}, 60(3):550--591.

\bibitem[Perrin et~al., 2007]{perrin2007effect}
Perrin, D., McMahon, N., Crane, M., Ruskin, H.~J., Crane, L., and Hurley, B. (2007).
\newblock The effect of a meteorological tower on its top-mounted anemometer.
\newblock {\em Applied Energy}, 84(4):413--424.

\bibitem[Platteeuw et~al., 2008]{platteeuw2008uncertainty}
Platteeuw, P., Loeven, G., and Bijl, H. (2008).
\newblock Uncertainty quantification applied to the k-epsilon model of turbulence using the probabilistic collocation method.
\newblock In {\em 49th AIAA/ASME/ASCE/AHS/ASC Structures, Structural Dynamics, and Materials Conference, 16th AIAA/ASME/AHS Adaptive Structures Conference, 10th AIAA Non-Deterministic Approaches Conference, 9th AIAA Gossamer Spacecraft Forum, 4th AIAA Multidisciplinary Design Optimization Specialists Conference}, page 2150.

\bibitem[Poroseva and Iaccarino, 2001]{poroseva2001simulating}
Poroseva, S. and Iaccarino, G. (2001).
\newblock Simulating separated flows using the ke model.
\newblock {\em Annual Research Briefs}, 2001:375--384.

\bibitem[Ramon et~al., 2020]{ramon2020tall}
Ramon, J., Lled{\'o}, L., P{\'e}rez-Zan{\'o}n, N., Soret, A., and Doblas-Reyes, F.~J. (2020).
\newblock The tall tower dataset: a unique initiative to boost wind energy research.
\newblock {\em Earth System Science Data}, 12(1):429--439.

\bibitem[Ray et~al., 2018]{ray2018learning}
Ray, J., Lefantzi, S., Arunajatesan, S., and Dechant, L. (2018).
\newblock Learning an eddy viscosity model using shrinkage and bayesian calibration: A jet-in-crossflow case study.
\newblock {\em ASCE-ASME Journal of Risk and Uncertainty in Engineering Systems, Part B: Mechanical Engineering}, 4(1):011001.

\bibitem[Richards and Hoxey, 1993]{richards1993appropriate}
Richards, P. and Hoxey, R. (1993).
\newblock Appropriate boundary conditions for computational wind engineering models using the k-epsilon turbulence model.
\newblock {\em Journal of wind engineering and industrial aerodynamics}, 46:145--153.

\bibitem[Richards and Norris, 2019]{richards2019appropriate}
Richards, P.~J. and Norris, S.~E. (2019).
\newblock Appropriate boundary conditions for computational wind engineering: Still an issue after 25 years.
\newblock {\em Journal of Wind Engineering and Industrial Aerodynamics}, 190:245--255.

\bibitem[Schalkwijk et~al., 2015]{schalkwijk2015year}
Schalkwijk, J., Jonker, H.~J., Siebesma, A.~P., and Bosveld, F.~C. (2015).
\newblock A year-long large-eddy simulation of the weather over cabauw: An overview.
\newblock {\em Monthly Weather Review}, 143(3):828--844.

\bibitem[Schmelzer et~al., 2020]{schmelzer2020discovery}
Schmelzer, M., Dwight, R.~P., and Cinnella, P. (2020).
\newblock Discovery of algebraic reynolds-stress models using sparse symbolic regression.
\newblock {\em Flow, Turbulence and Combustion}, 104:579--603.

\bibitem[Semaan, 2017]{semaan2017optimal}
Semaan, R. (2017).
\newblock Optimal sensor placement using machine learning.
\newblock {\em Computers \& Fluids}, 159:167--176.

\bibitem[Shin and Hong, 2011]{shin2011intercomparison}
Shin, H.~H. and Hong, S.-Y. (2011).
\newblock Intercomparison of planetary boundary-layer parametrizations in the wrf model for a single day from cases-99.
\newblock {\em Boundary-Layer Meteorology}, 139:261--281.

\bibitem[Shin et~al., 2024]{shin2024multi}
Shin, Y., Chan, M., Wang, J., Zahtila, T., Gorle, C., Iaccarino, G., and Howland, M. (2024).
\newblock Multi-fidelity modeling and uncertainty quantification of heterogeneous roughness.
\newblock {\em Proceedings of the Summer Program, Center for Turbulence Research, Stanford University}.

\bibitem[Singh and Duraisamy, 2016]{singh2016using}
Singh, A.~P. and Duraisamy, K. (2016).
\newblock Using field inversion to quantify functional errors in turbulence closures.
\newblock {\em Physics of Fluids}, 28(4).

\bibitem[Sommerfeld et~al., 2019]{sommerfeld_improving_2019}
Sommerfeld, M., Dörenkämper, M., Steinfeld, G., and Crawford, C. (2019).
\newblock Improving mesoscale wind speed forecasts using lidar-based observation nudging for airborne wind energy systems.
\newblock {\em Wind Energy Science}, 4(4):563--580.

\bibitem[Stoll et~al., 2020]{stoll2020large}
Stoll, R., Gibbs, J.~A., Salesky, S.~T., Anderson, W., and Calaf, M. (2020).
\newblock Large-eddy simulation of the atmospheric boundary layer.
\newblock {\em Boundary-Layer Meteorology}, 177:541--581.

\bibitem[Stull, 2012]{stull2012introduction}
Stull, R.~B. (2012).
\newblock {\em An introduction to boundary layer meteorology}.
\newblock Springer Science \& Business Media.

\bibitem[Sullivan, 2015]{sullivan2015introduction}
Sullivan, T.~J. (2015).
\newblock {\em Introduction to uncertainty quantification}, volume~63.
\newblock Springer.

\bibitem[Temel and van Beeck, 2017]{temel2017two}
Temel, O. and van Beeck, J. (2017).
\newblock Two-equation eddy viscosity models based on the monin--obukhov similarity theory.
\newblock {\em Applied Mathematical Modelling}, 42:1--16.

\bibitem[Tennekes, 1973]{tennekes1973similarity}
Tennekes, H. (1973).
\newblock Similarity laws and scale relations in planetary boundary layers.
\newblock In {\em Workshop on Micrometeorology}, volume 177, page 216. Amer. Meteor. Soc Boston.

\bibitem[Tierney, 1994]{tierney1994markov}
Tierney, L. (1994).
\newblock Markov chains for exploring posterior distributions.
\newblock {\em the Annals of Statistics}, pages 1701--1728.

\bibitem[Trevi{\~n}o and Andreas, 2000]{trevino2000averaging}
Trevi{\~n}o, G. and Andreas, E.~L. (2000).
\newblock Averaging intervals for spectral analysis of nonstationary turbulence.
\newblock {\em Boundary-layer meteorology}, 95:231--247.

\bibitem[Troen and Mahrt, 1986]{troen1986simple}
Troen, I. and Mahrt, L. (1986).
\newblock A simple model of the atmospheric boundary layer; sensitivity to surface evaporation.
\newblock {\em Boundary-Layer Meteorology}, 37(1):129--148.

\bibitem[Trucano et~al., 2006]{trucano2006calibration}
Trucano, T.~G., Swiler, L.~P., Igusa, T., Oberkampf, W.~L., and Pilch, M. (2006).
\newblock Calibration, validation, and sensitivity analysis: What's what.
\newblock {\em Reliability Engineering \& System Safety}, 91(10-11):1331--1357.

\bibitem[Turgeon et~al., 2001]{turgeon2001application}
Turgeon, {\'E}., Pelletier, D., and Borggaard, J. (2001).
\newblock Application of a sensitivity equation method to the k-epsilon model of turbulence.
\newblock In {\em 15th AIAA computational fluid dynamics conference}, page 2534.

\bibitem[Turutoglu and Cadirci, 2023]{turutoglu2023improvement}
Turutoglu, C. and Cadirci, S. (2023).
\newblock Improvement of standard k-epsilon turbulence model for round free jets by adjusting closure coefficients.
\newblock In {\em ASME International Mechanical Engineering Congress and Exposition}, volume 87660, page V009T10A032. American Society of Mechanical Engineers.

\bibitem[van~de Schoot et~al., 2021]{van2021bayesian}
van~de Schoot, R., Depaoli, S., King, R., Kramer, B., M{\"a}rtens, K., Tadesse, M.~G., Vannucci, M., Gelman, A., Veen, D., Willemsen, J., et~al. (2021).
\newblock Bayesian statistics and modelling.
\newblock {\em Nature Reviews Methods Primers}, 1(1):1.

\bibitem[Van~de Wiel et~al., 2003]{van2003intermittent}
Van~de Wiel, B., Moene, A., Hartogensis, O., De~Bruin, H., and Holtslag, A. (2003).
\newblock Intermittent turbulence in the stable boundary layer over land. part iii: A classification for observations during cases-99.
\newblock {\em Journal of the atmospheric sciences}, 60(20):2509--2522.

\bibitem[van~der Laan et~al., 2017]{van2017new}
van~der Laan, M.~P., Kelly, M.~C., and S{\o}rensen, N.~N. (2017).
\newblock A new k-epsilon model consistent with monin--obukhov similarity theory.
\newblock {\em Wind Energy}, 20(3):479--489.

\bibitem[Wehrle et~al., 2024]{wehrle2024introducing}
Wehrle, J., Jung, C., Giometto, M., Christen, A., and Schindler, D. (2024).
\newblock Introducing new morphometric parameters to improve urban canopy air flow modeling: A cfd to machine-learning study in real urban environments.
\newblock {\em Urban Climate}, 58:102173.

\bibitem[Weng and Taylor, 2003]{weng2003modelling}
Weng, W. and Taylor, P.~A. (2003).
\newblock On modelling the one-dimensional atmospheric boundary layer.
\newblock {\em Boundary-layer meteorology}, 107:371--400.

\bibitem[Wu et~al., 2018]{wu2018physics}
Wu, J.-L., Xiao, H., and Paterson, E. (2018).
\newblock Physics-informed machine learning approach for augmenting turbulence models: A comprehensive framework.
\newblock {\em Physical Review Fluids}, 3(7):074602.

\bibitem[Wyngaard, 2010]{wyngaard2010turbulence}
Wyngaard, J.~C. (2010).
\newblock {\em Turbulence in the Atmosphere}.
\newblock Cambridge University Press.

\bibitem[Xiao et~al., 2019]{xiao2019reduced}
Xiao, D., Heaney, C., Mottet, L., Fang, F., Lin, W., Navon, I., Guo, Y., Matar, O., Robins, A., and Pain, C. (2019).
\newblock A reduced order model for turbulent flows in the urban environment using machine learning.
\newblock {\em Building and Environment}, 148:323--337.

\bibitem[Xiao and Cinnella, 2019]{xiao2019quantification}
Xiao, H. and Cinnella, P. (2019).
\newblock Quantification of model uncertainty in rans simulations: A review.
\newblock {\em Progress in Aerospace Sciences}, 108:1--31.

\bibitem[Xu and Taylor, 1997]{xu1997turbulence}
Xu, D. and Taylor, P.~A. (1997).
\newblock On turbulence closure constants for atmospheric boundary-layer modelling: neutral stratification.
\newblock {\em Boundary-Layer Meteorology}, 84:267--287.

\bibitem[Yang et~al., 2009]{yang2009new}
Yang, Y., Gu, M., Chen, S., and Jin, X. (2009).
\newblock New inflow boundary conditions for modelling the neutral equilibrium atmospheric boundary layer in computational wind engineering.
\newblock {\em Journal of Wind Engineering and Industrial Aerodynamics}, 97(2):88--95.

\bibitem[Zhang et~al., 2020]{zhang2020evaluation}
Zhang, C., Wang, Y., and Xue, M. (2020).
\newblock Evaluation of an e--$\varepsilon$ and three other boundary layer parameterization schemes in the wrf model over the southeast pacific and the southern great plains.
\newblock {\em Monthly Weather Review}, 148(3):1121--1145.

\bibitem[Zhao et~al., 2022]{zhao2022generalizability}
Zhao, R., Liu, S., Liu, J., Jiang, N., and Chen, Q. (2022).
\newblock Generalizability evaluation of k-$\varepsilon$ models calibrated by using ensemble kalman filtering for urban airflow and airborne contaminant dispersion.
\newblock {\em Building and Environment}, 212:108823.

\bibitem[Zilitinkevich et~al., 2007]{zilitinkevich2007further}
Zilitinkevich, S., Esau, I., and Baklanov, A. (2007).
\newblock Further comments on the equilibrium height of neutral and stable planetary boundary layers.
\newblock {\em Quarterly Journal of the Royal Meteorological Society: A journal of the atmospheric sciences, applied meteorology and physical oceanography}, 133(622):265--271.

\bibitem[Zonato et~al., 2022]{zonato2022new}
Zonato, A., Martilli, A., Jimenez, P.~A., Dudhia, J., Zardi, D., and Giovannini, L. (2022).
\newblock A new k--$\varepsilon$ turbulence parameterization for mesoscale meteorological models.
\newblock {\em Monthly Weather Review}, 150(8):2157--2174.

\end{thebibliography}

\end{document}